\documentclass[preprint,showpacs,titlepage,aps,prd,tightenlines,byrevtex,nofootinbib]{revtex4}

\usepackage{graphicx, amsmath}

\begin{document}


\title{Parameter Degeneracy in Neutrino Oscillation \\ 
--- Solution Network and Structural Overview ---}

\author{Hisakazu Minakata}
\email{minakata@tmu.ac.jp}
\author{Shoichi Uchinami}
\email{uchinami@phys.metro-u.ac.jp}
\affiliation{Department of Physics, Tokyo Metropolitan University \\
1-1 Minami-Osawa, Hachioji, Tokyo 192-0397, Japan}

\date{April 27, 2010}

\vglue 1.4cm

\begin{abstract}

It is known that there is a phenomenon called ``parameter degeneracy'' in 
neutrino oscillation measurement of lepton mixing parameters;  
A set of the oscillation probabilities, e.g., 
$P(\nu_{\mu} \rightarrow \nu_e)$ and its CP-conjugate 
$P(\bar{\nu}_{\mu} \rightarrow  \bar{\nu}_e )$ at a particular neutrino energy 
does not determine uniquely the values of $\theta_{13}$ and $\delta$. 
With use of the approximate form of the oscillation probability \'a la Cervera {\it et al.}, 
a complete analysis of the eightfold parameter degeneracy is presented. 
We propose a unified view of the various types of the degeneracy as invariance 
of the oscillation probabilities under discrete mappings of the mixing parameters. 
Explicit form of the mapping is obtained either by symmetry argument, or by deriving 
exact analytic expressions of all the degeneracy solutions for a given true solution. 
Due to the one-to-one mapping structure the degeneracy solutions are shown to 
form a network. 
We extend our analysis into the parameter degeneracy in T- and CPT-conjugate 
measurement as well as to the setup with the golden and the silver channels, 
$P(\nu_e  \rightarrow \nu_{\mu})$ and $P(\nu_e  \rightarrow \nu_{\tau})$. 
Some characteristic features of the degeneracy solutions in CP-conjugate 
measurement, in particular their energy dependences, 
are illuminated by utilizing the explicit analytic solutions.  

\end{abstract}

\maketitle

\section{Introduction}

After establishing the neutrino masses and the lepton flavor mixing \cite{MNS} 
by the atmospheric \cite{atm}, the solar \cite{solar}, and the reactor 
experiments \cite{reactor}, 
which is further supported by the accelerator experiments \cite{K2K,MINOS}, 
there seems to exist a consensus that the next step is to measure $\theta_{13}$ 
and $\delta$, the remaining unknowns in the MNS matrix, and to determine 
the neutrino mass hierarchy. 
It was proposed that if $\theta_{13}$ is relatively large an intense neutrino beam 
from nuclear reactors can be used to measure it by using the near-far two-detector 
setting \cite{Kr2Det,MSYIS03}.
Alternatively, or complimentarily, the accelerator search for nonzero $\theta_{13}$ 
has advantage of potential possibility of extending it to CP violation search. 
The reactor \cite{DCHOOZ,Daya-Bay,RENO,reactor-white} and 
the accelerator experiments \cite{T2K,NOVA} 
are either ongoing or in construction to look for effects of nonzero $\theta_{13}$. 

It is well known, however, that detection of CP violation due to the lepton 
Kobayashi-Maskawa phase $\delta$ \cite{KM}, being the genuine three flavor effect, 
is suppressed by the two small factors, the ratio $\Delta m^2_{21} / \Delta m^2_{31}$ 
\cite{KamLAND,SNO,K2K,MINOS,SKatm}
and the value of $\theta_{13}$ bounded from above 
\cite{CHOOZ,Palo-Verde,K2K-bound,MINOS-bound}. 
Therefore, high precision experiments are inevitably required to measure 
CP violating phase $\delta$. 
Once precision measurement becomes the necessity, the experiment is better 
characterized as a simultaneous determination of $\theta_{13}$ and $\delta$. 
It is because even though one enjoys prior crude knowledges of magnitude of 
$\theta_{13}$ (assuming it relatively large), which certainly propels CP measurement, 
the required precision for detecting tiny effects of $\delta$ necessitates simultaneous measurement of $\theta_{13}$ in a precision far beyond the previously achieved ones.

\begin{figure}[bhtp]
\begin{center}
\vglue 0.3cm
\includegraphics[bb=0 0 240 237 , clip, width=0.36\textwidth]{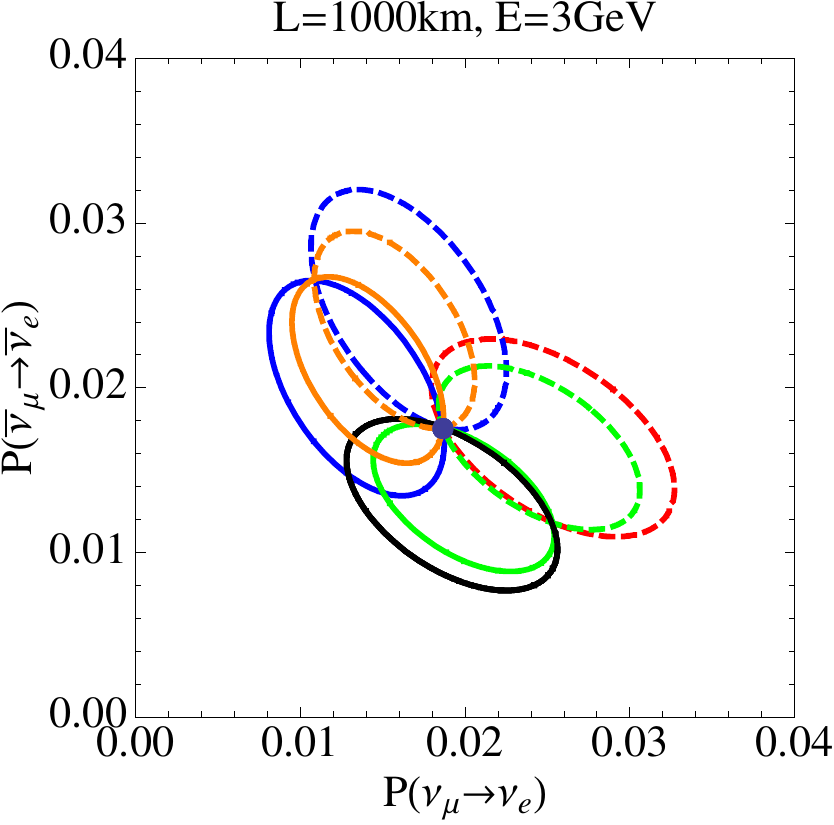}
\hspace{2mm}
\includegraphics[bb=0 0 240 172 , clip, width=0.40\textwidth]{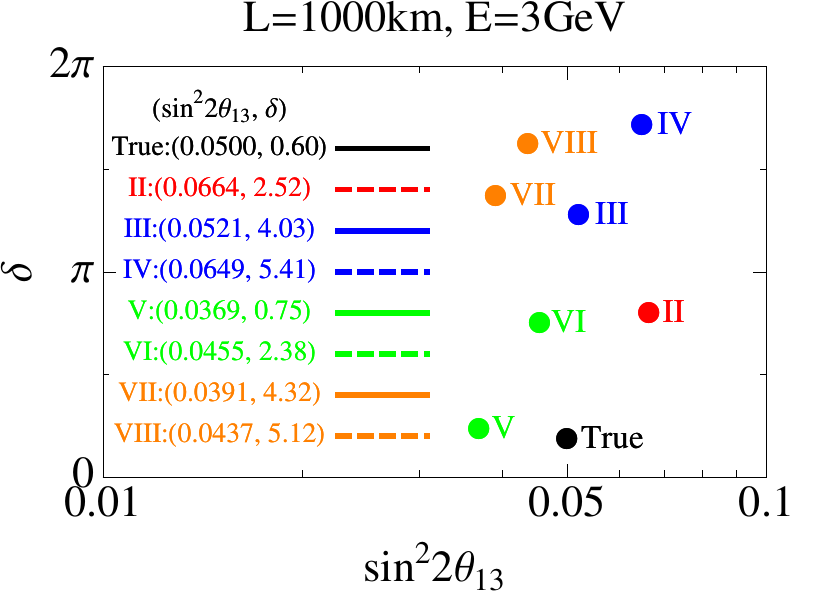}
\end{center}
\vglue -0.5cm
\caption{
Left panel: 
An illustrative example of the eightfold degeneracy is represented pictorially 
(as first appeared in \cite{MNtaup01}) in terms of the bi-probability plot in 
$P - P^{CP}$ space \cite{MNjhep01}. 
Right panel: 
The parameters ($\sin^2 2\theta_{13}$, $\delta$) of the true solutions 
and the clone ones II$-$VIII are presented as numbers and also by the dots in 
$\sin^2 2\theta_{13}- \delta$ space. 
The correspondence between the ellipses and the solution labels are made 
manifest by using the same color lines/symbols in both panels. 
}
\label{biP-plot-8fold}
\end{figure}

It is known that a set of measurement of the oscillation probabilities, e.g., 
$P(\nu_{\mu} \rightarrow \nu_e)$ and its CP conjugate 
$P(\bar{\nu}_{\mu} \rightarrow  \bar{\nu}_e )$ at a particular neutrino energy, 
no matter how accurate, does not determine uniquely the values of 
$\theta_{13}$ and $\delta$, the problem of parameter degeneracy \cite{intrinsic,MNjhep01,octant}. 
The nature of the degeneracy can be understood as the so called intrinsic 
degeneracy \cite{intrinsic} duplicated by the unknown sign of $\Delta m^2_{31}$ 
\cite{MNjhep01} and $\theta_{23}$ octant \cite{octant}, which entails the 
total eightfold degeneracy if $\theta_{23} \neq \pi/4$.
The feature can be seen in Fig.~\ref{biP-plot-8fold}. 
Some features of the degeneracy were further discussed in 
\cite{BMW02,KMN02,MNP2}, 
whose first two noticed special features that appear in the vacuum oscillation maximum. 
The notorious feature of the degeneracy is that difference between the true and 
the fake solutions can be so small that their distinction is extremely difficult, 
rendering resolution of completely different physical pictures, e.g., 
the mass hierarchies, untenable. 
Or, in the other cases, the difference between the true and the fake values 
of $\delta$ is so large to confuse CP violation with CP conservation.
%

It is the purpose of this paper to give a complete analysis of the parameter 
degeneracy in neutrino oscillations. 
To achieve a unified understanding of the phenomenon, we present and advance 
a new view of the degeneracy as an invariance of the oscillation probabilities 
under discrete mappings of the mixing parameters. 
With use of the approximate form of the oscillation probability obtained in \cite{golden} 
we present a self-contained derivation (re-derivation in CP conjugate case) 
of the analytic expressions of all the degeneracy solutions as functions of the 
true solutions, which supplies the explicit form of the mapping. 
Having the analytic solutions of the eightfold degeneracy with the proper 
convention at hand, we demonstrate that they form a solution network, 
the one-to-one correspondence structure between solutions, which will be 
pictorially represented in Fig.~\ref{eightfold-relation} in Sec.~\ref{structure}.

We use the analytic solutions to make plots of the differences 
between the true and the fake solutions to illuminate the global overview of 
the degeneracy. 
We illuminate, by using the plots, the characteristic features of the degeneracy 
and reveal the reasons why and how the sign-$\Delta m^2_{31}$ 
and the $\theta_{23}$ octant degeneracies are robust against the spectrum analysis. 
We note that the first attempt toward analytic solutions of the degeneracy 
was pursuit by the authors of \cite{BMW02} who obtained the solution for the 
intrinsic degeneracy. 
Then, the similar analysis was extended in \cite{donini03} to include the 
degeneracy solutions which involve the $\Delta m^2_{31}$-sign and/or 
the $\theta_{23}$ octant flips.

One may ask;  Why is the parameter degeneracy defined as above way so relevant? 
Mathematically speaking, the degeneracy is easy to solve; 
Repeating measurement at one more energy (or baseline), or adding a 
different oscillation channel immediately solves degeneracy. 
Or, if the spectrum information is available it may be more powerful to resolve 
the degeneracy. 
Despite these valid reasonings, unfortunately, at least some type of the 
degeneracy is shown to be robust and survives in varying experimental settings. 
One of the reasons for it is, as we will see in Sec.~\ref{overview}, that 
the energy dependence of difference between the true and the fake solutions 
is so mild that spectrum information is not powerful enough to resolve the degeneracy.

We emphasize that need for resolution of the degeneracy is not only because 
precision measurement is always desirable, but also because, far more importantly, 
it leaves e.g., the neutrino mass hierarchy undetermined after huge 
experimental efforts.
We want to warn the readers that we will not try to discuss how the degeneracy 
can be lifted by assuming concrete experimental settings.\footnote{
The analysis presented in this paper may be regarded as 
``pathological analysis of neutrino oscillation''. 
It would not tell us directly the experimental method for solving the degeneracy, 
but as in the case of pathology of human body, the clearer understanding of 
the disease may ultimately provide with us the correct recipe for its resoluion.  
}
%
Rather, we focus in on a complete understanding of structure of the degeneracy. 
The explicit analytic solutions and knowledges of structure of the degeneracy 
should serve for clearer understanding of the experimental data taken in 
precision measurement in the future. 
It will be definitely called for if future neutrino experiments reveal features that may 
not fit in into the standard three-neutrino mixing to clearly discriminate confusion 
by the degeneracy from new effects beyond the standard three-flavor mixing. 
For example, if they are so powerful to detect neutrino's nonstandard interactions (NSI) 
\cite{wolfenstein,valle,petcov,grossman,berezhiani} 
(see \cite{NOVE09-mina} for further references therein), 
the event structure will be modified by the new ingredients and enriched with 
new type of the degeneracies \cite{NSI-perturbation,NSI-2phase}. 
For the importance of lifting the degeneracy, quite naturally, a great amount of 
efforts were devoted to investigate how it can be done. 
The references \cite{burguetC02,huber02,huber03,huber05,donini-nufact03,autiero04,donini04,burguetC04, mena04,mena05,mena06,MEMPHYS,BNL,T2KK-1st,T2KK-2nd,nufact-lowE1,nufact-lowE2,BNL-fermi,resolve23,peres-smi23,concha23,choubey23,kajita-atm23,huber23,meloni08}  
are nothing but only a small subset of them.

In the next section, we define our machinery and introduce a new view of the 
degeneracy as invariance under the discrete transformations. 
Then, in the following two sections (\ref{CP-intrinsic-sign} and \ref{CP-octant}), 
we first give a complete treatment of the parameter 
degeneracy with CP-conjugate measurement. 
In Sec.~\ref{structure}, we present the explicit form of the discrete mapping 
and complete our understanding of the structure of the degeneracy. 
It will be supplemented by the discussions of degeneracy in 
T-conjugate (Sec.~\ref{T-conjugate}), 
the golden-silver (Sec.~\ref{Golden-Silver}), and 
CPT-conjugate (Sec.~\ref{CPT-conjugate}) 
measurement combining 
the $\nu_{e} \rightarrow \nu_{\mu}$ (golden) and the 
$\nu_{e} \rightarrow \nu_{\tau}$ (silver) channels.

\section{Neutrino Oscillation Probability and Its Invariance }
\label{probability}

In our analysis in this paper, we rely on the approximate formula 
for the appearance oscillation probability derived by Cervera {\it et al.} 
\cite{golden}. 
A simple way of deriving the formula is to use perturbative framework in which 
$s_{13}$ is assumed to be of order 
$\frac{ \Delta m^2_{21} } { \Delta m^2_{31} } \equiv \epsilon$ 
and keep the terms up to second order in $\epsilon$. 
For a review of this method, see e.g., \cite{NSI-perturbation}. 
Here, $\Delta m^2_{ji} \equiv m^2_{j} - m^2_{i}$ $ (i, j = 1, 2, 3)$. 
However, in this paper, we take an attitude to utilize the formula as far as 
it is reasonably accurate, even outside the region of validity of the perturbative 
ansatz. In fact, it is known that the formula gives a reasonable description 
of the oscillation probability even with larger values of $s_{13}$ \cite{munich04}.

\subsection{Approximate formula of the neutrino oscillation probability}

To present the formula for the $\nu_{e}$ appearance probability in a compact 
way we use the simplified notations. 
We summarize them together with their magnitudes for convenience of the readers:   
\begin{eqnarray} 
s &\equiv& s_{13} 
\nonumber \\
\Delta_{ji} &\equiv& \biggl | \frac{ \Delta m^2_{ji} L  }{4E} \biggr | 
=  1.27   
\left(\frac{|\Delta m^2_{ji}|}{10^{-3}\text{eV}^2}\right)
\left(\frac{L}{1000 \text{km}}\right)
\left(\frac{E}{1\text{GeV}}\right)^{-1}
\hspace{6mm} 
(i, j = 1, 2, 3), 
\nonumber \\
A &\equiv& \frac{ aL } {4E} 
= 0.27 
\left(\frac{\rho}{2.8 \text{g/cm}^3}\right)
\left(\frac{L}{1000 \text{km}}\right), 
\label{def-s-Delta-A}
\end{eqnarray}
where $a \equiv 2 \sqrt{2} G_{F} N_{e} E$,
the well known coefficient related to the index of refraction of 
neutrinos in matter \cite{wolfenstein}. 
$G_F$ is the Fermi constant, $\rho$  and $N_e \equiv \rho/m_{N}$ 
with $m_{N}$ being the nucleon mass denote, respectively, the averaged 
matter and the electron number densities along the neutrino trajectory in the earth, 
and we have assumed that the electron fraction $Y_{e}$ is 0.5. 
By using the definition of $\Delta_{ji}$ as positive definite quantities 
we choose to display explicitly the sign of $\Delta m^2_{31}$ as 
$\pm$ signs (sometimes called as the hierarchy signs) in the equations.

It may be useful to remember the ratio between the vacuum and the matter 
parameters for understanding the feature of the degeneracy solutions in 
Sec.~\ref{overview}: 
\begin{eqnarray}
\frac{ A }{ \Delta_{31} } = 0.085 
\left(\frac{|\Delta m^2_{ji}|}{2.5 \times 10^{-3}\text{eV}^2}\right)^{-1}
\left(\frac{\rho}{2.8 \text{g/cm}^3}\right)
\left(\frac{E}{1\text{GeV}}\right).
\label{A/Dm2}
\end{eqnarray}
Therefore, in typical low-energy superbeam \cite{MNplb00,sato,richter} 
experiments the ratio is small, $A / \Delta_{31} \lesssim 0.1$, 
whereas in neutrino factory \cite{Geer,De-Rujula}
with baseline of several thousand kilometers 
the ratio is large, $A / \Delta_{31} \sim 3-6$.

The oscillation probabilities of the neutrino flavor conversion processes 
$\nu_{\mu} \rightarrow \nu_e$, 
its CP-conjugate channel $\bar{\nu}_{\mu} \rightarrow  \bar{\nu}_e$, 
the T-conjugate channel $\nu_{e} \rightarrow \nu_{\mu}$, and 
the CPT-conjugate channel $\bar{\nu}_{e} \rightarrow  \bar{\nu}_{\mu}$,
in matter are given under the constant matter density approximation as \cite{golden}
\begin{eqnarray}
P \equiv
P(\nu_{\mu} \rightarrow \nu_e) &=&
X_{\pm}s^2 + 
Y_{\pm} s \cos {\left( \delta \pm \Delta_{31} \right)} + Z 
\label{Pmue} \\
P^{CP} \equiv
CP[P(\nu_{\mu} \rightarrow \nu_e)] &=&
P(\bar{\nu}_{\mu} \rightarrow  \bar{\nu}_e ) = 
\bar{X}_{\pm} s^2 + 
\bar{Y}_{\pm} s \cos {\left(  \delta \mp \Delta_{31}  \right)} + Z 
\nonumber \\
&=&
X_{\mp}s^2 - 
Y_{\mp} s \cos {\left(  \delta \mp \Delta_{31}  \right)} + Z 
\label{PmueCP} \\
P^{T} \equiv
T[P(\nu_{\mu} \rightarrow \nu_e)] &=&
P(\nu_e \rightarrow \nu_{\mu}) = 
X_{\pm}s^2 + 
Y_{\pm} s \cos {\left(  \delta \mp \Delta_{31}  \right)} + Z, 
\label{PmueT} \\
P^{CPT} \equiv
CPT[P(\nu_{\mu} \rightarrow \nu_e)] &=&
P(\bar{\nu}_e \rightarrow   \bar{\nu}_{\mu} ) = 
\bar{X}_{\pm} s^2 + 
\bar{Y}_{\pm} s \cos {\left(  \delta \pm \Delta_{31}  \right)} + Z 
\nonumber \\
&=&
X_{\mp}s^2 - 
Y_{\mp} s \cos {\left(  \delta \pm \Delta_{31}  \right)} + Z
\label{PmueCPT} 
\end{eqnarray}
where $\pm$ indicates the mass hierarchy, namely, 
the normal and the inverted mass hierarchies for 
the positive and the negative $\Delta m^2_{31}$, respectively.
The functions $X_{\pm}$, $Y_{\pm}$, and $Z$ are defined by
%
\begin{eqnarray}
X_{\pm} &=& 4 s^2_{23}
\left[ \frac{\Delta_{31}\sin({\Delta_{31} \mp A})}{(\Delta_{31} \mp A)} \right]^2, 
\nonumber  \\
Y_{\pm} &=& \pm 2\sqrt{X_\pm P_\odot} = 
\pm 4 \sin{2\theta_{12}} c_{23}s_{23}
\left[ \frac{\Delta_{31}\sin({\Delta_{31} \mp A})}{(\Delta_{31} \mp A)} \right]
\left[ \frac{\Delta_{21}\sin{ {A} }}{A}\right], 
\nonumber\\
Z & = & c^2_{23} \sin^2{2\theta_{12}} 
\left[ \frac{\Delta_{21}\sin{ {A} }}{A}\right]^2. 
\label{XYZ}
\end{eqnarray}
Their forms imply that the oscillation probability can be written as 
$P = |\sqrt{X} s + \mbox{e}^{i(\delta \pm \Delta_{31})} \sqrt{Z}|^2$, 
which allows simple interpretation of the $\delta$-sensitive 
term as an interference between the atmospheric and the solar scale oscillations.
$\bar{X}$ and $\bar{Y}$ in (\ref{PmueCP}) and (\ref{PmueCPT})  are related to 
$X$ and $Y$ as 
\begin{eqnarray}
\bar{X}_{\pm} (a) &=& X_{\pm} (- a) = X_{\mp} (a) 
\nonumber \\
\bar{Y}_{\pm} (a) &=& Y_{\pm} (- a)  = - Y_{\mp} (a). 
\label{bar-nobar} 
\end{eqnarray}
In our discussions in this paper, it is crucial to note the relation \cite{MNP2} 
between the coefficients $X_{\pm}$ and $Y_{\pm}$:
\begin{equation}
\frac{Y_+}{\sqrt{X_+}}
= - ~\frac{Y_-}{\sqrt{X_-}} 
\label{identity}
\end{equation}
which follows from the definitions. 
Notice that (\ref{bar-nobar}) means that the same relation as (\ref{identity}) 
holds also for $\bar{X}$ and $\bar{Y}$.

In this paper, our emphasis is placed on the oscillation channels between 
$\nu_{\mu}$ and $\nu_{e}$ and their anti-particles. 
To have a clearer view of the structure of parameter degeneracy, however,  
we will include the $\nu_{e} \rightarrow \nu_{\tau}$ 
appearance channel, which is sometimes called the ``silver channel'' \cite{silver}. 
See Sec.~\ref{Golden-Silver}. 
The oscillation probability $P(\nu_{e} \rightarrow \nu_{\tau})$ is given by 
\begin{eqnarray}
P^{S} \equiv P(\nu_{e} \rightarrow \nu_{\tau}) =
\cot^2{\theta_{23}} X_{\pm} s^2 - 
Y_{\pm} s \cos {\left( \delta \mp \Delta_{31} \right)} + \tan^2{\theta_{23}} Z. 
\label{Petau} 
\end{eqnarray}

\subsection{Parameter degeneracy as an invariance of the oscillation probability under discrete mapping}
\label{invariance}

It is not so well recognized that the appearance oscillation probability in matter 
under the Cervera {\it et al.} approximation has an invariance. 
Namely, it is easy to show that the oscillation probabilities $P$, $P^{T}$, and $P^{S}$  
defined in (\ref{Pmue}), (\ref{PmueT}) and (\ref{Petau}), respectively, 
with positive $\Delta m^2_{31}$ is invariant under the transformation 
\begin{eqnarray} 
\Delta m^2_{31} &\rightarrow& - \Delta m^2_{31},  
\nonumber \\
s &\rightarrow& \sqrt{  \frac{ X_{+} }{ X_{-} }  } s,  
\nonumber \\
\delta &\rightarrow& \pi - \delta.
\label{transformation}
\end{eqnarray}
Notice that the first transformation transforms $X_{+}$ and $Y_{+}$ into 
$X_{-}$ and $Y_{-}$, respectively, and use has been made of the key 
relation (\ref{identity}). 
Similarly, the oscillation probability with negative $\Delta m^2_{31}$ has 
an invariance similar to (\ref{transformation}), replacing the second one with 
$s \rightarrow \sqrt{  \frac{ X_{-} }{ X_{+} }  } s$. 
It is nothing but a generalization of the invariance of the oscillation probability 
in vacuum to that in matter, which was used to show the existence of degeneracy 
solutions with differing sign of $\Delta m^2_{31}$ \cite{MNjhep01}.

Then, it immediately follows that there exists the sign-$\Delta m^2_{31}$ 
degeneracy in measurement which combines any two of $P$, $P^{T}$, and $P^{S}$, 
and the explicit form of the degenerate solution can be obtained by the symmetry alone.
Later in Sec.~\ref{T-conjugate} and Sec.~\ref{Golden-Silver} we will 
explicitly verify it by working out explicit solutions. 


\begin{figure}[bhtp]
\begin{center}
\vglue 0.2cm
\includegraphics[bb=0 0 240 231 , clip, width=0.36\textwidth]{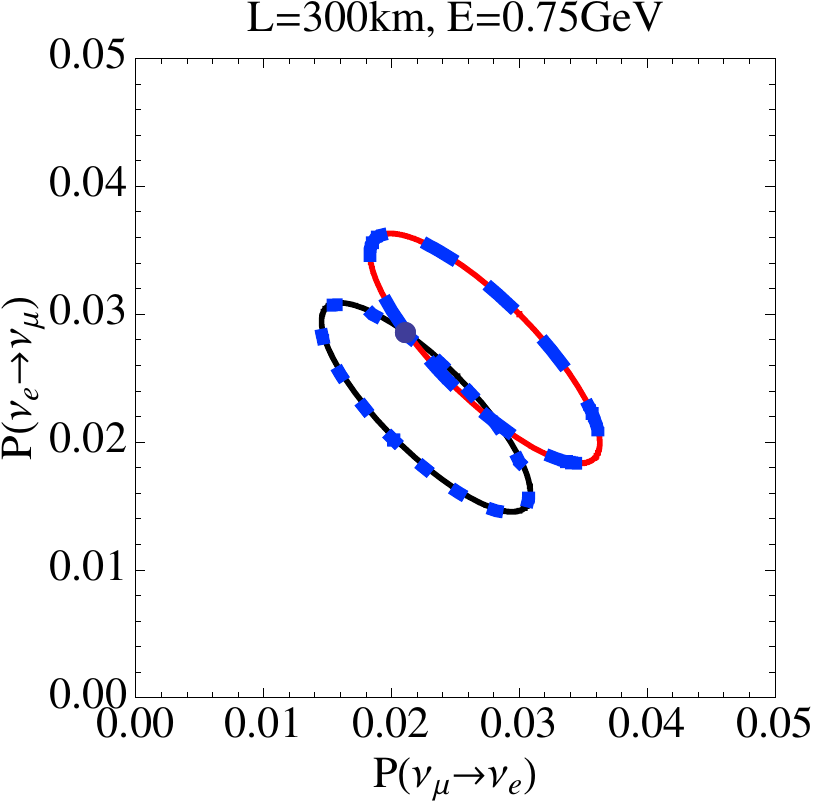}
\end{center}
\vglue -0.5cm
\caption{
$P - P^{T}$ bi-probability plot. 
$P - P^{T}$ bi-probability plot by using the approximate formulas given in 
(\ref{Pmue}) and (\ref{PmueT}).  
$\sin^2 2\theta_{13} = $  0.05 and NH (black solid), 
0.061 and NH (red solid), 0.058 and IH (blue dotted), 
0.069 and IH (blue dashed). 
}
\label{biP-plot-T}
\end{figure}

A better understanding of the implication of the invariance may be achieved 
by drawing the bi-probability plot \cite{MNjhep01}.
Here, we take the particular one in $P - P^{T}$ space \cite{MNP1} as shown in 
Fig.~\ref{biP-plot-T}. 
The simultaneous invariance of $P=P(\nu_{\mu} \rightarrow \nu_e)$ and 
$P^{T} = P(\nu_{e} \rightarrow \nu_{\mu})$ for any values of $\theta_{13}$ and 
$\delta$ means that one can find always two completely overlapping ellipses, 
one with positive and the other negative $\Delta m^2_{31}$. 
The $\delta$ label, if placed onto the ellipses, are different between the two ellipses, and they are related by $\delta_{-} = \pi - \delta_{+}$, where $\delta_{\pm}$ denote 
the $\delta$ for the respective hierarchies. 
More comments on the degeneracy in T-conjugate measurement will follow 
in Sec.~\ref{T-conjugate}. 


Now, let us focus on the degeneracy with CP-conjugate measurement.
Unfortunately, the similar simple-minded symmetry argument does not go 
through in the settings with CP and CPT conjugate measurement; 
The positive $\Delta m^2_{31}$ CP and CPT conjugate probabilities 
$P^{CP}$ and $P^{CPT}$ defined in (\ref{PmueCP}) and (\ref{PmueCPT}) 
are not simultaneously invariant under (\ref{transformation}). 
That is, $P^{CP}$ and $P^{CPT}$ are invariant under a transformation 
$s \rightarrow \sqrt{  \bar{X}_{+} / \bar{X}_{-} }  s$ 
accompanied with the other transformations in (\ref{transformation}). 
But, the transformation cannot be identical with (\ref{transformation}) in matter.
It is also true that the similar symmetry argument does not go through 
for the intrinsic and the $\theta_{23}$ octant degeneracies.

However, we will show by the discussions throughout the following two sections and 
Sec.~\ref{CPT-conjugate} that there is a similar invariance of a pair of 
the oscillation probabilities, e.g., $P$ and $P^{CP}$ in the CP-conjugate measurement, 
under the transformation 
\begin{eqnarray} 
s_{1} &\rightarrow& s_{ \text{N} } = \xi_{ \text{N} } (s_{1}, \delta_{1}, \theta_{23}^{\text{true}} ), 
\nonumber \\
\delta_{1} &\rightarrow& \delta_{ \text{N} } = \eta_{ \text{N} } (s_{1}, \delta_{1}, \theta_{23}^{\text{true}} ), 
\nonumber \\
\Delta m^2_{31} &\rightarrow& + \Delta m^2_{31}, ~\text{or}~ -\Delta m^2_{31},  
\nonumber \\
\theta_{23}^{\text{true}}  &\rightarrow& \theta_{23}^{\text{true}}, ~\text{or}~\theta_{23}^{\text{false}}. 
\label{transformation2}
\end{eqnarray}
In (\ref{transformation2}), $s_{1}$, $\delta_{1}$, and $\theta_{23}^{\text{true}}$ 
is the true parameters with subscript ``$1$'', the unique case with Arabic numerals. 
The degeneracy solutions are labeled by using Roman subscripts N= II$-$VIII.\footnote{
We note that each type of degeneracy is only two-fold merely because of the 
approximate form of the probabilities we use, or in other word, due to the 
smallness of $\theta_{13}$. 
Existence of more solutions may be signaled e.g., by the $s_{13}^3$ terms in the 
oscillation probabilities in a large-$\theta_{13}$ perturbation theory \cite{large-theta-P}.
}
%
Alternative choices in the last two transformations determine whether the degeneracy is of 
the type involving the sign change of $\Delta m^2_{31}$, or the $\theta_{23}$ octant flip. 
In this sense, the parameter degeneracy is nothing but the statement of invariance 
of $P$ and $P^{CP}$ under the discrete transformation (\ref{transformation2}).
Furthermore, what is to be really stressed is that the mapping in 
(\ref{transformation2}) can be constructed by the basic three mappings. 
See Sec.~\ref{structure}.

\section{Intrinsic and Sign-$\Delta m^2_{31}$ Degeneracies; CP-conjugate Measurement }
\label{CP-intrinsic-sign}

\subsection{Preliminary remarks}
\label{preliminary}

Since this is the first section to actually solve the degeneracy problem to 
obtain the clone solutions we make some preliminary remarks. 
In the rest of this paper, we analyze the structure of parameter degeneracy from 
various viewpoints, in particular, explicit analytic solutions, the symmetry aspect 
(as already mentioned), and use of the bi-probability plot to illuminate the 
respective characteristic features of the degeneracy. 
While the solutions were presented in a condensed way in \cite{donini03}, 
we present a step-by-step derivation of the degeneracy solutions because 
it is reader friendly and makes the understanding of structure of degeneracy 
much easier. 
We emphasize that the analytic solution with the proper convention for its 
definition {\em is} the integral part of our discussion of solution network to be given 
in Sec.~\ref{structure}.

In most part of this paper, we confine ourselves into the oscillation channels 
between $\nu_{\mu}$ and $\nu_{e}$ and their antiparticles. In particular, we focus on  
$\nu_{e}$ (and $\bar{\nu}_{e}$) appearance channel which will be available 
in conventional $\nu_{\mu}$ superbeam in this and the next sections. 
On the other hand, the T-conjugate channels, 
$\nu_{e} \rightarrow \nu_{\mu}$ and $\bar{\nu}_{e} \rightarrow \bar{\nu}_{\mu}$, 
would be provided by neutrino factory and the beta beam \cite{beta1,beta2}. 
Given understanding the degeneracy of the former channels, 
the corresponding informations on the latter may be obtained by regarding 
$\delta$ by $2\pi - \delta$. 
We include $\nu_{\tau}$ appearance channel in Sec.~\ref{Golden-Silver}.

In this paper we take the method for obtaining the degeneracy solutions 
for a given set of true parameters ($s_{1}, \delta_{1}$). 
Equivalently, one can choose an alternative way of obtaining the degeneracy 
solutions as a function of ``observable'', e.g., $P$ and $P^{CP}$, 
as pursued in \cite{MNP2}. 
If one want to take this attitude, one can simply do it (at least numerically) 
by regarding that both the clone solutions and ($P$, $P^{CP}$) are 
parametrically represented by ($s_{1}, \delta_{1}$). 

\vspace{2mm}

\noindent
{\bf  Notational comment: }
We denote the mass hierarchy of the true solution by the $\pm$ signs 
($+$ for the normal and $-$ for the inverted) 
to make the hierarchy choice always explicit. 
The relationship between the degeneracy solutions with input true mass 
hierarchies will be further discussed in Sec.~\ref{structure}.

Now, let us start our discussion of parameter degeneracy by taking CP-conjugate measurement. 
The setting seems to be the most promising one experimentally in the near future. 
In this section we confine ourselves to the degeneracy solutions which 
have the same $\theta_{23}$ octant, though we treat generically the case 
of arbitrary values of $\theta_{23}$. 
The degeneracy solutions across octants of $\theta_{23}$ will be discussed 
in the next section.

\subsection{The intrinsic degeneracy in CP-conjugate measurement}
\label{CP-intrinsic}

With expression of the oscillation probabilities in (\ref{Pmue}) 
and (\ref{PmueCP}), 
the intrinsic degeneracy solutions ($s_{i}$, $\delta_{i}$) (i=1, 2) 
in CP-conjugate measurement 
are defined with $\nu_{\mu} \rightarrow \nu_e$ channel by 
\begin{eqnarray} 
P - Z &=& 
X_{\pm}s_{1}^2 + 
Y_{\pm} s_{1} \left( 
\cos \delta_{1} \cos \Delta_{31} \mp \sin \delta_{1} \sin \Delta_{31} 
\right), 
\nonumber \\
P - Z &=& 
X_{\pm}s_{2}^2 + 
Y_{\pm} s_{2} \left( 
\cos \delta_{2} \cos \Delta_{31} \mp \sin \delta_{2} \sin \Delta_{31} 
\right), 
\label{CP-intrinsic-def1}
\end{eqnarray}
and in CP-conjugate channel by 
\begin{eqnarray} 
P^{CP} - Z &=& 
X_{\mp}s_{1}^2 - 
Y_{\mp} s_{1} \left( 
\cos \delta_{1} \cos \Delta_{31} \pm \sin \delta_{1} \sin \Delta_{31} 
\right), 
\nonumber \\
P^{CP} - Z &=& 
X_{\mp}s_{2}^2 - 
Y_{\mp} s_{2} \left( 
\cos \delta_{2} \cos \Delta_{31} \pm \sin \delta_{2} \sin \Delta_{31} 
\right). 
\label{CP-intrinsic-def2}
\end{eqnarray}
%
By subtracting two equations in (\ref{CP-intrinsic-def1}) and 
(\ref{CP-intrinsic-def2}) respectively, we obtain 
\begin{eqnarray} 
(s_{1}^2 - s_{2}^2) + 
\frac{ Y_{\pm} }{X_{\pm} } \cos \Delta_{31} 
( s_{1} \cos \delta_{1} - s_{2} \cos \delta_{2} ) \mp 
\frac{ Y_{\pm} }{X_{\pm} }  \sin \Delta_{31} 
( s_{1} \sin \delta_{1} - s_{2} \sin \delta_{2} ) &=& 0, 
\nonumber \\
(s_{1}^2 - s_{2}^2) - 
\frac{ Y_{\mp} }{X_{\mp} } \cos \Delta_{31} 
( s_{1} \cos \delta_{1} - s_{2} \cos \delta_{2} ) \mp 
\frac{ Y_{\mp} }{X_{\mp} }  \sin \Delta_{31} 
( s_{1} \sin \delta_{1} - s_{2} \sin \delta_{2} ) &=& 0. 
\label{CP-intrinsic1}
\end{eqnarray}
From (\ref{CP-intrinsic1}) we can obtain the expressions of 
$\cos \delta_{2}$ and $\sin \delta_{2}$ as 
\begin{eqnarray} 
s_{2} \cos \delta_{2} &=& 
s_{1} \cos \delta_{1} \pm 
\frac{ 2 }{ \cos \Delta_{31} } \frac{ C^{(-)} }{ \left( C^{(+)} \right)^2  - \left( C^{(-)} \right)^2 } 
\left( s_{2}^2 -  s_{1}^2 \right), 
\nonumber \\
s_{2} \sin \delta_{2} &=& 
s_{1} \sin \delta_{1} \pm 
\frac{ 2 }{ \sin \Delta_{31} } \frac{ C^{(+)} }{ \left( C^{(+)} \right)^2  - \left( C^{(-)} \right)^2 } 
\left( s_{2}^2 -  s_{1}^2 \right), 
\label{CP-intrinsic-delta-solution}
\end{eqnarray}
where $C^{(\pm)}$ is defined by
\begin{eqnarray} 
C^{(\pm)} \equiv 
\frac{ Y_{+} }{X_{+} }  \pm \frac{ Y_{-} }{X_{-} }. 
\label{Cpm-def}
\end{eqnarray}

Inserting (\ref{CP-intrinsic-delta-solution}) into $\cos^2 \delta_{2} + \sin^2 \delta_{2} = 1$ gives 
a quartic equation for $s_{2}$ as 
$( s_{2}^2 -  s_{1}^2 )( s_{2}^2 -  s_{ \text{II} }^2 ) = 0$. 
Of course, we obtain the trivial solution $s_{2}=s_{1}$, the situation unique to 
discussions of the intrinsic degeneracy.
The genuine intrinsic degeneracy solution is given by 
\begin{eqnarray} 
s_{ \text{II} } = 
\left[ 
s_{1}^2 \pm  
\frac{1 }{ (1 + R^2)  }
\left( s_{1} \cos \delta_{1} + R  s_{1} \sin \delta_{1} \right) 
\left( C^{(-)}  \cos \Delta_{31} - R C^{(+)} \sin \Delta_{31} \right) 
\right.
\nonumber \\
&&\hspace*{-80mm} {} +
\left.
\frac{1}{ 4 (1 + R^2) } 
\left( C^{(-)}  \cos \Delta_{31} - R C^{(+)} \sin \Delta_{31} \right)^2 
\right]^{ 1/2 }, 
\label{CP-intrinsic-s-solution}
\end{eqnarray}
where $R$ is defined by\footnote{
We note the relationship between the notations in this paper and 
in the previous papers \cite{intrinsic,MNP2}, denoted here as MNP: 
$C^{(\pm)} \vert_{ \text{this} } = 2 C^{(\mp)} \vert_{ \text{MNP} }$, and 
$R \vert_{ \text{this} }  = z^{-1} \vert_{ \text{MNP} } $. 
}
%
\begin{eqnarray} 
R \equiv 
\frac{ C^{(+)}  }{C^{(-)}  }  \cot{  \Delta_{31}  }.
\label{R-def}
\end{eqnarray}
Notice that the $\pm$ sign in (\ref{CP-intrinsic-s-solution}) represents the 
mass hierarchy of the true solution.
By using (\ref{CP-intrinsic-s-solution}) into (\ref{CP-intrinsic-delta-solution}) 
we obtain the solution of $\delta_{ \text{II} } $. 
From (\ref{CP-intrinsic-delta-solution}) one can obtain the following expressions: 
\begin{eqnarray} 
s_{ \text{II} }  \sin \left( \delta_{1} + \delta_{ \text{II} }  \right) 
&=& 
- \frac{ 2 R s_{1}  }{ 1 + R^2 } 
\mp \frac{ 1 }{ 2(1 + R^2) } 
\left( \sin \delta_{1} + R \cos \delta_{1}  \right) 
\left( C^{(-)}  \cos \Delta_{31} - R C^{(+)} \sin \Delta_{31} \right),  
\nonumber \\
s_{ \text{II} }  \cos \left( \delta_{1} + \delta_{ \text{II} }  \right) &=& 
- \left( \frac{ 1 - R^2 }{ 1 + R^2 } \right) s_{1} 
\nonumber \\
&\mp& \frac{ 1 }{ 2(1 + R^2) } 
\left( \cos \delta_{1} - R \sin \delta_{1}  \right) 
\left( C^{(-)}  \cos \Delta_{31} - R C^{(+)} \sin \Delta_{31} \right). 
\label{sum-delta-intrinsic}
\end{eqnarray}

If we further expand the solution in (\ref{CP-intrinsic-s-solution}) by the 
solar-atmospheric ratio $\frac{ \Delta m^2_{21} } { \Delta m^2_{31} } $, we obtain 
\begin{eqnarray} 
s_{ \text{II} } = s_{1} + 
\frac{ s_{1} }{ 2 (1 + R^2)  }
\left( \cos \delta_{1} + R \sin \delta_{1} \right) 
\left( C^{(-)}  \cos \Delta_{31} - R C^{(+)} \sin \Delta_{31} \right). 
\label{CP-s-solution-1st}
\end{eqnarray}
Similarly, the equation (\ref{sum-delta-intrinsic}) also simplifies to 
$\cos \left( \delta_{1} + \delta_{2} \right) = - (1 - R^2) / (1 + R^2 ) $. 
These expressions  reproduce the ones derived in \cite{intrinsic,MNP2}.

\subsection{The sign-$\Delta m^2$ degeneracy in CP-conjugate measurement}
\label{CP-signdm2}

We turn to the flipped $\Delta m^2$-sign degeneracy in CP-conjugate measurement. 
The true input solution ($s_{1}, \delta_{1}$) and the opposite sign clone solution 
($s_{3}, \delta_{3}$) satisfy the following equations.  In the neutrino channel, 
\begin{eqnarray} 
P - Z &=& 
X_{\pm}s_{1}^2 + 
Y_{\pm} s_{1} \left( 
\cos \delta_{1} \cos \Delta_{31} \mp \sin \delta_{1} \sin \Delta_{31} 
\right), 
\nonumber \\
P - Z &=& 
X_{\mp}s_{3}^2 + 
Y_{\mp} s_{3} \left( 
\cos \delta_{3} \cos \Delta_{31} \pm \sin \delta_{3} \sin \Delta_{31} 
\right), 
\label{CP-flipped-sign-def1}
\end{eqnarray}
and in CP-conjugate channel 
\begin{eqnarray} 
P^{CP} - Z &=& 
X_{\mp}s_{1}^2 - 
Y_{\mp} s_{1} \left( 
\cos \delta_{1} \cos \Delta_{31} \pm \sin \delta_{1} \sin \Delta_{31} 
\right), 
\nonumber \\
P^{CP} - Z &=& 
X_{\pm}s_{3}^2 - 
Y_{\pm} s_{3} \left( 
\cos \delta_{3} \cos \Delta_{31} \mp \sin \delta_{3} \sin \Delta_{31} 
\right), 
\label{CP-flipped-sign-def2}
\end{eqnarray}
By combining the first and the second equations in (\ref{CP-flipped-sign-def1}) 
and (\ref{CP-flipped-sign-def2}) we obtain 
\begin{eqnarray} 
\frac{ X_{\pm} }{ X_{\mp} } s_{1}^2 - s_{3}^2 
&+& \cos \Delta_{31} 
\left( \frac{Y_{\pm} }{ X_{\mp} } s_{1} \cos \delta_{1} -  \frac{Y_{\mp} }{ X_{\mp} } s_{3} \cos \delta_{3} \right) 
\nonumber \\
&\mp& \sin \Delta_{31} 
\left( \frac{Y_{\pm} }{ X_{\mp} } s_{1} \sin \delta_{1} + \frac{Y_{\mp} }{ X_{\mp} } s_{3} \sin \delta_{3} \right) = 0,
\nonumber \\
\frac{ X_{\mp} }{ X_{\pm} } s_{1}^2 - s_{3}^2 
&-& \cos \Delta_{31} 
\left( \frac{ Y_{\mp} }{ X_{\pm} } s_{1} \cos \delta_{1} -  \frac{ Y_{\pm} }{ X_{\pm} } s_{3} \cos \delta_{3} \right) 
\nonumber \\ 
&\mp& \sin \Delta_{31} 
\left( \frac{ Y_{\mp} }{ X_{\pm} } s_{1} \sin \delta_{1} + \frac{ Y_{\pm} }{ X_{\pm} } s_{3} \sin \delta_{3} \right) = 0. 
\label{CP-flipped-sign3}
\end{eqnarray}
Using $C^{(\pm)}$ defined in (\ref{Cpm-def}) we can simplify the equations. 
By subtracting and adding two equations in (\ref{CP-flipped-sign3}) 
we obtain 
\begin{eqnarray} 
T_{1 \pm}^{CP} 
&-& C^{(+)} s_{3} \cos \delta_{3}  \cos \Delta_{31} 
+ C^{(-)} s_{3} \sin \delta_{3}  \sin \Delta_{31} = 0, 
\nonumber \\
T_{2 \pm}^{CP} - 2 s_{3}^2 
&\pm& C^{(-)} s_{3} \cos \delta_{3}  \cos \Delta_{31} 
\mp C^{(+)} s_{3} \sin \delta_{3}  \sin \Delta_{31} = 0, 
\label{CP-flipped-sign5}
\end{eqnarray}
In (\ref{CP-flipped-sign5}), 
$T_{1 \pm}^{CP}$ and $T_{1 \pm}^{CP}$ are defined as: 
\begin{eqnarray} 
T_{1 \pm}^{CP} &\equiv& 
\pm E^{(-)} s_{1}^2  + D^{(+)} s_{1} \cos \delta_{1} \cos \Delta_{31} - D^{(-)} s_{1} \sin \delta_{1} \sin \Delta_{31}, 
\nonumber \\
T_{2 \pm}^{CP} &\equiv& 
E^{(+)} s_{1}^2  \pm D^{(-)} s_{1} \cos \delta_{1} \cos \Delta_{31} \mp D^{(+)} s_{1} \sin \delta_{1} \sin \Delta_{31}, 
\label{T1T2-def}
\end{eqnarray}
where we have introduced the new notations $D^{(\pm)}$ and $E^{(\pm)}$ as 
\begin{eqnarray} 
D^{(\pm)} \equiv 
\frac{ Y_{+} }{X_{-} }  \pm \frac{ Y_{-} }{X_{+} }, 
\hspace{10mm} 
E^{(\pm)} \equiv 
\frac{ X_{+} }{X_{-} }  \pm \frac{ X_{-} }{X_{+} }. 
\label{DEpm-def}
\end{eqnarray}
From (\ref{CP-flipped-sign5}) we obtain the expressions of $\cos \delta_{3}$ 
and $\sin \delta_{3}$ as  
\begin{eqnarray} 
s_{3} \cos \delta_{3} &=& 
\frac{ 1 }{ \cos \Delta_{31} } 
\frac{ 1 }{ \left( C^{(+)} \right)^2  - \left( C^{(-)} \right)^2 } 
\left[
C^{(+)} T_{1 \pm}^{CP} \pm 
C^{(-)} \left( T_{2 \pm}^{CP} - 2 s_{3}^2 \right)
\right], 
\nonumber \\
s_{3} \sin \delta_{3} &=& 
\frac{ 1 }{ \sin \Delta_{31} } 
\frac{ 1 }{ \left( C^{(+)} \right)^2  - \left( C^{(-)} \right)^2 } 
\left[
C^{(-)} T_{1 \pm}^{CP} \pm 
C^{(+)} \left( T_{2 \pm}^{CP} - 2 s_{3}^2 \right)
\right]. 
\label{CP-sign-delta-solution}
\end{eqnarray}
We insert (\ref{CP-sign-delta-solution}) into $\cos^2 \delta_{3} + \sin^2 \delta_{3} =1$ 
we obtain the quartic equation for $s_{3}$ as 
\begin{eqnarray} 
4 \left( 1 + R^2 \right) s_{3}^4 - 4 U_{\pm} s_{3}^2 + V_{\pm} = 0, 
\label{CP-s3-equation}
\end{eqnarray}
where 
\begin{eqnarray} 
U_{\pm} \equiv \frac{1}{4} \cos^2 \Delta_{31}  ( C^{(-)} )^2 
\left[ 1 - \left( \frac{ C^{(+)} }{ C^{(-)} } \right)^2 \right]^2 
\pm \frac{ 1 }{ \sin^2 \Delta_{31} } \left( \frac{ C^{(+)} }{ C^{(-)} } \right) T_{1 \pm}^{CP} 
+  \left( 1 + R^2 \right) T_{2 \pm}^{CP}, 
\label{F-def}
\end{eqnarray}
\begin{eqnarray} 
V_{\pm} \equiv 
\left[ \cot^2 \Delta_{31} + \left( \frac{ C^{(+)} }{ C^{(-)} } \right)^2 \right]  (T_{1 \pm}^{CP})^2 
\pm \frac{ 2 }{ \sin^2 \Delta_{31} } \left( \frac{ C^{(+)} }{ C^{(-)} } \right) T_{1 \pm}^{CP} T_{2 \pm}^{CP} 
+ 
\left( 1 + R^2 \right) (T_{2 \pm}^{CP})^2. 
\label{G-def}
\end{eqnarray}
Notice that $R$ is defined in (\ref{R-def}). 
Equation (\ref{CP-s3-equation}) has the obvious solutions 
\begin{eqnarray} 
s_{3}^2 = \frac{1}{ 2 \left( 1 + R^2 \right) } \left[ U_{\pm} 
\hspace{1mm} [\pm]^* \hspace{1mm} 
\sqrt{ U_{\pm}^2 - \left( 1 + R^2 \right) V_{\pm} } \right] 
\label{CP-sign-s-solution}
\end{eqnarray}
where $[\pm]^*$ denotes a temporary sign which is independent of the hierarchy sign.  
We discuss immediately below (Sec.~\ref{convention}) the way how to determine 
the sign convention.
One can easily show by using Cauchy-Schwarz inequality that $V_{\pm} \geq 0$, 
and hence the right-hand-side of (\ref{CP-sign-s-solution}) is non-negative.  
Therefore, there are four real solutions of $s_{3}$ and the two positive ones are physical; 
The sign-$\Delta m^2_{31}$ degeneracy is two-fold.

\begin{figure}[bhtp]
\begin{center}
\vglue 0.2cm
\includegraphics[bb=0 0 256 91 , clip, width=0.8\textwidth]{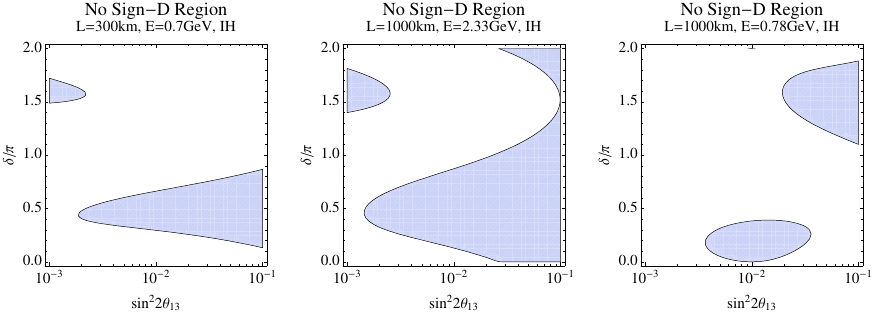}
\end{center}
\vglue -3mm
\caption{
Depicted as the shaded areas in the $\sin^2 2\theta_{13} - \delta/\pi$ space are 
the regions where no sign-$\Delta m^2_{31}$ degeneracy solution exists for 
the same three set of the baselines and the neutrino energies as in 
Figs.~\ref{R3-R4} and \ref{D3-D4}. 
The true mass hierarchy is taken to be the inverted one, and it may help to 
understand the relationship between the degeneracy solutions with different 
true mass hierarchies to be discussed in Sec.~\ref{normal-inverted}. 
}
\vglue -3mm
\label{no-sign-sol-region}
\end{figure}

The region specified by 
$D^{ \text{sign} }_{\pm} \equiv U_{\pm}^2 - \left( 1 + R^2 \right) V_{\pm} \leq 0$ 
defines the region in which there is no sign-$\Delta m^2_{31}$ degeneracy solution.
The region of no solution is displayed in Fig.~\ref{no-sign-sol-region} 
by taking the three typical setups for superbeam type experiments, 
SB1, MB1, and MB2 settings, 
which will be defined in Sec.~\ref{overview-variable}.\footnote{
Apparently, a complete description of the no solution region of the 
sign-$\Delta m^2_{31}$ and the $\theta_{23}$ octant degeneracies 
seems to be lacking in the literature.  
}
%
In this figure the true mass hierarchy is taken to be the inverted one.
If we take the input normal hierarchy we must have the figure with $\delta$ 
shifted by $\pi$, as one can confirm by looking at the white region in 
Figs.~\ref{R3-R4} and \ref{D3-D4} in Sec.~\ref{overview-sign}. 
This is expected by the general discussion to be given in 
Sec.~\ref{normal-inverted}.

A notable feature is that the region of absent solution occupies mostly around 
$\delta \sim \pi/2$ in the left two panels in Fig.~\ref{no-sign-sol-region} where 
$L/E$ taken are at around the first oscillation maximum. 
The region is farthest to the ``central region'' populated by the both 
$\Delta m^2_{31}$-sign ellipses in the bi-probability space, and hence it is 
the region  of lucky resolution of the sign-$\Delta m^2_{31}$ degeneracy 
\cite{MNnufact01} for the inverted ($\delta \sim 3\pi/2$ for the normal) mass hierarchy. 
Generally speaking the no sign-degeneracy region grows for longer baseline, 
and the tendency continues to e.g., $L=4000$ km and $E=20$ GeV 
(Figs.~\ref{RN-nufact} and \ref{DN-nufact} in Sec.~\ref{overview-nufact}). 
However, the feature changes for a region of the second oscillation 
maximum as seen in the third panel in Fig.~\ref{no-sign-sol-region}. 
Because of the dynamic behavior of the bi-probability ellipses 
(see e.g., Fig.~2 in \cite{T2KK-1st}) there are much better chance of having 
the sign-$\Delta m^2_{31}$ degeneracy.

\subsection{Problem of convention of labeling the degenerate solution} 
\label{convention}

We denote the two solutions in (\ref{CP-sign-s-solution}) 
as $s_{ \text{III} }$ and $s_{ \text{IV} }$. 
It is a highly nontrivial issue how to define these two solutions. 
In principle there are two ways: 

\vspace{3mm}
\noindent
{\bf Convention A}: 
One can take the convention such that always $ s_{ \text{IV} } \geq s_{ \text{III} }$. 
That is, the plus and the minus signs in (\ref{CP-sign-s-solution}) correspond to 
$s_{ \text{IV} }$ and $s_{ \text{III} }$, respectively. 

\vspace{3mm}
\noindent
{\bf Convention B}: 
One may choose the other convention such that the vacuum limit of the degenerate solutions can be taken smoothly. 

\vspace{3mm}
\noindent 
For reasons explained below we adopt the convention B.
We note that $D^{ \text{sign} }_{\pm}$ defined as 
$D^{ \text{sign} }_{\pm} \equiv U_{\pm}^2 - \left( 1 + R^2 \right) V_{\pm}$ 
can be written as 
\begin{eqnarray} 
&& D^{ \text{sign} }_{\pm} = 
\left\{ 1 - \left( \frac{ C^{(+)} }{ C^{(-)} } \right)^2 \right\}^2 
\biggl[ 
\frac{1}{16} \cos^4 \Delta_{31} (C^{(-)})^4 
\left\{ 1 - \left( \frac{ C^{(+)} }{ C^{(-)} } \right)^2 \right\}^2 
\nonumber \\
&& - \cot^2 \Delta_{31} (T_{1 \pm}^{CP})^2 
\pm \frac{1}{2} \cot^2 \Delta_{31} C^{(+)} C^{(-)} T_{1 \pm}^{CP} 
+ \frac{1}{2} \cos^2 \Delta_{31}  (C^{(-)})^2 \left( 1 + R^2 \right) T_{2 \pm}^{CP} 
\biggr].
\label{D-sign-pm}
\end{eqnarray}
In the vacuum oscillation limit, $a \rightarrow 0$, 
$C^{(+)}$, $D^{(+)}$, $R$, and $E^{(-)}$ all vanish, and $E^{(+)} = 2$ and 
$D^{(-)} = C^{(-)} =  C^{(-)}_{ \text{vac} }$ hold, where 
\begin{eqnarray} 
C^{(-)}_{ \text{vac} } = \lim_{a \rightarrow 0} \frac{ 2 Y_{+} }{ X_{\pm} }
= \frac{ 2 \Delta_{21} }{ \sin \Delta_{31} } 
\sin 2 \theta_{12} \cot \theta_{23}. 
\label{C-vac}
\end{eqnarray}
Then, $D^{ \text{sign} }_{\pm}$ has a vacuum limit 
$D^{ \text{oct-vac} }_{\pm} \equiv \lim_{ a \rightarrow 0 } D^{ \text{sign} }_{\pm} 
\equiv (d_{\pm}^{ \text{sign} } )^2$ where 
\begin{eqnarray} 
%
d_{\pm}^{ \text{sign} } = 
C^{(-)}_{ \text{vac} } \cos \Delta_{31} 
\left(  \frac{1}{4} C^{(-)}_{ \text{vac} } \cos \Delta_{31} \pm s_{1} \cos \delta_{1}  \right). 
\label{dpm-sign-vac}
\end{eqnarray}
The smooth limit to the sign-$\Delta m^2_{31}$ degenerate solution in vacuum 
can be achieved by taking the sign convention 
\begin{eqnarray} 
s_{ \text{III} }^2 = \frac{1}{ 2 \left( 1 + R^2 \right) } \left[ U_{\pm} - 
d_{\pm}^{ \text{sign} } 
\sqrt{ \frac{ D^{ \text{sign} }_{\pm}  }{ (d_{\pm}^{ \text{sign} } )^2 } } \right], 
\nonumber \\
s_{ \text{IV} }^2 = \frac{1}{ 2 \left( 1 + R^2 \right) } \left[ U_{\pm} + 
d_{\pm}^{ \text{sign} } 
\sqrt{ \frac{ D^{ \text{sign} }_{\pm}  }{ (d_{\pm}^{ \text{sign} } )^2 } } \right]. 
\label{CP-sign-s-solution-convention}
\end{eqnarray}
In (\ref{CP-sign-s-solution-convention}) we have taken the convention such that 
in $a \rightarrow 0$ limit $s_{ \text{III} }$ and $s_{ \text{IV} }$ smoothly tend to 
$s_{ \text{III} }^{ \text{vac} }$ and $s_{ \text{IV} }^{ \text{vac} }$, respectively, 
in vacuum defined in Sec.~\ref{vacuum}. 
Once the solutions of $s_{3}$ are specified with the well defined convention 
the solutions $\delta_{ \text{III} }$ and $\delta_{ \text{IV} }$ can be obtained 
by inserting $s_{ \text{III} }$ and $s_{ \text{IV} }$, respectively, 
into (\ref{CP-sign-delta-solution}).

We take the convention B because of number of desirable features. 
The matter perturbation theory \cite{AKS,MNprd98} can be formulated only 
with the convention because it requires the existence of smooth limit 
$a \rightarrow 0$ in each solution.
More importantly, the convention B makes the structure of the degenerate 
solutions transparent. 
That is, if we denote $s_{ \text{III} }$ in (\ref{CP-sign-s-solution-convention}) and 
$\delta_{ \text{III} }$ in (\ref{CP-sign-delta-solution}) in an abstract fashion as 
\begin{eqnarray} 
s_{ \text{III} } = \xi_{\pm}^{ \text{CP sign} } (s_{1}, \delta_{1} ), 
\hspace{10mm}
\delta_{ \text{III} } = \eta_{\pm}^{ \text{CP sign} }  (s_{1}, \delta_{1} ), 
\label{correspondence1}
\end{eqnarray}
%
then, one can show that 
\begin{eqnarray} 
s_{ \text{IV} } = \xi_{\pm}^{ \text{CP sign} }  (s_{ \text{II} }, \delta_{ \text{II} } ), 
\hspace{10mm}
\delta_{ \text{IV} } = \eta_{\pm}^{ \text{CP sign} }  (s_{ \text{II} }, \delta_{ \text{II} } ). 
\label{correspondence2}
\end{eqnarray}
%
%
%
In this sense there is the one-to-one correspondence between the two intrinsic 
degeneracy solutions and the sign-$\Delta m^2_{31}$ degeneracy solutions, 
a charming property which one can enjoy only with the convention B.\footnote{
Notice that $s_{ \text{II} }  \geq s_{1}$, or $s_{ \text{II} }  \leq s_{1}$, 
depending upon the region of experimental parameters. 
Therefore, once the correspondence relations (\ref{correspondence1}) 
and (\ref{correspondence2}) are established, we cannot take the convention A 
which implies that always $s_{ \text{IV} } \geq s_{ \text{III} }$. 
If we try to solve the problem of obtaining the eightfold degeneracy solutions for a 
given set of $(P, P^{CP})$. it would be possible to take the convention A. 
}
%
The relations will be further extended into the other types pf degeneracies 
and completed in Sec.~\ref{structure}.

However, there exists a somewhat disturbing feature of this convention; 
The solutions have discontinuity as a function of $E$, the neutrino energy. 
It by no means, however, that the solutions are unphysical. 
It merely implies that  the two solutions interchange themselves at the 
discontinuous point.

\section{$\theta_{23}$ Octant Degeneracy in CP-conjugate Measurement}
\label{CP-octant}

In this section we address the parameter degeneracy solutions, 
assuming that $\theta_{23} \neq \pi/4$, 
which have the different $\theta_{23}$ octant from the true one.  
We denote the octant in which the quantity lives by the superscript 
``true'' or ``false'' where the true $\theta_{23}$ can be in either 
the first or the second octants. 
Our treatment of the octant degeneracy will be done under the approximation 
that the two solutions of $\theta_{23}$ has the same value of $\sin 2 \theta_{23}$. 
In this approximation $Y_{\pm}$ is independent of the octant, 
but $X_{\pm}$ and $Z$ in (\ref{XYZ}) have to have additional superscripts 
such as $X_{\pm}^{\text{true}}$ or $X_{\pm}^{\text{false}}$ to indicate in 
which octant they live. 
Using the definition of $X_{\pm}$ and $Z_{\pm}$ given in (\ref{XYZ}) they are 
related with each other as 
\begin{eqnarray} 
X_{\pm}^{\text{false}} = \cot^2 \theta_{23}^{ \text{true} } X_{\pm}^{\text{true}}, 
\hspace{8mm}
Z_{\pm}^{\text{false}} = \tan^2 \theta_{23}^{ \text{true} } Z_{\pm}^{\text{true}}.
\label{Xtrue-Xfalse}
\end{eqnarray}

\subsection{Intrinsic degeneracy in the false $\theta_{23}$ octant}
\label{octant-intrinsic}

We first discuss the case in which the intrinsic degeneracy solutions exist 
in a $\theta_{23}$ octant different from the true one. 
The input solution ($s_{1}, \delta_{1}$) and the different octant clone solution 
($s_{5}, \delta_{5}$) satisfy the following equations  
\begin{eqnarray} 
P &=& 
X_{\pm}^{\text{true}} s_{1}^2 + 
Y_{\pm} s_{1} \left( 
\cos \delta_{1} \cos \Delta_{31} \mp \sin \delta_{1} \sin \Delta_{31} 
\right) + Z^{\text{true}},
\nonumber \\
P &=& 
X_{\pm}^{\text{false}} s_{5}^2 + 
Y_{\pm} s_{5} \left( 
\cos \delta_{5} \cos \Delta_{31} \mp \sin \delta_{5} \sin \Delta_{31} 
\right) + Z^{\text{false}}, 
\label{CP-octant-def1}
\end{eqnarray}
in the neutrino channel, and 
\begin{eqnarray} 
P^{CP} &=& 
X_{\mp}^{\text{true}} s_{1}^2 - 
Y_{\mp} s_{1} \left( 
\cos \delta_{1} \cos \Delta_{31} \pm \sin \delta_{1} \sin \Delta_{31} 
\right) + Z^{\text{true}}, 
\nonumber \\
P^{CP} &=& 
X_{\mp}^{\text{false}} s_{5}^2 - 
Y_{\mp} s_{5} \left( 
\cos \delta_{5} \cos \Delta_{31} \pm \sin \delta_{5} \sin \Delta_{31} 
\right) + Z^{\text{false}}, 
\label{CP-octant-def2}
\end{eqnarray}
in the CP-conjugate channel. 
The way we obtain $s_{5}$ and $\delta_{5}$ follows exactly the one 
for the sign-$\Delta m^2_{31}$ degeneracy solutions in Sec .~\ref{CP-signdm2}. 
Therefore, we can skip many equations and just say that 
(\ref{CP-octant-def1}) and (\ref{CP-octant-def2}) lead to 
\begin{eqnarray} 
s_{5} \cos \delta_{5} &=& 
\frac{ 1 }{ \cos \Delta_{31} } 
\frac{ 1 }{ (C_{2}^{(+)} )^2  - (C_{2}^{(-)} )^2 } 
\left[
C_{2}^{(+)} T_{3 \pm}^{CP} \mp 
C_{2}^{(-)} \left( T_{4 \pm}^{CP} - 2 s_{5}^2 \right)
\right], 
\nonumber \\
s_{5} \sin \delta_{5} &=& 
\frac{ 1 }{ \sin \Delta_{31} } 
\frac{ 1 }{ (C_{2}^{(+)} )^2  - (C_{2}^{(-)} )^2 } 
\left[
C_{2}^{(-)} T_{3 \pm}^{CP} \mp 
C_{2}^{(+)} \left( T_{4 \pm}^{CP} - 2 s_{5}^2 \right)
\right]. 
\label{CP-oct-intr-delta-solution}
\end{eqnarray}
In (\ref{CP-oct-intr-delta-solution}), we have used the new notations which 
essentially is a generalization of the previous one as well as new ones: 
\begin{eqnarray} 
C_{2}^{(\pm)} &\equiv& 
\frac{ Y_{+} }{X_{+}^{\text{false}} }  \pm \frac{ Y_{-} }{X_{-}^{\text{false}} }, 
\hspace{7mm} 
F^{(\pm)} \equiv 
\frac{ X_{+}^{\text{true}} }{X_{+}^{\text{false}} }  \pm \frac{ X_{-}^{\text{true}} }{X_{-}^{\text{false}} }, 
\nonumber \\
G^{(\pm)} &\equiv& 
\left( \frac{ 1 }{X_{+}^{\text{false}} }  \pm \frac{ 1 }{X_{-}^{\text{false}} } \right) \left( Z^{\text{true}} - Z^{\text{false}} \right). 
\label{CFG-def}
\end{eqnarray}
and defined $T_{3 \pm}^{CP}$ and $T_{4 \pm}^{CP}$, in parallel with 
$T_{1 \pm}^{CP}$ and $T_{2 \pm}^{CP}$ in (\ref{T1T2-def}), as 
\begin{eqnarray} 
T_{3 \pm}^{CP} &\equiv& 
\pm F^{(-)} s_{1}^2  \pm G^{(-)} + C_{2}^{(+)} s_{1} \cos \delta_{1} \cos \Delta_{31} - C_{2}^{(-)} s_{1} \sin \delta_{1} \sin \Delta_{31}, 
\nonumber \\
T_{4 \pm}^{CP} &\equiv& 
F^{(+)} s_{1}^2 + G^{(+)} \pm C_{2}^{(-)} s_{1} \cos \delta_{1} \cos \Delta_{31} \mp C_{2}^{(+)} s_{1} \sin \delta_{1} \sin \Delta_{31}. 
\label{T3T4-def}
\end{eqnarray}
Inserting (\ref{CP-oct-intr-delta-solution}) into $\cos^2 \delta_{5} + \sin^2 \delta_{5} =1$ 
we obtain the quartic equation for $s_{5}$ as 
\begin{eqnarray} 
4 \left( 1 + R_{2}^2 \right) s_{5}^4 - 4 H_{\pm} s_{5}^2 + I_{\pm} = 0
\label{s2-octant-intrinsic}
\end{eqnarray}
which is actually a quadratic equation of $s^2_{13}$ because of the 
quadratic dependence on $s_{13}$ of $s_{5} \cos \delta_{5}$ and 
$s_{5} \sin \delta_{5}$. 
Thus, there exist only two physical (i.e., positive) solutions, which 
implies that the octant degeneracy is two-fold. 
In (\ref{s2-octant-intrinsic}) $H$ and $I$ are defined as 
\begin{eqnarray} 
H_{\pm} = \frac{1}{4} \cos^2 \Delta_{31}  ( C_{2}^{(-)} )^2 
\left\{ 1 - \left( \frac{ C_{2}^{(+)} }{ C_{2}^{(-)} } \right)^2 \right\}^2 
\mp \frac{ 1 }{ \sin^2 \Delta_{31} } \left( \frac{ C_{2}^{(+)} }{ C_{2}^{(-)} } \right) T_{3 \pm}^{CP} 
+  \left( 1 + R_{2}^2 \right) T_{4 \pm}^{CP}, 
\label{H-def}
\end{eqnarray}
\begin{eqnarray} 
I_{\pm} = 
\left[ \cot^2 \Delta_{31} + \left( \frac{ C_{2}^{(+)} }{ C_{2}^{(-)} } \right)^2 \right]  (T_{3 \pm}^{CP})^2 
\mp \frac{ 2 }{ \sin^2 \Delta_{31} } \left( \frac{ C_{2}^{(+)} }{ C_{2}^{(-)} } \right) T_{3 \pm}^{CP} T_{4 \pm}^{CP} 
+ 
\left( 1 + R_{2}^2 \right) (T_{4 \pm}^{CP})^2. 
\label{I-def}
\end{eqnarray}
where $R_{2} \equiv \frac {C_{2}^{(+)} }{ C_{2}^{(-)} } \cot \Delta_{31}$. 
Then, the octant degeneracy solution $s_{5}$ is given by 
\begin{eqnarray} 
s_{5}^2 = \frac{1}{ 2 \left( 1 + R_{2}^2 \right) } 
\left[ H_{\pm} 
\hspace{1mm} [\pm]^* \hspace{1mm} 
\sqrt{ D^{ \text{oct-intr} }_{\pm} } 
\right], 
\label{CP-oct-intr-s-solution}
\end{eqnarray}
where $[\pm]^*$ is the temporary sign to be specified below, 
and $D^{ \text{oct-intr} }_{\pm}$ is defined as 
\begin{eqnarray} 
D^{ \text{oct-intr} }_{\pm} &=& 
\left\{ 1 - \left( \frac{ C_{2}^{(+)} }{ C_{2}^{(-)} } \right)^2 \right\}^2 
\biggl[ 
\frac{1}{16} \cos^4 \Delta_{31} (C_{2}^{(-)})^4 
\left\{ 1 - \left( \frac{ C_{2}^{(+)} }{ C_{2}^{(-)} } \right)^2 \right\}^2 
\nonumber \\
&-& \cot^2 \Delta_{31} (T_{3 \pm}^{CP})^2 
\mp \frac{1}{2} \cot^2 \Delta_{31} C_{2}^{(+)} C_{2}^{(-)} T_{3 \pm}^{CP} 
+ \frac{1}{2} \cos^2 \Delta_{31}  (C_{2}^{(-)})^2 \left( 1 + R_{2}^2 \right) T_{4 \pm}^{CP}
\biggr]. 
\nonumber \\
\label{Dpm-oct-intr}
\end{eqnarray}
The region defined by $D^{ \text{oct-intr} }_{\pm} \leq 0$ defines the region in which 
there is no intrinsic degeneracy solution in an octant different from the true $\theta_{23}$. 
Once there is a solution, it must be obvious that $s_{5}^2$ is positive definite 
following the similar argument as in Sec.~\ref{CP-signdm2}.

The region of no $\theta_{23}$ octant degeneracy solution is displayed as the white regions in the top two panels in Figs.~\ref{R5-R6-R7-R8} and \ref{D5-D6-D7-D8} for 
superbeam type settings, SB1, MB1, and MB2 
(to be defined in Sec.~\ref{overview-variable}). 
The corresponding informations for neutrino factory setting NF are given in 
Figs.~\ref{RN-nufact} and \ref{DN-nufact}. 
In fact, one can observe that the feature of no octant degeneracy region is insensitive 
to the baseline and energies, but depend on in which $\theta_{23}$ octant 
the true solution exists. 
If $\theta_{23}^{\text{true}}$ is in the first octant, no-degeneracy regions 
are around $\delta \sim \pi/2$ and $3\pi/2$, 
whereas if it is in the second octant it is confined into the small $\theta_{13}$ region, 
$\sin^2 2\theta_{13} \lesssim 10^{-3}$. 
Both of the features can be easily understood by drawing the bi-probability plot; 
The region of no degeneracy solution is the one spanned only by the ellipses 
with a single octant $\theta_{23}$ \cite{uchinami-thesis}.



Now, we have to revisit the issue of convention to define unambiguously 
the octant degeneracy solutions. 
We take the following new convention: 

\vspace{3mm}
\noindent
{\bf Convention C}: 
We define ($s_{ \text{V} }, \delta_{ \text{V} }$) and ($s_{ \text{VI} }, \delta_{ \text{VI} }$) 
such that they have a smooth limit to the intrinsic degeneracy solutions 
($s_{1}, \delta_{1}$) and ($s_{ \text{II} }, \delta_{ \text{II} }$), respectively, 
when the maximum $\theta_{23}$ limit $\theta_{23} \rightarrow \pi/4$ is taken.
It can be understood as a consistency condition.

%

\vspace{3mm}
\noindent 
The convention allows us to formulate perturbative framework with use of the 
small expansion parameter $\theta_{23} - \pi/4$, 
as we will do in Appendix~\ref{octant-perturb}. 
Notice that the degenerate solutions 
($s_{ \text{II} }, \delta_{ \text{II} }$), 
($s_{ \text{III} }, \delta_{ \text{III} }$), and 
($s_{ \text{IV} }, \delta_{ \text{IV} }$) themselves obtained 
in Sec.~\ref{CP-intrinsic-sign} are valid independent of the value of $\theta_{23}$.  

In the maximum $\theta_{23}$ limit, $F^{(-)}$ and $G^{(\pm)}$ all vanish, and 
$F^{(+)} = 2$, $C_{2}^{(\pm)} = C^{(\pm)}$ and $R_{2} = R$ hold. 
Then, $D^{ \text{oct-intr} }_{\pm}$ has the maximum $\theta_{23}$ limit 
\begin{eqnarray} 
D^{ \text{oct-intr-max} }_{\pm} \equiv \lim_{ \theta_{23} \rightarrow \pi/4 }
D^{ \text{oct-intr} }_{\pm} \equiv (d^{ \text{oct-intr} }_{\pm})^2
\label{Dpm-oct-intr2}
\end{eqnarray}
where
\begin{eqnarray} 
d^{ \text{oct-intr} }_{\pm} &=&  
\cos \Delta_{31} 
\left\{ 1 - \left( \frac{ C^{(+)} }{ C^{(-)} } \right)^2 \right\} 
\nonumber \\
&\times&
\left[ 
\frac{1}{4}  \cos \Delta_{31} (C^{(-)})^2 
\left\{ 1 - \left( \frac{ C^{(+)} }{ C^{(-)} } \right)^2 \right\}
\pm C^{(-)} \left( s_{1} \cos \delta_{1}  +  R s_{1} \sin \delta_{1}  \right) 
\right]. 
\label{dpm-def}
\end{eqnarray}
The smooth limit to the same-octant intrinsic degeneracy solution can be 
achieved by taking the sign convention 
\begin{eqnarray} 
s_{ \text{V} }^2 &=& \frac{1}{ 2 \left( 1 + R_{2}^2 \right) } 
\left[
H_{\pm} 
-
d^{ \text{oct-intr} }_{\pm} 
\sqrt{ \frac{ D^{ \text{oct-intr} }_{\pm} }{ (d^{ \text{oct-intr} }_{\pm})^2 } } 
\right], 
\nonumber \\
s_{ \text{VI} }^2 &=& \frac{1}{ 2 \left( 1 + R_{2}^2 \right) } 
\left[
H_{\pm} 
+ 
d^{ \text{oct-intr} }_{\pm} 
\sqrt{ \frac{ D^{ \text{oct-intr} }_{\pm} }{ (d^{ \text{oct-intr} }_{\pm})^2 } } 
\right]. 
\label{CP-oct-intr-s5s6-solution}
\end{eqnarray}
One can easily verify that in the maximum $\theta_{23}$ limit 
$s_{ \text{V} }$ and $s_{ \text{VI} }$ tend to 
$s_{1}$ and $s_{ \text{II} }$, respectively. 
%

\subsection{Sign-$\Delta m^2_{31}$ degeneracy in the false $\theta_{23}$ octant}
\label{octant-sign}

We next discuss the sign-$\Delta m^2_{31}$ degeneracy in a $\theta_{23}$ octant 
different from the true one. 
The input solution ($s_{1}, \delta_{1}$) and the false octant clone solution 
($s_{7}, \delta_{7}$) satisfy the following equations  
\begin{eqnarray} 
P &=& 
X_{\pm}^{\text{true}} s_{1}^2 + 
Y_{\pm} s_{1} \left( 
\cos \delta_{1} \cos \Delta_{31} \mp \sin \delta_{1} \sin \Delta_{31} 
\right) + Z^{\text{true}},
\nonumber \\
P &=& 
X_{\mp}^{\text{false}} s_{7}^2 + 
Y_{\mp} s_{7} \left( 
\cos \delta_{7} \cos \Delta_{31} \pm \sin \delta_{7} \sin \Delta_{31} 
\right) + Z^{\text{false}}, 
\label{CP-octant-sign-def1}
\end{eqnarray}
in the neutrino channel, and 
\begin{eqnarray} 
P^{CP} &=& 
X_{\mp}^{\text{true}} s_{1}^2 - 
Y_{\mp} s_{1} \left( 
\cos \delta_{1} \cos \Delta_{31} \pm \sin \delta_{1} \sin \Delta_{31} 
\right) + Z^{\text{true}}, 
\nonumber \\
P^{CP} &=& 
X_{\pm}^{\text{false}} s_{7}^2 - 
Y_{\pm} s_{7} \left( 
\cos \delta_{7} \cos \Delta_{31} \mp \sin \delta_{7} \sin \Delta_{31} 
\right) + Z^{\text{false}}. 
\label{CP-octant-sign-def2}
\end{eqnarray}
in the CP-conjugate channel. 
Proceeding along the same way as in Secs.~\ref{CP-signdm2} and \ref{octant-intrinsic} 
we obtain 
\begin{eqnarray} 
s_{7} \cos \delta_{7} &=& 
\frac{ 1 }{ \cos \Delta_{31} } 
\frac{ 1 }{ (C_{2}^{(+)} )^2  - (C_{2}^{(-)} )^2 } 
\left[
C_{2}^{(+)} T_{5 \pm}^{CP} \pm 
C_{2}^{(-)} \left( T_{6 \pm}^{CP} - 2 s_{7}^2 \right)
\right], 
\nonumber \\
s_{7} \sin \delta_{7} &=& 
\frac{ 1 }{ \sin \Delta_{31} } 
\frac{ 1 }{ (C_{2}^{(+)} )^2  - (C_{2}^{(-)} )^2 } 
\left[
C_{2}^{(-)} T_{5 \pm}^{CP} \pm 
C_{2}^{(+)} \left( T_{6 \pm}^{CP} - 2 s_{7}^2 \right)
\right]. 
\label{CP-oct-sign-delta-solution}
\end{eqnarray}
where we have defined 
\begin{eqnarray} 
T_{5 \pm}^{CP} &\equiv& 
\pm E_{2}^{(-)} s_{1}^2 \mp G^{(-)} + D_{2}^{(+)} s_{1} \cos \delta_{1} \cos \Delta_{31} - D_{2}^{(-)} s_{1} \sin \delta_{1} \sin \Delta_{31}, 
\nonumber \\
T_{6 \pm}^{CP} &\equiv& 
E_{2}^{(+)} s_{1}^2 + G^{(+)} \pm D_{2}^{(-)} s_{1} \cos \delta_{1} \cos \Delta_{31} \mp D_{2}^{(+)} s_{1} \sin \delta_{1} \sin \Delta_{31}. 
\label{T5T6-def}
\end{eqnarray}
We have used the notations 
\begin{eqnarray} 
D_{2}^{(\pm)} &\equiv& 
\frac{ Y_{+} }{X_{-}^{\text{false}} }  \pm \frac{ Y_{-} }{X_{+}^{\text{false}} }, 
\hspace{7mm} 
E_{2}^{(\pm)} \equiv 
\frac{ X_{+}^{\text{true}} }{X_{-}^{\text{false}} }  \pm \frac{ X_{-}^{\text{true}} }{X_{+}^{\text{false}} }, 
\nonumber \\
G^{(\pm)} &\equiv& 
\left( \frac{ 1 }{X_{+}^{\text{false}} }  \pm \frac{ 1 }{X_{-}^{\text{false}} } \right) \left( Z^{\text{true}} - Z^{\text{false}} \right). 
\label{DEG-def}
\end{eqnarray}
Inserting (\ref{CP-oct-sign-delta-solution}) into 
$\cos^2\delta_{7} + \sin^2\delta_{7} = 1$
we obtain the quartic equation of $s_{13}$, 
\begin{eqnarray} 
4 \left( 1 + R_{2}^2 \right) s_{7}^4 - 4 J_{\pm} s_{7}^2 + K_{\pm} = 0, 
\label{s2-octant-sign}
\end{eqnarray}
which is actually a quadratic equation of $s^2_{13}$ because of the 
quadratic dependence on $s_{13}$ of $s_{7} \cos \delta_{7}$ and 
$s_{7} \sin \delta_{7}$. 
Thus, there exist only two physical (i.e., positive) solutions, which 
implies that the octant degeneracy is two-fold. 
In (\ref{s2-octant-sign}) $J$ and $K$ are defined as 
\begin{eqnarray} 
J_{\pm} = \frac{1}{4} \cos^2 \Delta_{31}  ( C_{2}^{(-)} )^2 
\left[ 1 - \left( \frac{ C_{2}^{(+)} }{ C_{2}^{(-)} } \right)^2 \right]^2 
\pm \frac{ 1 }{ \sin^2 \Delta_{31} } \left( \frac{ C_{2}^{(+)} }{ C_{2}^{(-)} } \right) T_{5 \pm}^{CP} 
+  \left( 1 + R_{2}^2 \right) T_{6 \pm}^{CP}, 
\label{J-def}
\end{eqnarray}
\begin{eqnarray} 
K_{\pm} = 
\left[ \cot^2 \Delta_{31} + \left( \frac{ C_{2}^{(+)} }{ C_{2}^{(-)} } \right)^2 \right]  (T_{5 \pm}^{CP})^2 
\pm \frac{ 2 }{ \sin^2 \Delta_{31} } \left( \frac{ C_{2}^{(+)} }{ C_{2}^{(-)} } \right) T_{5 \pm}^{CP} T_{6 \pm}^{CP} 
+ 
\left( 1 + R_{2}^2 \right) (T_{6 \pm}^{CP})^2. 
\label{K-def}
\end{eqnarray}
Then, the sign-$\Delta m^2_{31}$ degeneracy solution across 
$\theta_{23}$ octant is given by 
\begin{eqnarray} 
s_{7}^2 = \frac{1}{ 2 \left( 1 + R_{2}^2 \right) } 
\left[ J_{\pm} 
\hspace{1mm} [\pm]^* \hspace{1mm} 
\sqrt{ D^{ \text{oct-sign} }_{\pm} } 
\right], 
\label{CP-oct-sign-s-solution}
\end{eqnarray}
where $[\pm]^*$ is the temporary sign to be specified below and 
$D^{ \text{oct-sign} }_{\pm}$ is defined as 
\begin{eqnarray} 
D^{ \text{oct-sign} }_{\pm} &=& 
\left\{ 1 - \left( \frac{ C_{2}^{(+)} }{ C_{2}^{(-)} } \right)^2 \right\}^2 
\biggl[ 
\frac{1}{16} \cos^4 \Delta_{31} (C_{2}^{(-)})^4 
\left\{ 1 - \left( \frac{ C_{2}^{(+)} }{ C_{2}^{(-)} } \right)^2 \right\}^2 
\nonumber \\
&-& \cot^2 \Delta_{31} (T_{5 \pm}^{CP})^2 
\pm \frac{1}{2} \cot^2 \Delta_{31} C_{2}^{(+)} C_{2}^{(-)} T_{5 \pm}^{CP} 
+ \frac{1}{2} \cos^2 \Delta_{31}  (C_{2}^{(-)})^2 \left( 1 + R_{2}^2 \right) T_{6 \pm}^{CP}
\biggr]. 
\nonumber \\
\label{Dpm-oct-sign}
\end{eqnarray}
The region defined by $D^{ \text{oct-sign} }_{\pm} \leq 0$ defines the region of no 
sign-$\Delta m^2_{31}$ degeneracy solutions in the false $\theta_{23}$ octant. 
The region of no sign-octant degeneracy solution is displayed as the white regions 
in the bottom two panels in Figs.~\ref{R5-R6-R7-R8} and \ref{D5-D6-D7-D8} for 
superbeam type settings, SB1, MB1, and MB2. 
The corresponding informations for neutrino factory setting NF are given in 
Figs.~\ref{RN-nufact} and \ref{DN-nufact}. 
Once there is a solution, the same argument as before assures that the solutions 
for $s_{7}^2$ in (\ref{CP-oct-sign-s-solution}) are positive definite.

To define unambiguously the sign-$\Delta m^2_{31}$ degeneracy solutions 
in the false $\theta_{23}$ octant we need the following new convention. 
That is, we need to take both the maximum $\theta_{23}$ and the vacuum limits.

\vspace{3mm}
\noindent
{\bf Convention D}: 
We take the convention such that 
($s_{ \text{VII} }, \delta_{ \text{VII} }$) and ($s_{ \text{VIII} }, \delta_{ \text{VIII} }$) tend to 
($s_{ \text{III} }^{ \text{vac} }, \delta_{ \text{III} }^{ \text{vac} }$) and ($s_{ \text{IV} }^{ \text{vac} }, \delta_{ \text{IV} }^{ \text{vac} }$), respectively, 
in the simultaneous maximum-$\theta_{23}$ and the vacuum limit.

\vspace{3mm}
\noindent 
One can easily show in the combined limit that 
$\lim_{ \theta_{23} \rightarrow \pi/4, \text{vacuum} }
D^{ \text{oct-sign} }_{\pm} = (d^{ \text{oct-sign} }_{\pm})^2$, 
where $d^{ \text{oct-sign} }_{\pm} = d^{ \text{sign} }_{\pm} (\theta_{23} = \pi/4)$ 
with $d^{ \text{sign} }_{\pm}$ being defined in (\ref{dpm-sign-vac}).
The smooth limit to the maximum $\theta_{23}$ of the sign-$\Delta m^2_{31}$ 
degeneracy solutions can be achieved by taking the sign convention D. 
The solutions read  
\begin{eqnarray} 
s_{ \text{VII} }^2 &=& \frac{1}{ 2 \left( 1 + R_{2}^2 \right) } 
\left[
J_{\pm} - 
d^{ \text{oct-sign} }_{\pm} 
\sqrt{ \frac{ D^{ \text{oct-sign} }_{\pm} }{ (d^{ \text{oct-sign} }_{\pm})^2 } } 
\right], 
\nonumber \\
s_{ \text{VIII} }^2 &=& \frac{1}{ 2 \left( 1 + R_{2}^2 \right) } 
\left[
J_{\pm} + 
d^{ \text{oct-sign} }_{\pm} 
\sqrt{ \frac{ D^{ \text{oct-sign} }_{\pm} }{ (d^{ \text{oct-sign} }_{\pm})^2 } } 
\right]. 
\label{CP-oct-sign-s7s8-solution}
\end{eqnarray}

\section{Structure of Parameter degeneracy}
\label{structure}

We now make the structure of parameter degeneracy transparent based on 
knowledges obtained in the previous two sections by using CP-conjugate 
measurement. 
First, we summarize the relationships between the true and the degeneracy solutions.

\subsection{Mappings between the true and the degeneracy solutions}
\label{mapping}

\begin{figure}[bhtp]
\begin{center}
\vglue 0.3cm
\includegraphics[bb=0 0 480 210 , clip, width=0.6\textwidth]{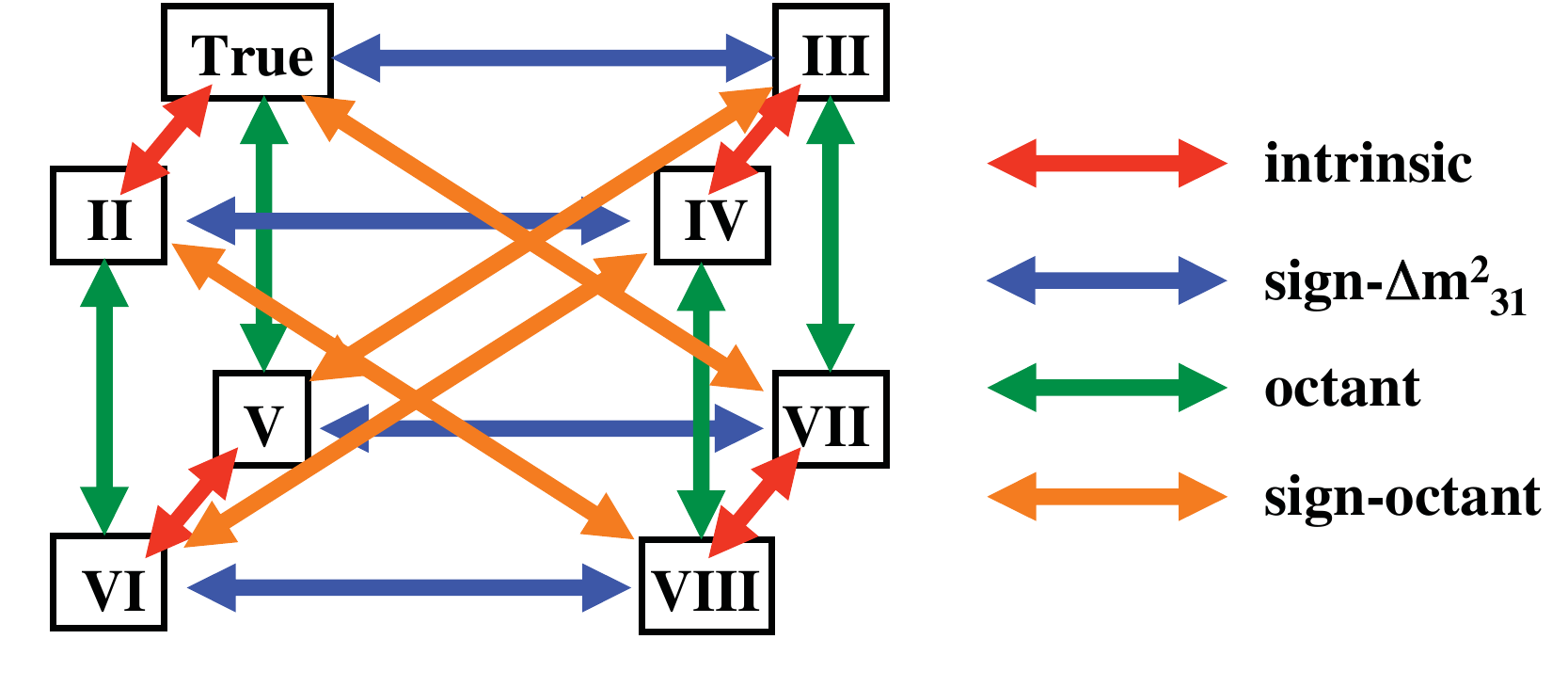}
\end{center}
\vglue -0.2cm
\caption{
The correspondence relations (\ref{correspondence3}) between the solutions, 
the true one and the clones II$-$VIII, are pictorially represented. 
}
\label{eightfold-relation}
\end{figure}

Let us start by putting the relationship between each intrinsic degeneracy 
pair of the solutions in order. 
If we denote the relationship between the intrinsic degeneracy solution 
derived in Sec.~\ref{CP-intrinsic} as 
$s_{ \text{II} } = \zeta_{\pm} ( s_{1}, \delta_{1}, \theta_{23}^{\text{true}} )$ and 
$\delta_{ \text{II} } =  \iota_{\pm} ( s_{1}, \delta_{1}, \theta_{23}^{\text{true}} )$ 
then the other intrinsic degeneracy 
pairs of the solutions satisfy 
\begin{eqnarray} 
s_{ \text{IV} } &=& \zeta_{\mp} ( s_{ \text{III} }, \delta_{ \text{III} }, \theta_{23}^{\text{true}} ), 
\hspace{16mm}
\delta_{ \text{IV} } = \iota_{\mp} ( s_{ \text{III} }, \delta_{ \text{III} }, \theta_{23}^{\text{true}} ), 
\nonumber \\
s_{ \text{VI} } &=& \zeta_{\pm} ( s_{ \text{V} }, \delta_{ \text{V} }, \pi/2 - \theta_{23}^{\text{true}}), 
\hspace{10mm}
\delta_{ \text{VI} } = \iota_{\pm} ( s_{ \text{V} }, \delta_{ \text{V} }, \pi/2 - \theta_{23}^{\text{true}} ), 
\nonumber \\
s_{ \text{VIII} } &=& \zeta_{\mp} ( s_{ \text{VII} }, \delta_{ \text{VII} }, \pi/2 - \theta_{23}^{\text{true}} ), 
\hspace{8mm}
\delta_{ \text{VIII} } = \iota_{\mp} ( s_{ \text{VII} }, \delta_{ \text{VII} }, \pi/2 - \theta_{23}^{\text{true}} ), 
\label{intrinsic-pair}
\end{eqnarray}
completing the one-to-one correspondence between them. 
We then summarize the one-to-one correspondence relations between 
the solutions which involve the $\Delta m^2_{31}$-sign and/or the $\theta_{23}$ 
octant flips: 
\begin{eqnarray} 
s_{ \text{III} } = \xi_{\pm}^{ \text{CP sign} } (s_{1}, \delta_{1}, \theta_{23}^{\text{true}} ), 
\hspace{10mm}
&& s_{ \text{IV} } = \xi_{\pm}^{ \text{CP sign} }  (s_{ \text{II} }, \delta_{ \text{II} }, \theta_{23}^{\text{true}} ), 
\nonumber \\
%
s_{ \text{V} } = \xi_{\pm}^{ \text{CP oct} } (s_{1}, \delta_{1}, \theta_{23}^{\text{true}} ), 
\hspace{10mm}
&& s_{ \text{VI} } = \xi_{\pm}^{ \text{CP oct} }  (s_{ \text{II} }, \delta_{ \text{II} }, \theta_{23}^{\text{true}} ),
\nonumber \\
%
s_{ \text{VII} }  = \xi_{\pm}^{ \text{CP oct-sign} } (s_{1}, \delta_{1}, \theta_{23}^{\text{true}}  ), 
\hspace{10mm}
&& s_{ \text{VIII} } = \xi_{\pm}^{ \text{CP oct-sign} }  (s_{ \text{II} }, \delta_{ \text{II} }, \theta_{23}^{\text{true}}  ), 
\nonumber \\
%
s_{ \text{VII} } = \xi_{\pm}^{ \text{CP sign} }  (s_{ \text{V} }, \delta_{ \text{V} }, \pi/2 - \theta_{23}^{\text{true}} ), 
\hspace{10mm}
&& s_{ \text{VIII} } = \xi_{\pm}^{ \text{CP sign} }  (s_{ \text{VI} }, \delta_{ \text{VI} }, \pi/2 - \theta_{23}^{\text{true}} ), 
\nonumber \\
%
s_{ \text{VII} } = \xi_{\mp}^{ \text{CP oct} } (s_{ \text{III} }, \delta_{ \text{III} }, \theta_{23}^{\text{true}}  ),
\hspace{10mm}
&& s_{ \text{VIII} } = \xi_{\mp}^{ \text{CP oct} }  (s_{ \text{IV} }, \delta_{ \text{IV} }, \theta_{23}^{\text{true}}  ), 
\nonumber \\
%
s_{ \text{V} } = \xi_{\mp}^{ \text{CP oct-sign} }  (s_{ \text{III} }, \delta_{ \text{III} }, \theta_{23}^{\text{true}}  ), 
\hspace{10mm}
&& s_{ \text{VI} } = \xi_{\mp}^{ \text{CP oct-sign} }  (s_{ \text{IV} }, \delta_{ \text{IV} }, \theta_{23}^{\text{true}}  ). 
%
\label{correspondence3}
\end{eqnarray}
The functional form of $\xi_{\pm}^{ \text{CP sign} }$, 
$\xi_{\pm}^{ \text{CP oct} }$, and $\xi_{\pm}^{ \text{CP oct-sign} }$ are defined in 
(\ref{correspondence1}) in Sec.~\ref{convention}, 
the first line in (\ref{CP-oct-intr-s5s6-solution}) in Sec.~\ref{octant-intrinsic}, and 
the first line in (\ref{CP-oct-sign-s7s8-solution}), respectively.  
There exist the similar relationships between $\delta$'s through the function 
$\eta_{\pm}^{\text{CP sign}} $, but we do not display them explicitly here. 
The correspondence relations (\ref{correspondence3}) between the true and 
the degeneracy solutions II$-$VIII are pictorially represented in 
Fig.~\ref{eightfold-relation}. 
The relationship between the degeneracy solutions in (\ref{intrinsic-pair}) and  (\ref{correspondence3}) as well as in Fig.~\ref{eightfold-relation} {\em is}
the precise meaning of the statement that nature of the parameter degeneracy 
is the intrinsic degeneracy duplicated respectively by the $\Delta m^2_{31}$-sign 
and the $\theta_{23}$ octant flips.

It is easy to prove (\ref{correspondence3}) by considering the original defining 
equations for the degeneracy. 
For example, it is easy to show that 
$\xi_{\pm}^{\text{CP sign}} (s_{\text{V}}, \delta_{\text{V}}, \pi/2 - \theta_{23}^{\text{true}}) \equiv s_{\text{X}}$ 
and 
$\eta_{\pm}^{\text{CP sign}} (s_{\text{V}}, \delta_{\text{V}}, \pi/2 - \theta_{23}^{\text{true}}) \equiv \delta_{\text{X}}$ are solutions of the equation 
\begin{eqnarray}
X_{\pm}^{\text{(false)}} s_{\text{V}}^2 + Y_{\pm} s_{\text{V}} \cos(\delta_{\text{V}} \pm \Delta_{31}) + Z^{\text{(false)}}
&=&
X_{\mp}^{\text{(false)}} s_{\text{X}}^2 + Y_{\mp} s_{\text{X}} \cos(\delta_{\text{X}} \mp \Delta_{31}) + Z^{\text{(false)}} \nonumber \\
X_{\mp}^{\text{(false)}} s_{\text{V}}^2 - Y_{\mp} s_{\text{V}} \cos(\delta_{\text{V}} \mp \Delta_{31}) + Z^{\text{(false)}}
&=&
X_{\pm}^{\text{(false)}} s_{\text{X}}^2 - Y_{\pm} s_{\text{X}} \cos(\delta_{\text{X}} \pm \Delta_{31}) + Z^{\text{(false)}}. \nonumber \\
\label{prove-relation}
\end{eqnarray}
Note that left hand side of Eq.(\ref{prove-relation}) can be expressed by $s_{1}$, $\delta_{1}$, and $\theta_{23}^{\text{true}}$ 
from Eq.(\ref{CP-octant-def1}) and (\ref{CP-octant-def2}), 
it is nothing but the equations of sign-octant degeneracy.

Though we do not present explicit formulas the similar structure exists in all the 
degeneracy solutions in other settings, T-conjugate, Golden-Silver, and 
CPT-conjugate measurement to be discussed in the following sections.

\subsection{Relation between cases of true normal vs. true inverted mass hierarchies}
\label{normal-inverted}

Here, we note an important property of the degeneracy solutions. 
Namely, if we know the degeneracy solutions for the input true normal mass hierarchy, 
then the solutions for the true inverted mass hierarchy can be immediately 
obtained from the former ones.\footnote{
In spite of this charming and useful property we have been denoting explicitly  
the input mass hierarchies as the $\pm$ signs in the subscript for clarity and 
simplicity of keep tracking of the true mass hierarchy we are working.
}
%
Since the statement can be confusing to the readers, we give an explicit proof as below.

Let us take the sign-$\Delta m^2_{31}$ degeneracy (without $\theta_{23}$ octant flip) 
for definiteness. 
For clarity, we denote the degeneracy solution for the case of true normal 
and true inverted mass hierarchies as 
($s_{3 \text{N} }, \delta_{3  \text{N} }$) and  ($s_{3  \text{I} }, \delta_{3  \text{I} }$), respectively, where the subscript ``3'' implies either III or IV. 
For a given set of the probabilities $P$ and $P^{CP}$, 
assuming that the true mass hierarchy is normal, the true solution 
($s_{1}, \delta_{1}$) and the fake one ($s_{3 \text{N} }, \delta_{3  \text{N} }$) satisfy 
\begin{eqnarray}
\left( P = \right)~ 
X_{+} s_{1}^2 + Y_{+} s_{1} \cos (\delta_{1} + \Delta_{31}) + Z
&=&
X_{-} s_{3 \text{N} }^2 + Y_{-} s_{3 \text{N} } \cos (\delta_{3 \text{N} } - \Delta_{31}) + Z,  
\nonumber \\
\left( P^{CP} = \right)~ 
X_{-} s_{1}^2 - Y_{-} s_{1} \cos (\delta_{1} - \Delta_{31}) + Z
&=&
X_{+} s_{3 \text{N} }^2 - Y_{+} s_{3 \text{N} } \cos (\delta_{3 \text{N} } + \Delta_{31}) + Z. 
\label{eq-NH}
\end{eqnarray}
If the true mass hierarchy is inverted, then the degeneracy solution satisfies 
a different set of equations as  
\begin{eqnarray}
\left( P = \right)~ 
X_{-} s_{1}^2 + Y_{-} s_{1} \cos (\delta_{1} - \Delta_{31}) + Z
&=&
X_{+} s_{3 \text{I} }^2 + Y_{+} s_{3 \text{I} } \cos (\delta_{3 \text{I} } + \Delta_{31}) + Z, 
\nonumber \\
\left( P^{CP} = \right)~ 
X_{+} s_{1}^2 - Y_{+} s_{1} \cos (\delta_{1} + \Delta_{31}) + Z
&=&
X_{-} s_{3 \text{I} }^2 - Y_{-} s_{3 \text{I} } \cos (\delta_{3 \text{I} } - \Delta_{31}) + Z. 
\label{eq-IH}
\end{eqnarray}
We define  $\delta_{1 \text{new} }$ and $\delta_{3 \text{Inew} }$   
as $\delta_{1} = \delta_{1 \text{new} }  + \pi$ and 
$\delta_{3 \text{I} } = \delta_{3 \text{Inew} }  + \pi$, respectively,  
and rewrite (\ref{eq-IH}) by using them. It reads 
\begin{eqnarray}
X_{-} s_{1}^2 - Y_{-} s_{1} \cos (\delta_{1 \text{new} } - \Delta_{31}) + Z 
&=&
X_{+} s_{3 \text{I} }^2 - Y_{+} s_{3 \text{I} } \cos (\delta_{3 \text{Inew} } + \Delta_{31}) + Z, 
\nonumber \\
X_{+} s_{1}^2 + Y_{+} s_{1} \cos (\delta_{1 \text{new} } + \Delta_{31}) + Z 
&=&
X_{-} s_{3 \text{I} }^2 + Y_{-} s_{3 \text{I} } \cos (\delta_{3 \text{Inew} } - \Delta_{31}) + Z. 
\label{eq-IH2}
\end{eqnarray}
Comparison between (\ref{eq-NH}) and (\ref{eq-IH2}) tells us that if the set 
($s_{3 \text{N} }, \delta_{3 \text{N} }$) is the solution to the sign-$\Delta m^2_{31}$ degeneracy equation for the true normal hierarchy, then the set 
($s_{3 \text{I} }, \delta_{3 \text{Inew} }$) is the solution 
for  the true inverted hierarchy, 
provided that the true value of $\delta (= \delta_{1})$ is replaced by $\delta_{1 \text{new} }$. 
Stated more explicitly, if the first two equations in (\ref{transformation2}) define the 
sign-$\Delta m^2_{31}$ degeneracy solution for the true normal hierarchy, 
\begin{eqnarray} 
s_{ 3 \text{N} } = \xi_{3} (s_{1}, \delta_{1}), 
\hspace{8mm}
\delta_{ 3 \text{N} } = \eta_{3} (s_{1}, \delta_{1}), 
\label{solution-normal}
\end{eqnarray}
then, the degeneracy solution for the true inverted hierarchy is given, 
if expressed in terms of the true input parameters, as 
%
%
%
\begin{eqnarray} 
s_{ 3 \text{I} } = \xi_{3} (s_{1}, \delta_{1} - \pi), 
\hspace{8mm}
\delta_{ 3 \text{I} } =  \eta_{3} (s_{1}, \delta_{1} - \pi) + \pi. 
\label{solution-inverted2}
\end{eqnarray}
In other word, the mapping functions $\xi_{\pm}$ and $\eta_{\pm}$ are 
related with each other as 
%
%
\begin{eqnarray} 
\xi_{-}^{\text{CP sign}}  (s_{1}, \delta_{1}) &=& 
\xi_{+}^{\text{CP sign}}  (s_{1}, \delta_{1} - \pi), 
\nonumber \\
\eta_{-}^{\text{CP sign}}  (s_{1}, \delta_{1}) &=& 
\eta_{+}^{\text{CP sign}}  (s_{1}, \delta_{1} - \pi) + \pi.  
\label{xi-eta-pm-relation}
\end{eqnarray}
Thus, the sign-$\Delta m^2_{31}$ degeneracy solutions for the true inverted 
hierarchy are essentially determined by the solutions for the true normal hierarchy. 
The relationship between the degeneracy solutions with the normal and the 
inverted hierarchies as the true solution may be understood better by comparing 
the no-slotuion regions given in Fig.~\ref{no-sign-sol-region} which are drawn 
with the inverted hierarchy as the input true solution to those of 
Figs.~\ref{R3-R4} and \ref{D3-D4}.

It can be easily seen that the same treatment goes through for other types 
of degeneracies, 
the intrinsic and the sign-$\Delta m^2_{31}$ degeneracies in the same or the 
different $\theta_{23}$ octants. 
Therefore, the relationship (\ref{xi-eta-pm-relation}) holds for the whole eightfold 
degeneracy.

\subsection{Asymptotic expansion}
\label{asymptotic}

At the end of this section, which is devoted to illuminate general properties of the 
degeneracy solutions, we make comments on high-energy behavior 
of the solutions. 
It would help clarifying some features of the energy dependence of the 
degeneracy solutions which are discussed in the next section.

First of all, it can be easily verified that all the degeneracy solutions $s_{ \text{N} }$ 
and $\delta_{ \text{N} }$ (N=II$-$VIII) derived in 
Secs.~\ref{CP-intrinsic-sign} and \ref{CP-octant} 
have finite asymptotic limit as $E \rightarrow \infty$. 
Then, we note an interesting property that they are invariant under the 
transformation $E \rightarrow - E$, or 
$\Delta_{j1} \rightarrow - \Delta_{j1}$ ($j = 2, 3$). 
The invariance, of course, stems from the one of the oscillation probabilities. 
Though the transformation is unphysical in nature, the transformation property is useful.  
It means that when we do asymptotic expansion of the degeneracy solutions as 
$s_{ \text{N} } = \sum_{n=0} a_{n}^{ \text{N} } (\Delta_{31} )^n$ 
the odd terms are absent: 
\begin{eqnarray} 
s_{ \text{N} } &=& a_{0}^{ \text{N} } + a_{2}^{ \text{N} } \Delta_{31}^2 + 
\mathcal{O} \left( \Delta_{31}^4 \right), 
\nonumber \\
\delta_{ \text{N} } &=& b_{0}^{ \text{N} } + b_{2}^{ \text{N} } \Delta_{31}^2 + 
\mathcal{O} \left( \Delta_{31}^4 \right). 
\label{expansion}
\end{eqnarray}
Absence of the first-order term in $1/E$ implies that onset to the high-energy 
asymptotic behavior of the degeneracy solutions is relatively fast, 
as we will confirm in the next section.

\section{Overview of the Eightfold Parameter Degeneracy in CP-Conjugate Measurement}
\label{overview}

In this section, we try to give an overview of the intrinsic, the sign-$\Delta m^2_{31}$, 
and the $\theta_{23}$ octant degeneracies. 
In fact, the features of the degeneracy solutions are quite different for differing 
baselines and neutrino energies, which makes the overview in a genuine 
sense extremely difficult. 
Therefore, we restrict ourselves in this paper into a few typical settings which may 
be relevant for the settings of future neutrino experiments discussed in the literatures. 
If the readers want to examine features of the degeneracy with some alternative 
experimental parameters, they can do it quite easily by using the analytic 
solutions presented in this paper.

We want to warn the readers that all of our comments to be made in this section 
are qualitative in nature. 
Therefore, when we say, ``spectrum analysis would resolve the degeneracy'' 
it actually means that it may be possible to resolve it {\em if} appropriate experimental 
settings are provided.\footnote{
Similarly, when we say ``the degeneracy A is easier to lift than the degeneracy B'' 
it actually means so {\it provided that} an appropriate experimental condition is 
prepared such that the similar sensitivities would be expected for both the 
solutions A and B. 
}
%
Yet, we try to be based the experiences gained in some previous analyses. 
The readers may still wonder whether the discussion of degeneracy based on 
the probability makes sense because the observable in the experiments must 
be obtained after convolution with neutrino fluxes and cross sections. 
However, this is {\em not} the only possible attitude to take. 
One can, in principe, obtain the ``experimental data of probability'' by 
de-convoluting the fluxes and cross sections as shown in \cite{kobayashi}.

We also remind the readers of the fact that limited statistics in measurement 
has a nontrivial impact on the features of the degeneracy to be observed in 
the actual experiments.
If two (or more) degeneracy solutions exist in nearby locations in the mixing 
parameter space they can merge together, producing an apparent single solution, 
which could be misinterpreted as no degeneracy. 
We have discussed in Secs.~\ref{CP-intrinsic-sign} and \ref{CP-octant} the regions 
of no degenerate solutions. However, there might be cases that 
the degeneracy solutions do exist in these regions because of lack of statistics and/or 
shift of the allowed regions due to systematic uncertainties. 
Discussion of these features is outside the scope of this paper.

\subsection{Variables used for display and baselines and neutrino energies adopted}
\label{overview-variable}

We try to illuminate some characteristic features of the degeneracy 
by presenting the differences between the true solution and the fake ones. 
To display the difference between the solutions we define the ratio $R_{ \text{N} } $ as 
\begin{eqnarray} 
R_{ \text{N} } \equiv \frac{ \sin^2 2\theta_{13}^{ \text{N} } -  \sin^2 2\theta_{13}^{\text{true}} }{ \sin^2 2\theta_{13}^{\text{true}}   }
\label{RN-def}
\end{eqnarray}
where ${ \text{N} }=\text{II-VIII}$ denote the degeneracy solution labels. 
We use the variable $\sin^2 2\theta_{13}$ because it is closer to the 
experimentally measured quantity.
Similarly, we define the quantity $D_{ \text{N} }$ to represent the differences 
between the true and the clone solutions. 
For this purpose, there are two appropriate ways to define it, the types $(1)$ and $(2)$; 
\begin{eqnarray} 
D_{ \text{N} }^{(1)} \equiv \frac{ (\delta^{ \text{N} } - \delta^{ \text{true} } ) }{\pi}, 
\hspace{10mm}
D_{ \text{N} }^{(2)} \equiv \frac{ \delta^{ \text{N} } - (\pi - \delta^{ \text{true} } ) }{\pi}. 
\label{DN-def}
\end{eqnarray}
We use either one of $D_{ \text{N} }^{(1)}$ or $D_{ \text{N} }^{(2)}$ whichever 
appropriate depending upon the degeneracy types.

As typical experimental settings, we use the following four cases of baselines 
and neutrino energies: 

\begin{itemize}

\item
SB1: $L=300$ km, $E=700$ MeV; A short baseline low energy $\nu_{\mu}$ 
(and $\bar{\nu}_{\mu}$) superbeam near the first oscillation maximum 

\item
MB1: $L=1000$ km, $E=2.33$ GeV; A medium baseline superbeam near the first oscillation maximum

\item
MB2: $L=1000$ km, $E=780$ MeV; A medium baseline superbeam near the second oscillation maximum

\item
NF: $L=4000$ km, $E=20$ GeV; A typical setting for neutrino factory 

\end{itemize}

\noindent
The first three settings, SB1, MB2, and MB1, are examined in the  following three subsections (Secs.~\ref{overview-intrinsic}, \ref{overview-sign}, 
and \ref{overview-octant}), while the last one, NF, in Sec.~\ref{overview-nufact}.
We should remark here that the energies corresponding to each baseline are 
chosen rather arbitrarily just for display, while their order of magnitudes are 
dictated by the baseline distances.\footnote{
To avoid the energy of perfect oscillation maximum of about 600 MeV at 300 km, 
where the features of the degeneracy are special, 
we tentatively added 100 MeV. 
Hence, $\epsilon$, the deviation of $\Delta_{31}$ from $\pi/2$ is given by 
$\Delta_{31} (L=300 \text{ km }, E=700 \text{ MeV})= \pi/2 - \epsilon$ from which 
$\epsilon$ is determined as 
$\epsilon=\frac{\pi}{2} \frac{1}{7} = 0.224$.
For the second oscillation maximum, we have arbitrarily chosen that 
$\Delta_{31} (L=1000 \text{ km }, E)= 3 (\pi/2 - \epsilon)$, 
that is $E=780$ MeV.
}
%
In some limited cases, the features of degeneracy solutions of the first two cases, 
SB1 and MB1, are so similar that we omit MB1 plots. 
Comparison between MB1 and MB2 settings, former (latter) being around the 
first (second) oscillation maximum, would be interesting to know physics 
behind the potential of the two-detector setting \cite{MNplb97} 
and/or the BNL-type wide band beam approach \cite{BNL}.

The key to resolve the degeneracy is to utilize spectrum informations. 
Therefore, we also present the energy dependence of the difference 
between the true and the degeneracy solutions. 
A point of interest is how the energy dependence differs among the three 
different types of the degeneracies.

We note that in all the figures presented in this section we take the normal mass 
hierarchy as the input true solution. 
If one wants to have the corresponding informations for the inverted mass hierarchy, 
one can do it just by changing the ordinate label of the figures as 
$\delta^{\text{true}} - \pi$ not $\delta^{\text{true}}$. 
This is discussed in detail in Sec.~\ref{normal-inverted}. 

We use the following values for the 
mixing parameters as summarized below: 
$\Delta m^2_{21} = 7.9 \times 10^{-5} \text{ eV}^2$, $\sin^2 \theta_{12} = 0.31$, and $\Delta m^2_{31} = 2.5 \times 10^{-3} \text{ eV}^2$. 
The matter density is taken as 
$\rho = 2.8 \text{g/cm}^3$ for SB1, MB1, and MB2 settings, and 
$\rho = 3.6 \text{g/cm}^3$ for NF setting.

\subsection{Intrinsic degeneracy in the true $\theta_{23}$ octant}
\label{overview-intrinsic}

\begin{figure}[bhtp]
\begin{center}
\vglue 0.3cm
\includegraphics[bb=0 0 211 101 , clip, width=0.64\textwidth]{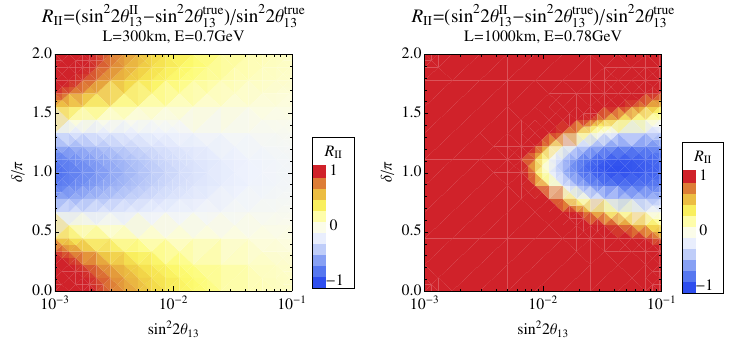}
%
\end{center}
\vglue -3mm
\caption{
The ratio 
$R_{ \text{II} } = 
[ \sin^2 2\theta_{13}^{ \text{N} } -  \sin^2 2\theta_{13}^{\text{true}}] / \sin^22\theta_{13}^{\text{true}} $ defined in (\ref{RN-def}) is presented 
in $\sin^2 2\theta_{13}^{ \text{true} } - \delta^{ \text{true} }/\pi$ space
for the two typical cases of energies and baselines, SB1 and MB2, defined in  
Sec.~\ref{overview-variable}. 
We do not present the same plots for the setting MB1 because they are 
so similar to those of SB1.  
%
%
}
\label{R2}
\end{figure}

\begin{figure}[bhtp]
\begin{center}
\includegraphics[bb=0 0 211 101 , clip, width=0.64\textwidth]{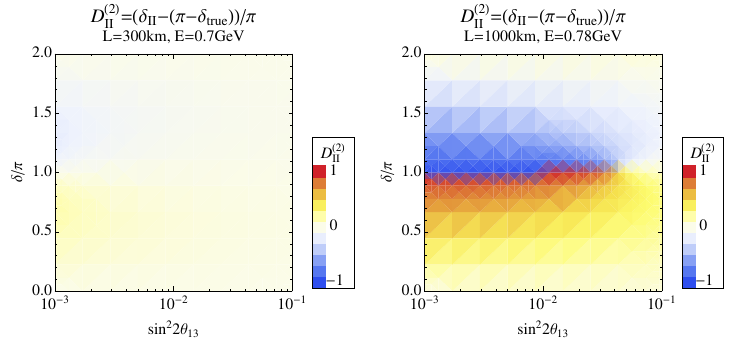}
\end{center}
\vglue -3mm
\caption{
The normalized difference 
$D_{ \text{II} }^{(2)} \equiv [\delta^{ \text{II} } - (\pi - \delta^{ \text{true} } ) ] / \pi$, 
defined in (\ref{DN-def}) is presented 
in $\sin^2 2\theta_{13}^{ \text{true} } - \delta^{ \text{true} }/\pi$ space
for the two typical settings SB1 and MB2 defined in  
Sec.~\ref{overview-variable}. 
We do not present the same plots for the setting MB1 because they are 
so similar to those of SB1.  
}
\label{D2}
\end{figure}

In Figs.~\ref{R2} and \ref{D2}, we present 
$R_{ \text{II} }$, the normalized difference of  $\sin^2 2\theta_{13}$, and 
$D_{ \text{II} }^{(2)}$, a difference of $\delta/\pi$ defined in (\ref{DN-def}), 
respectively, between the true and the intrinsic degeneracy solutions for 
the two typical cases of energies and baselines, SB1 and MB2. 
We do not present the same plots for MB1 setting because they are 
so similar to those of SB1.  
A color variation is used to clearly represent the ratio $R_{ \text{II} }$ 
and $D_{ \text{II} }^{(2)}$ in a visual way, 
which will be used later also to all of $R_{ \text{N} }$ and $D_{ \text{N} }^{(i)}$. 
From blue to red $R_{ \text{II} }$ and $D_{ \text{II} }^{(2)}$ vary from $-1$ to +1. 
The only exception to this rule is $R_{ \text{N} }$ at color graduation of the 
deepest red; It contains the region with $R_{ \text{N} }$ greater than 1. 
Notice that there is no region of $R_{ \text{N} } < - 1$ by definition in 
(\ref{RN-def}).\footnote{
Here we need to mention about some details of color variation. 
20 color graduation are used to draw $R_{ \text{N} }$ and 
$D_{ \text{N} }^{(i)}$ so that a single color graduation spans 5\% of the entire region.
}
%
We should note that in the case of $D_{ \text{N} }^{(i)}$, unlike $R_{ \text{N} }$, 
the deep blue region smoothly continues to the deep red because of the 
periodicity in $\delta$.

%
%
\begin{figure}[bhtp]
\begin{center}
\includegraphics[bb=0 0 240 269 , clip, width=0.30\textwidth]{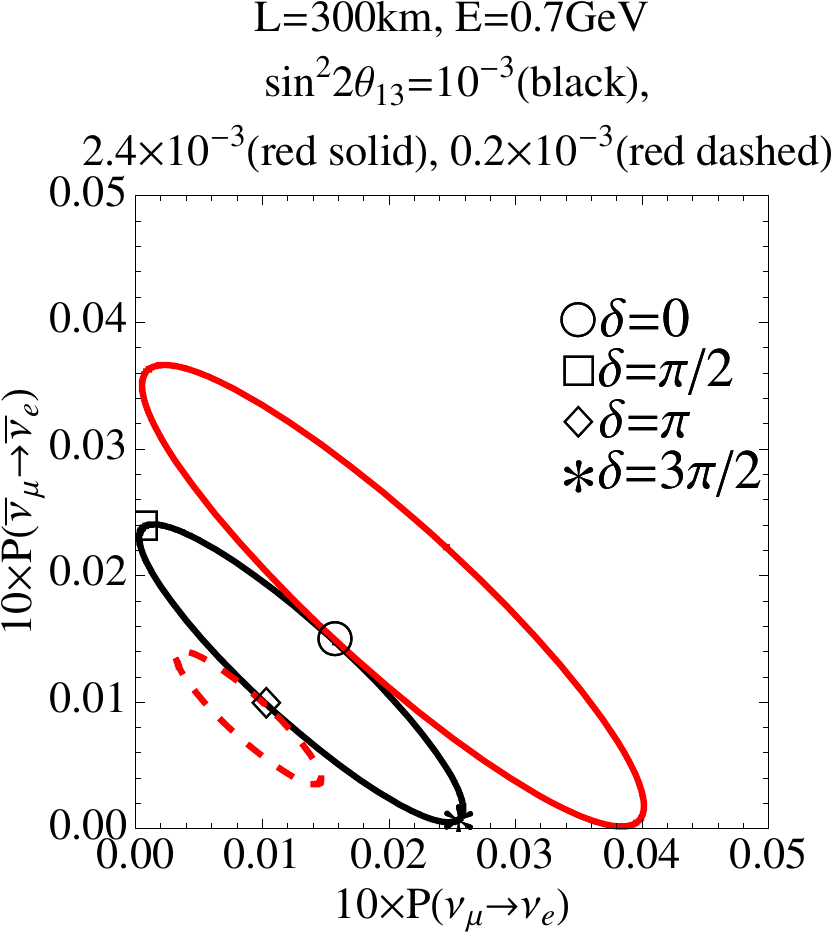}
\includegraphics[bb=0 0 240 269 , clip, width=0.30\textwidth]{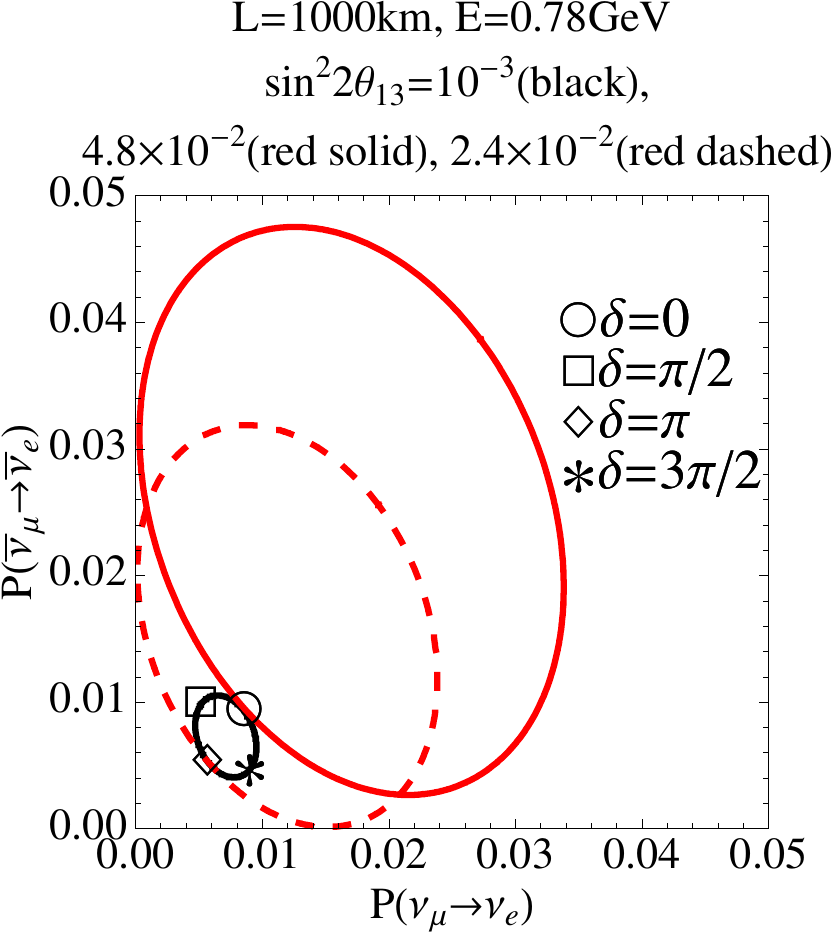}
\end{center}
\vglue -3mm
\caption{
Bi-probability plot for settings of SB1 (left panel) and MB2 (right panel). 
In both plots the true value of $\theta_{13}$ is taken to be 
$\sin^2 2\theta_{13}=10^{-3}$. 
The ellipses of the intrinsic degeneracy solutions are depicted by the red solid 
and red dashed lines.
}
\label{bi-P-SB1-MB2}
\end{figure}


One of the most notable features in Fig.~\ref{R2} is a clear difference between 
SB1 (left panel) and MB2 (right panel) settings. 
In large $\theta_{13}$ region in SB1 setting, $\sin^2 2\theta_{13} \gtrsim 10^{-2}$, 
$R_{ \text{II} }$ is small. 
At small $\theta_{13}$ in SB1 setting and at large $\theta_{13}$ in MB2 setting, 
$R_{ \text{II} }$ is large and positive (negative) at $\delta \sim 0$ ($\pi$). 
In small $\theta_{13}$ region in MB2 setting, $\sin^2 2\theta_{13} \lesssim 10^{-2}$, 
$R_{ \text{II} }$ is large and positive independent of $\delta$.\footnote{
At extremely small $\theta_{13}$, the left plot in Fig.~\ref{R2} 
for SB1 almost looks like the right plot for MB2, but with scale of 
$\sin^2 2\theta_{13}$ two orders of magnitude smaller than that of the 
right panel in Fig.~\ref{R2}. 
}
%
Let us understand these features.

We start from the above first feature. It can be understood by the analytic solution 
(\ref{CP-intrinsic-s-solution}). 
If $\theta_{13}$ is relatively large, $s_{13} \gg \Delta m_{21}^2/\Delta m_{31}^2$, 
$s_{1}$ is the dominant term in $s_{ \text{II} }$ in (\ref{CP-s-solution-1st}), and 
$R_{ \text{II} } \sim \mathcal{O} \left( s_{1}^{-1} \Delta m_{21}^2/\Delta m_{31}^2 \right)$. 
Therefore, $R_{ \text{II} }$ is small at large $\theta_{13}$ in SB1 setting. 
Now, the behavior of $R_{ \text{II} }$ at small $\theta_{13}$ can be easily understood 
by looking into the bi-probability plot, the left panel in Fig.~\ref{bi-P-SB1-MB2}. 
The degeneracy ellipse which shares the point around $\delta=0$ ($\delta=\pi$) 
of the true ellipse is the dashed (solid) one with considerably larger (smaller) 
$\theta_{13}$.
The similar consideration explains the feature of $R_{ \text{II} }$ 
at large $\theta_{13}$ in MB2 setting. 
The remaining feature that needs explanation is the large positive $R_{ \text{II} }$ 
at small $\theta_{13}$ in MB2 setting. 
At such small $\theta_{13}$ as $\sin^2 2\theta_{13} \sim 10^{-3}$ and the 
baseline $L=1000$ km, the oscillation probability is dominated by the solar term $Z$. 
Since it is independent of $\delta$ the probability ellipse shrinks to a small ``circle'', 
as can be seen in the right panel in Fig.~\ref{bi-P-SB1-MB2}. 
Then, the degeneracy solution ellipses are inevitably large as indicated by 
the red solid and dashed lines, resulting degenerate solutions much larger 
than the true $s_{13}$.

We observe for $D_{ \text{II} }^{(2)}$ plotted in Fig.~\ref{D2} that 
in SB1 (and MB1) setting $D_{ \text{II} }^{(2)}$ is small in the entire region 
of $\sin^2 2\theta_{13} - \delta/\pi$ space covered.\footnote{
Hereafter, when we talk about 
$\sin^2 2\theta_{13}$ and $\delta$ such as 
$\sin^2 2\theta_{13} - \delta/\pi$ space, 
it actually means the 
$\sin^2 2\theta_{13}^{ \text{true} }$ and $\delta^{ \text{true} }$, respectively. 
We use the simpler notation to avoid cumbersome 
superscript ``true''as much as possible. 
}
%
It means that the approximate formula 
$\delta^{ \text{II} } \simeq \pi - \delta^{ \text{true} }$ works well \cite{intrinsic}. 
It is a nice feature of measurement by SB1 setting because CP violation is 
unlikely to be confused with CP conservation. 
Whereas for MB2 setting $D_{ \text{II} }^{(2)}$ is small only in a limited region 
$-0.2 \lesssim \delta / \pi \lesssim 0.2$, and in 
 a small strip around $\delta \simeq 0$ at large $\theta_{13}$. 
The deviation from the approximation 
$\delta^{ \text{II} } \simeq \pi - \delta^{ \text{true} }$ is significant in the 
second and the third quadrants of $\delta$, in particular in region 
$\sin^2 2\theta_{13} \lesssim \text{a few} \times10^{-2}$ in MB2 setting. 
It is possible to understand this behavior of $D_{ \text{II} }^{(2)}$ qualitatively 
at very small $\theta_{13}$, $\sin^2 2\theta_{13} \sim 10^{-3}$ by 
using the bi-probability plot for MB2 setting (right panel in Fig.~\ref{bi-P-SB1-MB2}).  
For the true value of $\delta^{ \text{true} } \simeq 0$ the degeneracy ellipse is depicted by 
the solid line and $\delta_{ \text{II} } \simeq \pi$, which implies $D_{ \text{II} }^{(2)} \ll 1$. 
On the other hand, for $\delta^{ \text{true} } \simeq \pi$ the degeneracy ellipse 
depicted by the dashed line touches to the true ellipse also at around $\delta \simeq \pi$, 
hence $D_{ \text{II} }^{(2)}$ is of order unity. 
Notice again that the deep blue region smoothly continues to the deep red 
because of the periodicity in $\delta$.

\begin{figure}[bhtp]
\begin{center}
\vglue 0.2cm
\includegraphics[bb=0 0 281 102 , clip, width=0.7\textwidth]{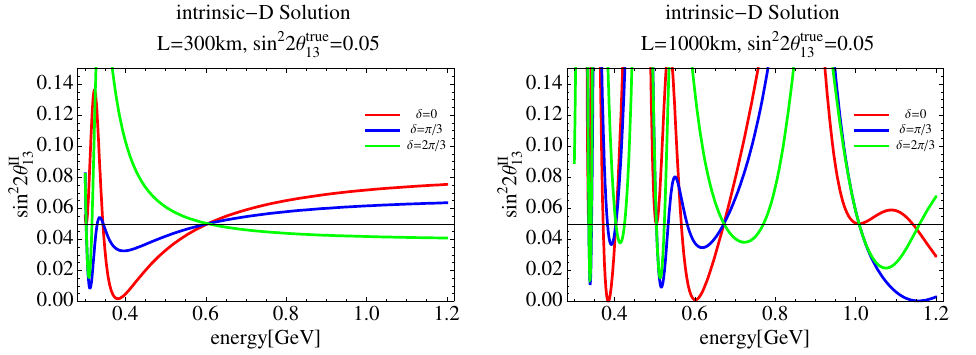}
\end{center}
\vglue -3mm
\caption{
The energy dependence of $\sin^2 2\theta_{13}^{ \text{II} }$ is 
plotted for the two typical settings SB1 (left panel) and MB2 (right panel) 
defined in  Sec.~\ref{overview-variable}. 
The true value of $\theta_{13}$ is taken as $\sin^2 2\theta_{13}=0.05$, which 
is indicated by the horizontal solid line in the figure. 
}
\label{energy-dep-s2}
\end{figure}
%
\begin{figure}[bhtp]
\begin{center}
\includegraphics[bb=0 0 281 103 , clip, width=0.7\textwidth]{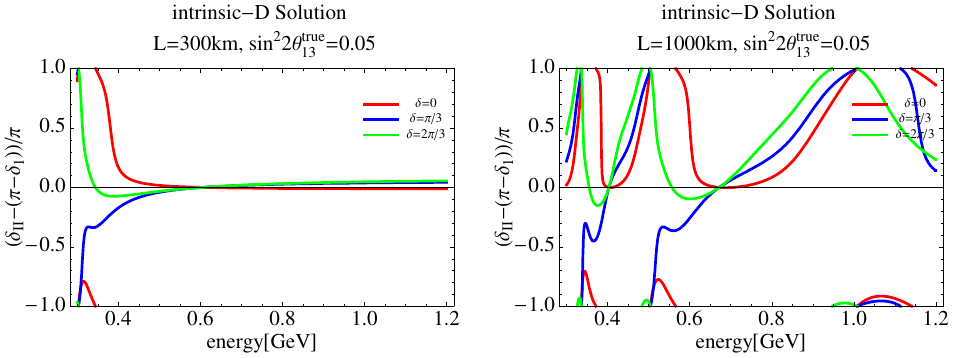}
\end{center}
\vglue -3mm
\caption{
The energy dependence of the ratio $D_{ \text{II} }^{(2)}$ defined in (\ref{DN-def}) is 
plotted for the two typical settings SB1 (left panel) and MB2 (right panel) 
defined in  Sec.~\ref{overview-variable}. 
The true value of $\theta_{13}$ is taken as $\sin^2 2\theta_{13}=0.05$. 
}
\label{energy-dep-delta2}
\end{figure}

One of the most important issues to find possible ways to resolve the degeneracy 
is to know the energy dependence of the difference between the true and the 
clone solutions. 
Therefore, we present in Fig.~\ref{energy-dep-s2} 
the energy dependence of $\sin^2 2\theta_{13}^{ \text{II} }$ for SB1 (left panel) 
and MB2 (right panel) settings, assuming the true value of 
$\sin^2 2\theta_{13}=0.05$. It may be regarded as a typical value for 
relatively large $\theta_{13}$ to which we will have an access by the 
ongoing experiments. 
Hereafter, whenever we present the similar figures of the energy dependence 
of the degeneracy solutions, we use only the three values of $\delta$,  
$\delta = 0, \pi/3$, and $2\pi/3$, for better visibility.
(We have tried $\delta = 0, 2\pi/3$, and $4\pi/3$, but the latter two curves tend to overlap.) 
Similarly, in Fig.~\ref{energy-dep-delta2} the energy dependence of 
$D_{ \text{II} }^{(2)}$ is plotted for the same settings, SB1 and MB2, 
with the same true value of $\theta_{13}$. 
Again, we do not present the case of MB1 setting because the plots are 
very similar to those of SB1 apart from minor differences at low energies, 
$E \lesssim 1$ GeV.\footnote{
For more extensive presentation of these plots including those of MB1 setting and 
energy dependence plots with other values of $\delta$, see \cite{uchinami-thesis}, 
and partly \cite{MU-version1}.
}

Clearly, there exists a significant energy dependence of $\sin^2 2\theta_{13}^{ \text{II} }$  even for SB1 setting. 
One can see that they vary by a factor of 2$-$4 (30\%$-$40\%), 
or more at low (high) energies depending upon $\delta$
in region of $E=0.4-1.2$ GeV for $L=300$ km. 
It must be contrasted to almost flat curves of energy dependence given in the 
following subsections, Figs.~\ref{energy-dep-s3-s4} (Sec.~\ref{overview-sign}) 
and \ref{energy-dep-s5-s6-s7-s8} (Sec.~\ref{overview-octant}) for the 
sign-$\Delta m^2_{31}$ and the octant degeneracies, respectively. 
Then, the spectrum analysis must be powerful in resolving the intrinsic degeneracy. 
It has been seen to be the case in the analysis of T2K II experiment \cite{T2K} 
done in \cite{T2KK-1st}. 
For $D_{ \text{II} }^{(2)}$ the energy dependence is significant only at low energies, 
below the first oscillation maximum, where usually the signal-to-background ratio 
is not helpful.
Therefore, potential power for the spectrum analysis relies more on the energy 
dependence of $\sin^2 2\theta_{13}^{ \text{II} }$ not on $\delta_{ \text{II} }$'s.

The energy dependences of $\sin^2 2\theta_{13}^{ \text{II} }$ and $D_{ \text{II} }^{(2)}$ plotted for MB2 setting magnify the low energy part of MB1 setting (not shown) 
at the same baseline of $L=1000$ km. 
The energy dependences are far more pronounced and depend sensitivity on $\delta$, 
and at some particular energies 
$\sin^2 2\theta_{13}^{ \text{II} }$ and $D_{ \text{II} }^{(2)}$ are pinned to 
$\sin^2 2\theta_{13}^{ \text{true} }$ and zero, respectively. 
To understand better these features, we first note that the oscillation probabilities 
show violent energy dependences below the first oscillation maximum. 
The $\delta$-dependent strong energy dependence of the degeneracy solutions, 
however, is under restriction by pinning to the true value or zero at the energies 
corresponding to the $n$-th oscillation maxima ($n = 1, 2, 3 $), as seen in the left 
and the right panels of Fig.~\ref{energy-dep-s2}.
In SB1 setting (300 km) the first oscillation maximum is reached at $E \simeq 600$ MeV,  
while for $L=1000$ km the first, second and the third minima are approximately at 
$E \simeq 2$ GeV, 700 MeV, and 400 MeV, respectively. 
The zeros of $\sin^2 2\theta_{13}^{ \text{II} } - \sin^2 2\theta_{13}^{ \text{true} }$ and 
$D_{ \text{II} }^{(2)}$ arise due to the special feature of the parameter degeneracy 
at the $n$-th oscillation maxima \cite{BMW02,KMN02}, 
which can be seen explicitly from the formulas in Sec.~\ref{CP-intrinsic}. 
It can also be intuitively understood by having a flattened ellipse in the bi-probaility 
diagram at the oscillation maxima \cite{MNjhep01}.\footnote{
The similar zero at $E=1$ GeV can be understood as shrinking the ellipse into 
a small size at the oscillation minimum and is not interesting to us. 
}
%
The regularity of alternating ``pinning to zero'' and violent energy dependence 
gives us at least clear picture of what is seen in energy dependence 
of the degeneracy solutions in MB2 setting. 
The similar features will be seen in many figures of energy dependence in MB2 
setting presented in the rest of this paper, which will allow analogous explanations.

Despite a possibility of confusion due to too complicated dependence on energy 
and $\delta$ at MB2 only setting, combination of MB2 at somewhat off the 
oscillation maxima with more quiet SB1 settings would be an ideal machinery for 
resolving the degeneracy. 
This was observed to occur in \cite{T2KK-1st,T2KK-2nd} which utilizes the informations 
at the second oscillation maximum by a Korean detector, and most probably 
gives an explanation for high sensitivity achievable in the BNL-type wide band 
beam strategy \cite{BNL,BNL-fermi}.

\subsection{Sign-$\Delta m^2_{31}$ degeneracy in the true $\theta_{23}$ octant}
\label{overview-sign}

\begin{figure}[bhtp]
\begin{center}
\vglue 0.3cm
\includegraphics[bb=0 0 280 180 , clip, width=0.9\textwidth]{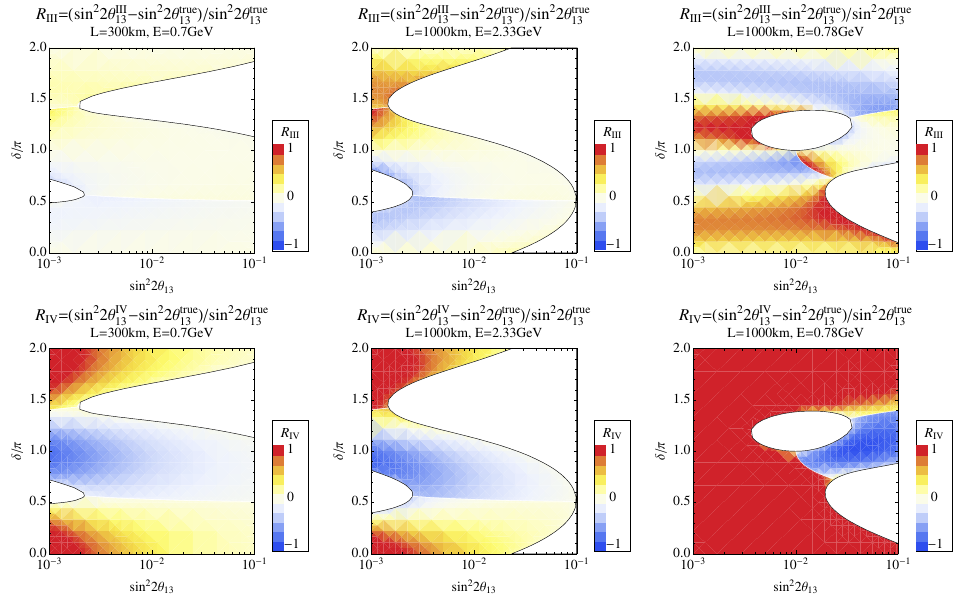}
\end{center}
\vglue -3mm
\caption{
The ratios 
$R_{ \text{III} } = 
[ \sin^2 2\theta_{13}^{ \text{III} } -  \sin^2 2\theta_{13}^{\text{true}}] / \sin^22\theta_{13}^{\text{true}} $ (upper three panels) and 
$R_{ \text{IV} }$ (lower  three panels) defined in (\ref{RN-def})  in 
$\sin^2 2\theta_{13} - \delta/\pi$ space is presented 
for the three typical settings 
SB1 (left panel), MB1 (middle panel), and MB2 (right panel) defined in 
Sec.~\ref{overview-variable}. 
The regions of white color denote the regions of no sign-degeneracy solution. 
}
\label{R3-R4}
\end{figure}

We now turn to the sign-$\Delta m^2_{31}$ degeneracy which exists in the 
same $\theta_{23}$ octant as the true one. 
Since there are two solutions, 
($s_{ \text{III} }$, $\delta_{ \text{III} }$) and ($s_{ \text{IV} }$, $\delta_{ \text{IV} }$), 
we present them in the same figures. 
In Fig.~\ref{R3-R4}, $R_{ \text{III} }$ and $R_{ \text{IV} }$ defined in (\ref{RN-def}) 
are plotted in $\sin^2 2\theta_{13} - \delta/\pi$ space. 
White region is the region of no degenerate solution as discussed in 
Sec.~\ref{CP-signdm2}. 
We now show also MB1 case because difference from SB1 setting 
becomes non-negligible, in particular in small $\theta_{13}$ region. 
Nevertheless, the difference is not so significant at large $\theta_{13}$ 
$\sin^2 2\theta_{13} \gtrsim 10^{-2}$ 
apart from the change in the no-solution region.

%
\begin{figure}[bhtp]
\begin{center}
\includegraphics[bb=0 0 340 162 , clip, width=0.9\textwidth]{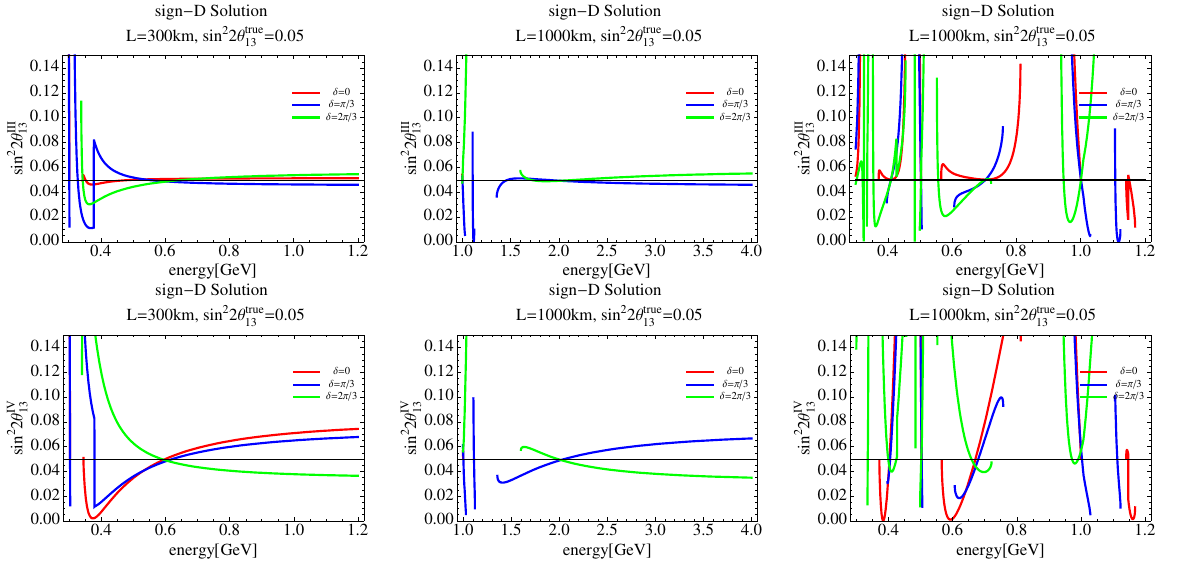}
\end{center}
\vglue -3mm
\caption{
The energy dependence of $\sin^2 2\theta_{13}^{ \text{III} }$  (upper three panels) 
and 
$\sin^2 2\theta_{13}^{ \text{IV} }$  (lower  three panels) are plotted 
for the three typical settings 
SB1 (left panel), MB1 (middle panel), and MB2 (right panel) defined in 
Sec.~\ref{overview-variable}. 
The true value of $\theta_{13}$ is taken as $\sin^2 2\theta_{13}=0.05$, which 
is indicated by the horizontal solid line in the figure. 
}
\label{energy-dep-s3-s4}
\end{figure}

By comparing Fig.~\ref{R3-R4} with Fig.~\ref{R2}, it is evident that 
the difference of $\sin^2 2\theta_{13}$ between the true solution 
$s_{1}$ and $s_{ \text{III} }$ is much smaller than the case of intrinsic 
degeneracy solutions for SB1 and MB1 settings. 
Furthermore, the energy dependence of $\sin^2 2\theta_{13}^{ \text{III} }$ 
is much milder than the case of solution ($s_{ \text{II} }$, $\delta_{ \text{II} }$) of 
the intrinsic degeneracy, as one can clearly see 
by comparing Fig.~\ref{energy-dep-s3-s4} with Fig.~\ref{energy-dep-s2}. 
These features make resolution of the sign-$\Delta m^2_{31}$ degeneracy 
much more difficult compared to the intrinsic degeneracy in these settings.

The difference between energy dependences of the intrinsic and the 
sign-$\Delta m^2_{31}$ degeneracies in SB1 and MB1 settings can be 
easily understood at least qualitatively. 
As we learned in Sec.~\ref{asymptotic}, by lacking $\sim 1/E$ terms, 
reach to high-energy asymptotic behavior is relatively fast. 
At low energies, $\sin^2 2\theta_{13}^{ \text{III} }$ is constrained to be small 
as one can show by the formulas based on the matter perturbation theory given 
in Appendix \ref{matter-perturb}. 
The first order correction term, from which the energy dependence comes in 
is small, of the order of $A/\Delta_{31} \simeq 0.06-0.07 (0.2)$ for SB1 and MB2 (MB1) settings. Whereas for $\sin^2 2\theta_{13}^{ \text{II} }$ there is no small parameter 
which forces it small. 
The mild energy dependence and the pinning to a small value makes 
$\sin^2 2\theta_{13}^{ \text{III} }$ small in the entire region of $E$. 
The relatively fast reach to the asymptotic behavior can be seen in most of the 
plots of energy dependence of the degeneracy solutions for SB1 and MB1 settings. 
Notice, however, that the asymptotic energy can be reached at much higher 
energies for MB2 setting.

Here are comments on the solution ($s_{ \text{IV} },  \delta_{ \text{IV} }$): 
$R_{ \text{IV} }$ essentially looks like $R_{ \text{II} }$ apart 
from the presence of no-solution regions. 
Given smallness of $R_{ \text{III} }$, $R_{ \text{IV} }$ must looks like 
$R_{ \text{II} }$ because they are the intrinsic degeneracy pairs, 
as discussed in Sec.~\ref{structure}. 
It is also true that the energy dependence of $\sin^2 2\theta_{13}^{ \text{IV} }$ 
is very similar to the behavior of $\sin^2 2\theta_{13}^{ \text{II} }$. 
Therefore, lifting degeneracy between $\theta_{13}^{ \text{IV} }$ and 
$\theta_{13}^{ \text{III} }$ can be done with spectrum analysis via a similar manner 
as in the case of intrinsic degeneracy. 
If powerful enough the spectrum informations would solve both 
the degeneracy between the true solution and $\theta_{13}^{ \text{II} }$, and the 
one between $\theta_{13}^{ \text{IV} }$ and $\theta_{13}^{ \text{III} }$ 
at the same time.

\begin{figure}[bhtp]
\begin{center}
\includegraphics[bb=0 0 280 180 , clip, width=0.9\textwidth]{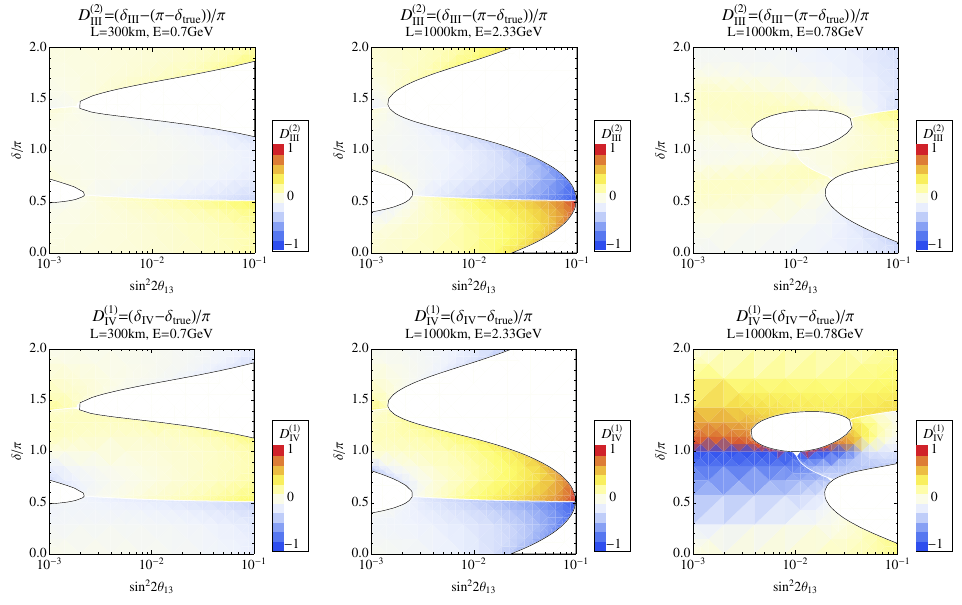}
\end{center}
\vglue -3mm
\caption{
$D_{ \text{III} }^{(2)} \equiv [\delta^{ \text{III} } - (\pi - \delta^{ \text{true} } ) ] / \pi$ 
(upper three panels) and 
$D_{ \text{IV} }^{(1)} \equiv (\delta^{ \text{IV} } - \delta^{ \text{true} } ) / \pi$ 
(lower three panels) 
defined in (\ref{DN-def}) is presented 
in $\sin^2 2\theta_{13} - \delta/\pi$ space 
for the three typical settings 
SB1 (left panel), MB1 (middle panel), and MB2 (right panel) defined in 
Sec.~\ref{overview-variable}. 
The regions of white color denote the regions of no sign-degeneracy solution. 
}
\label{D3-D4}
\end{figure}
%
\begin{figure}[bhtp]
\begin{center}
\includegraphics[bb=0 0 340 82 , clip, width=0.9\textwidth]{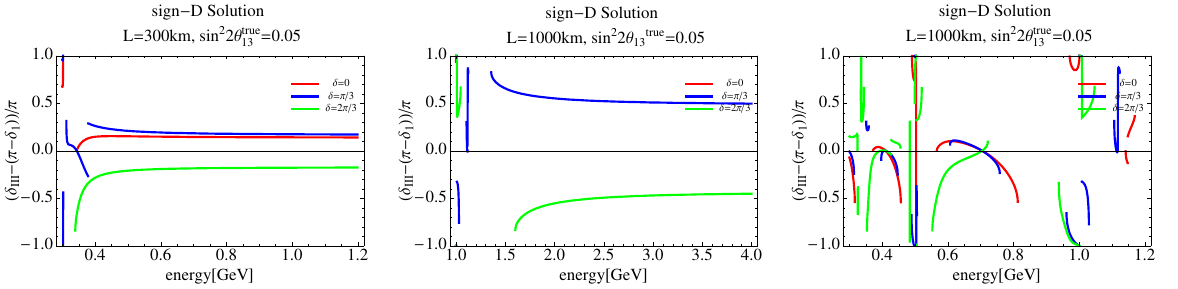}
\end{center}
\vglue -3mm
\caption{
The energy dependence of the ratios $D_{ \text{III} }^{(2)}$ 
defined in (\ref{DN-def}) are plotted for the three typical settings 
SB1 (left panel), MB1 (middle panel), and MB2 (right panel) defined in 
Sec.~\ref{overview-variable}. 
$D_{ \text{IV} }^{(1)}$ is not shown because the similarity in its energy dependence, 
apart from reversing the positive and negative regions of the ordinate. 
The true value of $\theta_{13}$ is taken as $\sin^2 2\theta_{13}=0.05$.
}
\label{energy-dep-D3-D4}
\end{figure}

Next, we discuss $D_{ \text{III} }^{(2)}$ and $D_{ \text{IV} }^{(1)}$ which are 
presented in the upper and lower three panels, respectively, in Fig.~\ref{D3-D4}. 
We note that they are small in SB1 setting, leaving the sign-$\Delta m^2_{31}$ 
degeneracy intact in this short baseline setting. 
Notice, however, that it is not all bad, because the smallness of 
$D_{ \text{III} }^{(2)}$ implies that no severe confusion takes place between 
CP violation and CP conservation. 
Now, the difference between SB1 and MB1 settings further develops in particular 
in large $\theta_{13}$ region. 
The clear distinction between SB1 and MB1 settings is also prominent 
in the energy dependence presented in Fig.~\ref{energy-dep-D3-D4}. 
Of course, it is basically due to larger matter effect in MB1 setting. 
It is interesting to observe that the difference shows up first in $\delta$, 
but not quite for $\theta_{13}$ at large $\theta_{13}$.

In the SB1 and MB1 settings, as can be seen in Fig.~\ref{D3-D4}, 
$D_{ \text{III} }^{(2)}$ and $D_{ \text{IV} }^{(1)}$ are largest in region of the largest 
possible $\theta_{13}$ for which the sign-$\Delta m^2_{31}$ degeneracy solution 
exist. 
In this region $\delta \sim \pi/2$. 
On the other hand, $R_{ \text{III} }$ and $R_{ \text{IV} }$ are small in the region 
as is seen in Fig.~\ref{R3-R4}. 
It is easy to understand these features. 
At around the largest value of $\theta_{13}$ which allows the sign-$\Delta m^2_{31}$ 
degeneracy the two ellipses, the ones with normal and inverted mass hierarchies, 
barely overlap with each other. See Fig.~8 in \cite{MNP2}. 
The general feature of the bi-probability plot \cite{MNjhep01} tells us that in the 
overlap regions of the two ellipses the point of $\delta \sim \pi/2$ in the positive 
$\Delta m^2_{31}$ ellipse is close to point of $\delta \sim 3\pi/2$ of the negative 
$\Delta m^2_{31}$ ellipse. 
Therefore, $D_{ \text{III} }^{(2)} \simeq 1$ and $D_{ \text{IV} }^{(1)} \simeq 1$ hold, 
explaining the above features. 
Because the center of the two ellipses are located at almost the same distances 
from the origin (which is equal to $s_{23}^2 \sin^2 2\theta_{13}$), 
$s_{ \text{III} } \simeq s_{ \text{IV} }  \simeq s_{1}$.

We want to note that the energy dependences of 
$D_{ \text{III} }^{(2)}$ and $D_{ \text{IV} }^{(1)}$ are quite mild in energy region 
above the first oscillation maximum for SB1 and MB1 settings.
Because the energy dependence of $D_{ \text{IV} }^{(1)}$ is similar to 
that of $D_{ \text{III} }^{(2)}$ in for SB1 and MB1 settings, 
apart from reversing the positive and negative regions of ordinate, 
we do not present it in Fig.~\ref{energy-dep-D3-D4}. 
(Hereafter, we just quote the reference either \cite{MU-version1} or 
\cite{uchinami-thesis} if the omitted figures are available in them.)
For MB2 setting qualitative features of rapid up and down are very similar 
in all the figures of energy dependences presented, or abbreviated. 
Considering the almost no energy dependence of 
$\sin^2 2\theta_{13}^{ \text{III} } - \sin^2 2\theta_{13}^{ \text{true} }$ 
as given in Fig.~\ref{energy-dep-s3-s4}, 
and noting that spectrum analysis is highly challenging at low energies, 
it would be difficult to resolve the sign-$\Delta m^2_{31}$ degeneracy by 
a single detector setting of either SB1 or MB1. 

We notice that the difference between SB1-MB1 and MB2 settings is always 
evident as can be seen in Figs.~\ref{R3-R4}, \ref{energy-dep-s3-s4}, \ref{D3-D4} 
and \ref{energy-dep-D3-D4}. 
Therefore, MB2 setting alone may have chance to resolve the 
sign-$\Delta m^2_{31}$ degeneracy \cite{nufact-lowE1, nufact-lowE2}.
Or, if the informations gained at around the second oscillation maximum 
can somehow be combined it would greatly help resolving the 
sign-$\Delta m^2_{31}$ degeneracy \cite{T2KK-1st,T2KK-2nd,BNL,BNL-fermi}. 
It may be expected even from our formulas obtained for a ``mono-energetic 
neutrino beam'' because the parameter regions with degeneracy solutions 
in SB1 and MB2 settings tend to ``repel'' (avoid to overlap) with each other 
at large $\theta_{13}$, though not completely.

Finally, we should note that abrupt termination of lines in the figures that appears in 
Figs.~\ref{energy-dep-s3-s4} and \ref{energy-dep-D3-D4} are either due to 
disappearance of the degeneracy solutions, or switching phenomenon between 
solutions that takes place due to our convention of labeling degeneracy solutions. 
The feature will be seen also in the foregoing subsections. 
See Secs.~\ref{convention} and \ref{CP-octant} for discussion on this point.

\subsection{Intrinsic and sign-$\Delta m^2_{31}$ degeneracies in the false 
$\theta_{23}$ octant}
\label{overview-octant}

\begin{figure}[bhtp]
\begin{center}
\vglue 0.3cm
\includegraphics[bb=0 0 280 360 , clip, width=0.9\textwidth]{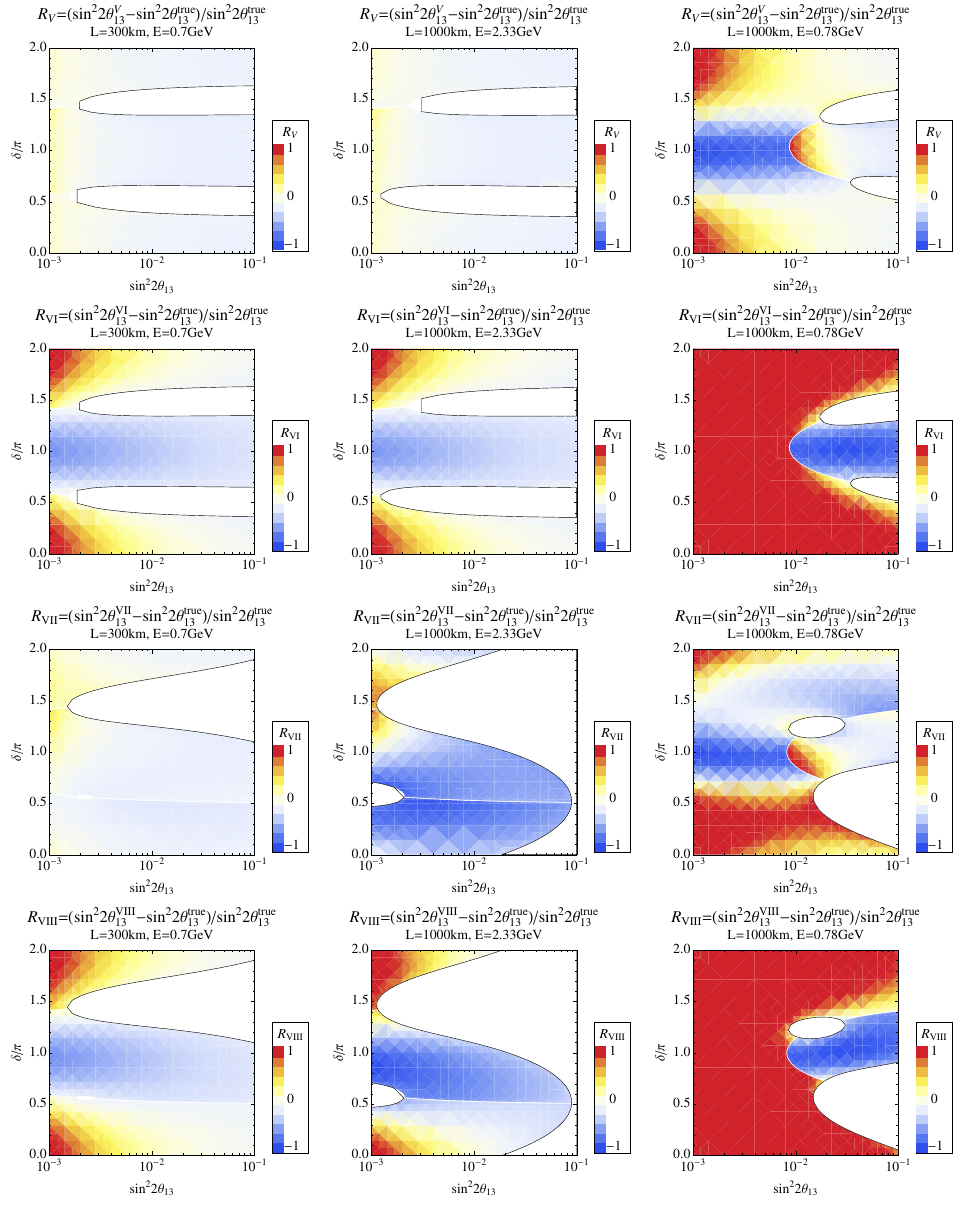}
\end{center}
\caption{
Plotted are the ratios 
$R_{ \text{V} } = 
[ \sin^2 2\theta_{13}^{ \text{V} } -  \sin^2 2\theta_{13}^{\text{true}}] / \sin^22\theta_{13}^{\text{true}} $ (top three panels), 
$R_{ \text{VI} }$ (next to top panels), 
$R_{ \text{VII} }$ (next next to top panels), and 
$R_{ \text{VIII} }$ (bottom three panels) defined in (\ref{RN-def}) in 
$\sin^2 2\theta_{13} - \delta/\pi$ space is presented 
for three typical cases of energies and baselines, 
SB1 (left panels), MB1 (middle panels), and MB2 (right panels) defined in 
Sec.~\ref{overview-variable}. 
The regions of white color denote the regions of no degeneracy solution. 
}
\label{R5-R6-R7-R8}
\end{figure}

\begin{figure}[bhtp]
\begin{center}
\includegraphics[bb=0 0 340 81 , clip, width=0.9\textwidth]{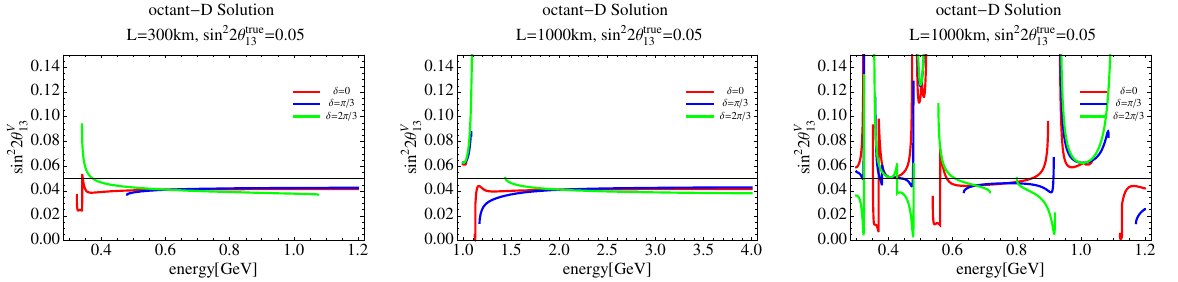}
\end{center}
\vglue -3mm
\caption{
The energy dependence of $\sin^2 2\theta_{13}^{ \text{V} }$ are plotted 
for the three typical settings 
SB1 (left panel), MB1 (middle panel), and MB2 (right panel) defined in 
Sec.~\ref{overview-variable}. 
For the behavior of the other solutions not shown, 
$\sin^2 2\theta_{13}^{ \text{VI} }$, 
$\sin^2 2\theta_{13}^{ \text{VII} }$, and  
$\sin^2 2\theta_{13}^{ \text{VIII} }$, 
see the text. 
The true value of $\theta_{13}$ is taken as $\sin^2 2\theta_{13}=0.05$, which 
is indicated by the horizontal solid line in the figure. 
The true value of $\theta_{23}$ is $42^{\circ}$. 
}
\label{energy-dep-s5-s6-s7-s8}
\end{figure}

Now, we turn to the $\theta_{23}$ octant degeneracy with solutions which 
lives in the different $\theta_{23}$ octant from the true solution.
Having the overview at hand, we present the intrinsic and the sign-$\Delta m^2_{31}$ degeneracy solutions at the same time. 
Presented in Fig.~\ref{R5-R6-R7-R8} in $\sin^2 2\theta_{13} - \delta/\pi$ space 
are the ratios 
$R_{ \text{V} } = 
\left( \sin^2 2\theta_{13}^{ \text{V} } -  \sin^2 2\theta_{13}^{\text{true}} \right) / \sin^22\theta_{13}^{\text{true}} $ (top three panels), 
$R_{ \text{VI} }$ (next to top panels), 
$R_{ \text{VII} }$ (next next to top panels), and 
$R_{ \text{VIII} }$ (bottom three panels) defined in (\ref{RN-def}) for three 
typical cases of energies and baselines, 
SB1 (left panels), MB1 (middle panels), and MB2 (right panels) defined in 
Sec.~\ref{overview-variable}. 
In Fig.~\ref{energy-dep-s5-s6-s7-s8}, 
the energy dependence of degeneracy solutions of 
$\sin^2 2\theta_{13}^{ \text{V} } $ is presented. 
%
We note that the energy dependence of $\sin^2 2\theta_{13}^{ \text{VII} } $ 
is quite similar to that of $\sin^2 2\theta_{13}^{ \text{V} } $. 
The relation between the energy dependences of 
$\sin^2 2\theta_{13}^{ \text{V} } $ and $\sin^2 2\theta_{13}^{ \text{VI} } $ 
($\sin^2 2\theta_{13}^{ \text{VII} } $ and $\sin^2 2\theta_{13}^{ \text{VIII} } $) 
is similar to the one between 
$\sin^2 2\theta_{13}^{ \text{III} } $ and $\sin^2 2\theta_{13}^{ \text{IV} } $ 
given in Fig.~\ref{energy-dep-s3-s4}. (See \cite{MU-version1,uchinami-thesis}.) 
It is natural because they are the intrinsic degeneracy partners, and hence 
they are not shown.

One of the most notable features in Fig.~\ref{R5-R6-R7-R8} is again quite 
distinct behaviors in the MB2 setting. 
In general, $R_{ \text{N} }$ are large 
(apart from the strips where $R_{ \text{N} }$ switches its sign) 
with notable exceptions of 
$R_{ \text{V} }$ for SB1 and MB1 settings, and  
$R_{ \text{VII} }$ for SB1 setting. 
It is also notable that behavior of $R_{ \text{VI} }$ and $R_{ \text{VIII} }$ is 
reminiscent of the one of $R_{ \text{II} }$ in Fig.~\ref{D2} in SB1 and MB1 
settings in small $\theta_{13}$ region, except for the presence of no-solution region. 
Considering the small values of $R_{ \text{V} }$ and $R_{ \text{VII} }$ 
(except for $R_{ \text{VII} }$ for MB1), it is quite natural to see the behavior 
given the fact that they are the intrinsic degeneracy partners. 
The feature that $R_{ \text{VI} }$ and $R_{ \text{VIII} }$ trace the behavior of 
their intrinsic degeneracy partners also applies to MB2 setting. 
The behavior of $R_{ \text{V} }$ and $R_{ \text{VII} }$ 
($R_{ \text{VI} }$ and $R_{ \text{VIII} }$) is somewhat similar, apart from the 
presence of no-solution region, to that of $R_{ \text{II} }$ for SB1 (MB2) setting. 
It may be understood by the similar consideration using the bi-probability plot. 
Therefore, we concentrate below on SB1 and MB1 settings.

%

As mentioned above $R_{ \text{V} }$ for SB1 and MB1 settings, and 
$R_{ \text{VII} }$ for SB1 setting are small in region $\sin^2 2\theta_{13} \gtrsim 10^{-2}$. 
The region of $\theta_{13}$, however, is nothing but a good target for superbeam experiments, and it will be a challenge for them to lift the degeneracy solutions. 
Here, we try to understand this feature, but in a wider perspective which includes 
the energy dependence of $\sin^2 2\theta_{13}^{ \text{V} }$ and $\sin^2 2\theta_{13}^{ \text{VII} }$ for SB1 and MB1 settings. 
In Fig.~\ref{energy-dep-s5-s6-s7-s8}, we observe that the difference between 
$\sin^2 2\theta_{13}^{ \text{V} }$  (or, $\sin^2 2\theta_{13}^{ \text{VII} }$) 
and $\sin^2 2\theta_{13}^{ \text{true} }$ is nonzero but energy independent 
in a wide region except for at very low energies, 
a somewhat unexpected behavior to see.

Now, we point out that the behaviors mentioned above can be understood 
by formulating the $\theta_{23}$ perturbation theory, 
as done in Appendix \ref{octant-perturb}. 
Namely, one can derive the perturbative expression of $\sin^2 2\theta_{13}^{ \text{N} }$ and other quantities by assuming that deviation of $\theta_{23}$ from $\pi/4$ is small, 
$\theta_{23} - \pi/4 \equiv \epsilon_{ \text{oct} } \ll 1$. 
One can expect that the expansion by $\epsilon_{ \text{oct} }$ is indeed 
a good approximation because e.g., $\epsilon_{ \text{oct} }=0.05$ 
for $\theta_{23}=42^{\circ}$. Then, we obtain 
\begin{eqnarray}
\sin^2 2\theta_{13}^{ \text{V} } - \sin^2 2\theta_{13}^{ \text{true} } = 
4 \epsilon_{ \text{oct} } 
\sin^2 2\theta_{13}^{ \text{true} } 
\label{diff-1-V-VII}
\end{eqnarray}
for which we have used the fact that the last term in (\ref{octant-s5-th23P}) 
is negligibly small as far as we remain in a region 
$\sin^2 2\theta_{13}^{ \text{true} } \gg Z$. 
In fact, we confirmed that the correction term becomes non-negligible 
at small $\theta_{13}$ around $\sin 2\theta_{13}^{ \text{true} } = 10^{-3}$. 
Certainly, the condition is fulfilled for the settings SB1 and MB1. 
The similar equation holds for $\theta_{13}^{ \text{VII} }$ but with replacement 
of $\theta_{13}^{ \text{true} }$ to $\theta_{13}^{ \text{III} }$ because their relation as 
the intrinsic degeneracy partner. 
Therefore, $R_{ \text{V} }$ and $R_{ \text{VII} }$ are small and the difference 
in (\ref{diff-1-V-VII}) is approximately energy independent. 
One may ask why the feature does not exist in MB2 setting with small $\theta_{13}$. 
As mentioned before, the solar term is dominant in this region. 
Therefore, if $\epsilon_{\text{oct}} Y_{\pm}$ is comparable to 
$s_{1}$ the difference $s_{ \text{V} } - s_{1}$ is no more small.

Another notable point is that region of no degeneracy solution is not additive, 
as can be seen by comparing Figs.~\ref{R3-R4} and \ref{R5-R6-R7-R8}.
That is, the region of no degeneracy solution with $\Delta m^2_{31}$-sign and 
$\theta_{23}$ octant flips (VII and VIII) is not the union of no-solution regions of 
the sign-$\Delta m^2_{31}$ (III) and the $\theta_{23}$ octant (V) degeneracies. 
It is simply because the degeneracy solution with both sign and octant flips can exist 
even in a region of $\theta_{13}$ and $\delta$ where e.g., the octant degeneracy 
solution does not exist.

\begin{figure}[bhtp]
\begin{center}
\vglue 0.3cm
\includegraphics[bb=0 0 280 360 , clip, width=0.9\textwidth]{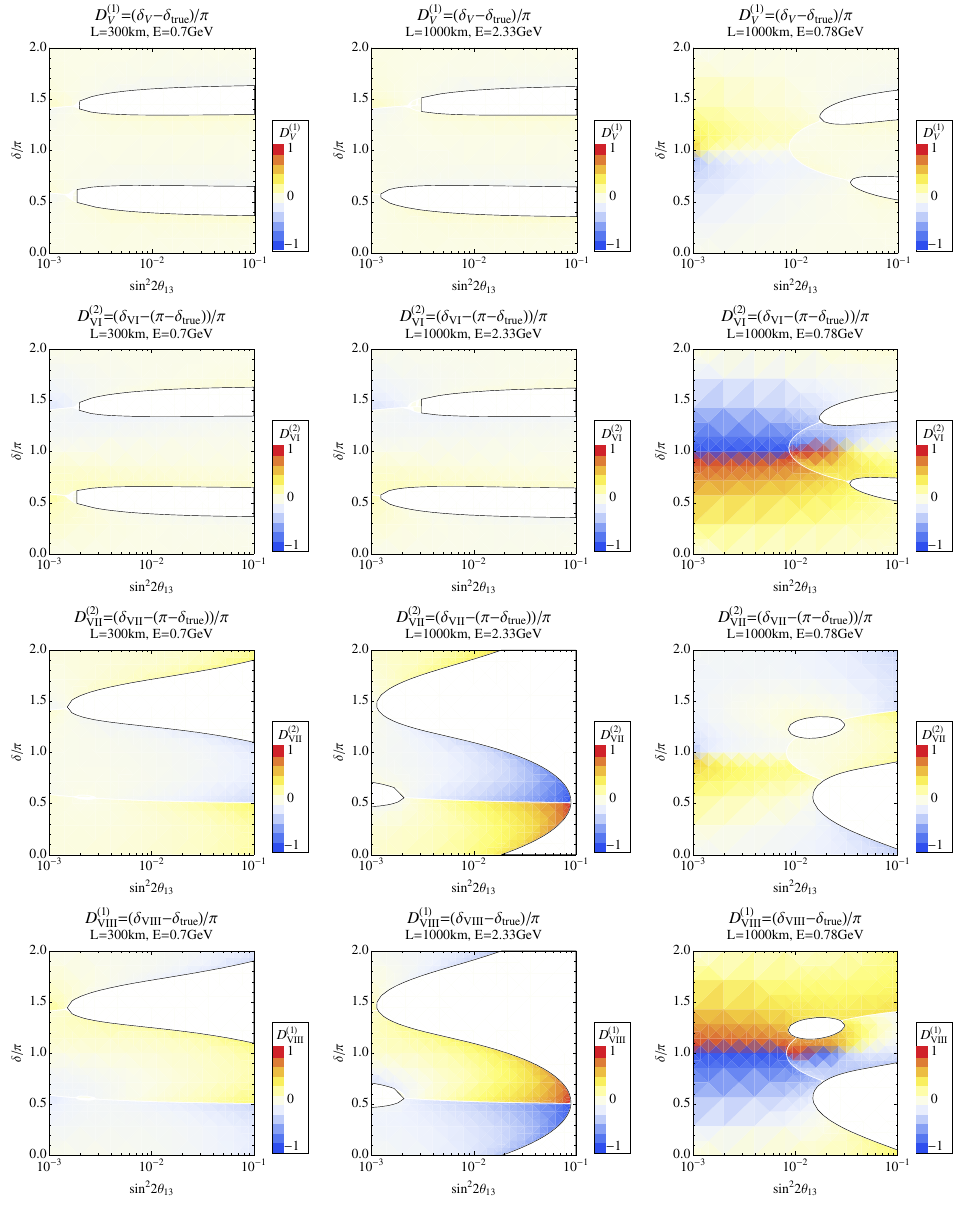}
\end{center}
\caption{
From top panels to bottom panels presented are in order: 
$D_{ \text{V} }^{(1)} \equiv (\delta^{ \text{V} } - \delta^{ \text{true} } ) / \pi$, 
$D_{ \text{VI} }^{(2)} \equiv [\delta^{ \text{VI} } - (\pi - \delta^{ \text{true} } ) ] / \pi$,  
$D_{ \text{VII} }^{(2)} \equiv [\delta^{ \text{VII} } - (\pi - \delta^{ \text{true} } ) ] / \pi$, and 
$D_{ \text{VIII} }^{(1)} \equiv (\delta^{ \text{VIII} } - \delta^{ \text{true} } ) / \pi$ 
defined in (\ref{DN-def}) for three typical cases of energies and 
baselines in $\sin^2 2\theta_{13}^{ \text{true} } - \delta^{ \text{true} }/\pi$ space. 
The regions of white color denote the regions of no sign-degeneracy solution. 
}
\label{D5-D6-D7-D8}
\end{figure}

\begin{figure}[bhtp]
\begin{center}
\includegraphics[bb=0 0 340 82 , clip, width=0.9\textwidth]{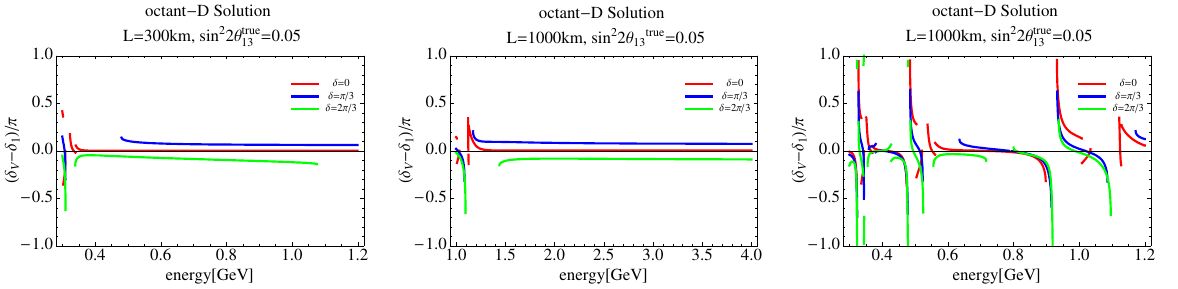}
\end{center}
\vglue -3mm
\caption{
The energy dependence of the ratio $D_{ \text{V} }^{(1)}$ 
defined in (\ref{DN-def}) are plotted for the three typical settings 
SB1 (left panel), MB1 (middle panel), and MB2 (right panel) defined in 
Sec.~\ref{overview-variable}. 
For the behavior of the other solutions not shown, 
$D_{ \text{VI} }^{(2)}$, $D_{ \text{VII} }^{(2)}$, and $D_{ \text{VIII} }^{(1)}$, 
see the text. 
The true value of $\theta_{13}$ is taken as $\sin^2 2\theta_{13}=0.05$.
}
\label{energy-dep-delta5-delta6-delta7-delta8}
\end{figure}

In Fig.~\ref{D5-D6-D7-D8}, from the top to the bottom, 
the normalized differences between the true and fake solutions of phases,  
$D_{ \text{V} }^{(1)}$, 
$D_{ \text{VI} }^{(2)}$, $D_{ \text{VII} }^{(2)}$, and $D_{ \text{VIII} }^{(1)}$ 
defined in (\ref{DN-def}) are presented for SB1 (left panels), MB1 (middle panels), 
MB2 (right panels) settings. 
In Fig.~\ref{energy-dep-delta5-delta6-delta7-delta8}, 
the energy dependence of $D_{ \text{V} }^{(1)}$ is plotted. 
The behavior of $D_{ \text{VI} }^{(2)}$ is similar to $D_{ \text{II} }^{(2)}$ given 
in Fig.~\ref{energy-dep-delta2}, while those of  
$D_{ \text{VII} }^{(2)}$ and $D_{ \text{VIII} }^{(1)}$ are very similar to the ones of 
$D_{ \text{III} }^{(2)}$ and $D_{ \text{IV} }^{(1)}$ 
(the latter not shown but the behavior explained) in Fig.~\ref{energy-dep-D3-D4}.  
See \cite{MU-version1,uchinami-thesis}. 
We notice that for SB1 setting $D_{ \text{N} }^{(i)}$ ($i$ either 1 or 2) is small in 
most of the regions of true values of $\delta$ for all the solutions V-VIII. 
For MB1 setting the same statement applies for the solutions V and VI.
A notable feature is that $D_{ \text{V} }^{(1)}$ (and $D_{ \text{VII} }^{(2)}$) is small 
in MB2 setting. It can also be understood from the $\theta_{23}$ perturbative formula 
for $\delta$ given in Appendix~\ref{octant-perturb}; 
The difference between $\delta_1$ and $\delta_{ \text{V} }$ is always 
suppressed by $\epsilon_{ \text{oct} }$.

As discussed above the energy dependence is very mild for most of the solutions 
V$-$VIII, except for at low energies, $E \lesssim 0.4$ GeV, in SB1 and MB1 settings. 
Therefore, it may be extremely challenging for experiments with the settings to 
lift the degeneracy. 
Because of the likely difficulty in resolving the $\theta_{23}$ octant degeneracy 
several methods have been proposed; 
the reactor-accelerator combined method \cite{MSYIS03,resolve23}, 
the various ways to detect solar $\Delta m^2_{21}$ scale oscillations, 
using atmospheric \cite{peres-smi23,concha23,choubey23,kajita-atm23} 
or accelerator neutrinos \cite{nufact-lowE1, nufact-lowE2,T2KK-2nd}, or both combined \cite{huber23,MEMPHYS}. 
The silver channel could be of help \cite{meloni08}. 
As in the previous cases the behavior of degeneracy solutions are far more violent 
in MB2 setting. It by itself might mean the great sensitivity to resolve the degeneracy. 
Or, it is a natural way of thinking to combine it with the measurement 
at the first oscillation maximum.

%

We give here a brief summary of the characteristic features of the 
degeneracy in superbeams, SB1, MB1, and MB2 settings.

\begin{itemize}

\item

A prominent difference between the true and the clone solutions at relatively 
large $\theta_{13}$, $\sin^2 2\theta_{13} \gtrsim 10^{-2}$, exists in 
$R_{ \text{N} }$ 
in the intrinsic degeneracy, while for the sign-$\Delta m^2_{31}$ degeneracy 
it is in the phase difference $D_{ \text{N} }^{(i)}$ for SB1 and MB1 settings. 

\item

The solutions III for the sign-$\Delta m^2_{31}$ degeneracy appears to be 
difficult to resolve for SB1 setting even if spectrum information is available, 
because energy dependences are so weak for both $R_{ \text{III} }$ and 
$D_{ \text{III} }^{(2)}$. 
The similar difficulty exists for MB1 if $\theta_{13}$ is large, 
$\sin^2 2\theta_{13} \gtrsim 10^{-2}$.
For the same reason, the solutions V and VII of the $\theta_{23}$ octant degeneracy 
is difficult to lift. 

\item

The short baseline SB1 option is unique among the three superbeam settings 
in the sense that it by itself may not be able to lift the sign-$\Delta m^2_{31}$ 
and the $\theta_{23}$ octant degeneracies, but can provide a clean discovery of 
CP violation without confusion with CP conservation. 
This is in accord with the basic motivation for low energy superbeam \cite{MNplb00}. 

\item

In comparison with SB1 and MB1 settings, the features of degeneracy solutions 
are always quite distinct at MB2 setting, where the energy region around the 
second oscillation maximum is explored. 
It by itself, or combined with other settings, would provide ways to help resolving 
the eightfold degeneracy.

\end{itemize}

\subsection{Parameter degeneracy in neutrino factory setting}
\label{overview-nufact}

In this subsection, we display the features of various degeneracy solutions 
by taking a setting which may be appropriate for neutrino factory. 
The two-detector setting with baselines $L = 3000-4000$ km and 
$L \sim 7000$ km seems to be considered as the ``standard'' one \cite{ISS-nufact} 
both for measurement of standard mixing parameters \cite{intrinsic,huber-winter}, 
possibly as well as for search for effects of NSI \cite{NSI-nufact,NSI-2phase,kopp3}. 
For the former purpose, the far detector at the ``magic baseline'' 
(as named by \cite{huber-winter}) plays a key role in resolving the degeneracy 
because of absence of $\delta$ dependence \cite{BMW02}. 
By restricting our purpose to illuminate the features of the degeneracy, 
we use the setup with just one detector at $L=4000$ km. 

Note that we have used set of probabilities 
$P(\nu_{\mu} \to \nu_{e})$ and $P(\bar{\nu}_{\mu} \to \bar{\nu}_{e})$ to obtain the 
degenerate solutions. 
Therefore, if you want to consider the more realistic setting of a neutrino factory 
in which T-conjugate (golden) channels will be used, please regard $\delta$ as 
$2 \pi - \delta$. 

\begin{figure}[bhtp]
\begin{center}
\vglue 0.3cm
\includegraphics[bb=0 0 340 160 , clip, width=0.9\textwidth]{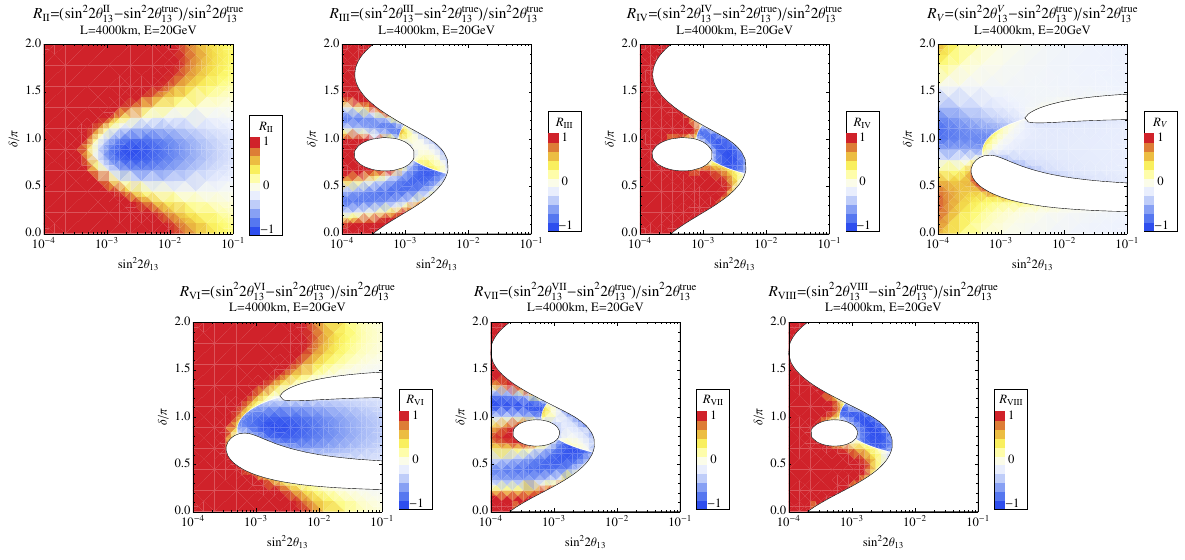}
\end{center}
\vglue -3mm
\caption{
The ratio $R_{ \text{II} } - R_{ \text{VIII} }$ are plotted in 
$\sin^2 2\theta_{13} - \delta/\pi$ space 
for a typical baseline and energy of neutrino factory setting. 
The ratio $R_{ \text{N} }$ is defined as 
$R_{ \text{N} } = 
[ \sin^2 2\theta_{13}^{ \text{N} } -  \sin^2 2\theta_{13}^{\text{true}}] / \sin^22\theta_{13}^{\text{true}} $ defined in (\ref{RN-def}) 
}
\label{RN-nufact}
\end{figure}

\begin{figure}[bhtp]
\begin{center}
\vglue 0.3cm
\includegraphics[bb=0 0 340 160 , clip, width=0.9\textwidth]{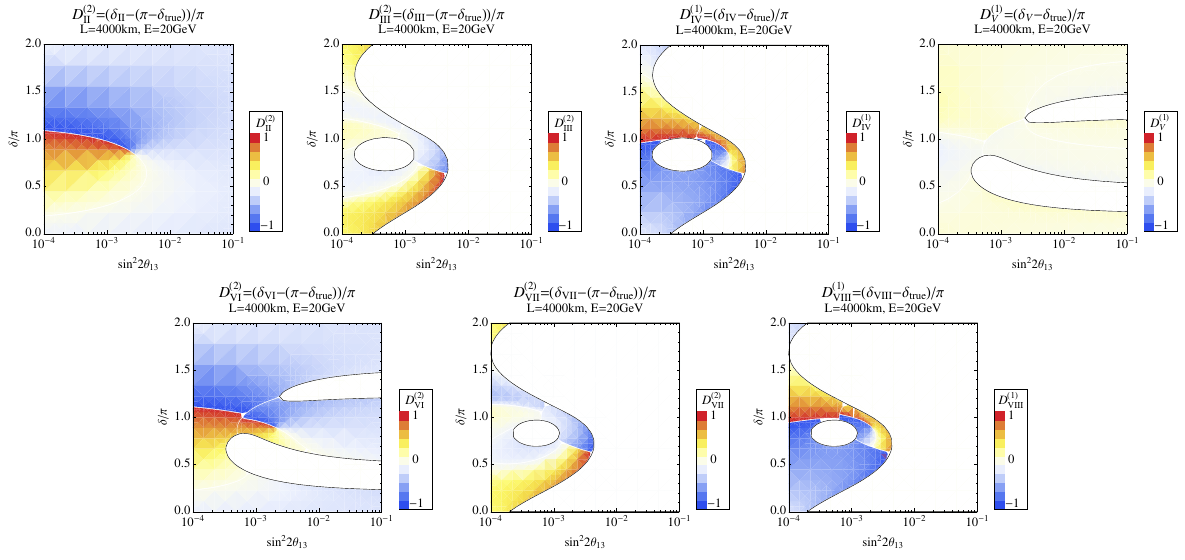}
\end{center}
\vglue -3mm
\caption{
The normalized differences 
$D_{ \text{II} }^{(2)}$, 
$D_{ \text{III} }^{(2)}$, 
$D_{ \text{IV} }^{(1)}$,
 $D_{ \text{V} }^{(1)}$, 
 $D_{ \text{VI} }^{(2)}$, 
 $D_{ \text{VII} }^{(2)}$, and 
 $D_{ \text{VIII} }^{(1)}$ are plotted in order in 
$\sin^2 2\theta_{13} - \delta/\pi$ space 
for a typical baseline and energy of neutrino factory setting. 
$D_{ \text{N} }$ is defined in (\ref{DN-def}) as 
$D_{ \text{N} }^{(2)} \equiv [\delta^{ \text{N} } - (\pi - \delta^{ \text{true} } ) ] / \pi$ and 
$D_{ \text{N} }^{(1)} \equiv [\delta^{ \text{N} } - \delta^{ \text{true} } ] / \pi$. 
}
\label{DN-nufact}
\end{figure}

In Figs.~\ref{RN-nufact} and \ref{DN-nufact}, the differences between the true 
solution and the fake degeneracy solutions, $R_{ \text{N} }$ and $D_{ \text{N} }^{(i)}$ 
(N=II$-$VIII, $i=1~\text{or}~2$), respectively, are plotted. 
As can be seen in these figures 
the differences between the true solution and the fake degeneracy solutions 
are generically larger than the cases of superbeam type settings discussed in the 
previous subsections. 
In accord with the expected higher sensitivities, we extend the region of  
$\sin^2 2\theta_{13}$ to $10^{-4}$.

%

\begin{figure}[bhtp]
\begin{center}
\includegraphics[bb=0 0 340 83 , clip, width=0.9\textwidth]{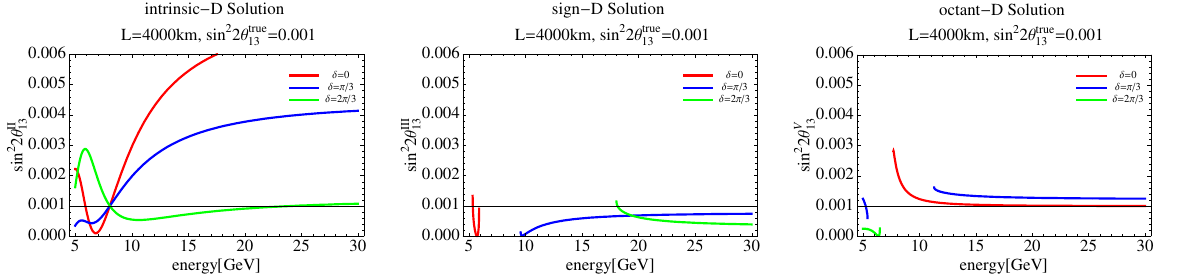}
\end{center}
\vglue -3mm
\caption{
The energy dependences of 
$\sin^2 2\theta_{13}^{ \text{II} }$ (left panel),  
$\sin^2 2\theta_{13}^{ \text{III} }$ (middle panel),  and 
$\sin^2 2\theta_{13}^{ \text{V} }$ (right panel) 
are plotted. 
The true value of $\theta_{13}$ is taken as $\sin^2 2\theta_{13}=0.001$, which 
is indicated by the horizontal solid line in the figure. 
}
\label{energy-dep-sN-nufact}
\end{figure}

\begin{figure}[bhtp]
\begin{center}
\includegraphics[bb=0 0 340 83 , clip, width=0.9\textwidth]{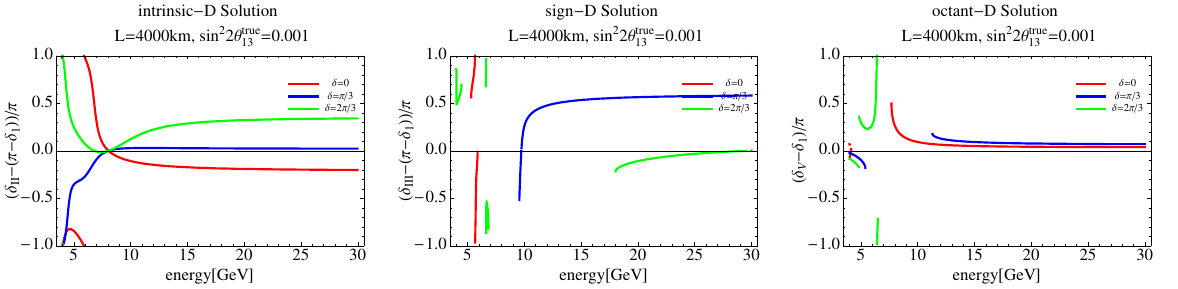}
\end{center}
\vglue -3mm
\caption{
The energy dependences of 
$D_{ \text{II} }^{(2)}$  (left panel), 
$D_{ \text{III} }^{(2)}$  (middle panel),  and  
 $D_{ \text{V} }^{(1)}$  (right panel) 
 are plotted. 
$D_{ \text{N} }^{(1)}$ and $D_{ \text{N} }^{(2)}$ are defined in (\ref{DN-def}). 
}
\label{energy-dep-deltaN-nufact}
\end{figure}

In Figs.~\ref{energy-dep-sN-nufact} and \ref{energy-dep-deltaN-nufact}, 
the energy dependences of $\sin^2 2\theta_{13}^{ \text{N} }$ and 
$D_{ \text{N} }^{(i)}$, respectively, are presented for three representative 
solutions, II, III, and V. 
For energy dependences of the other solutions see \cite{MU-version1,uchinami-thesis}. 
Generally speaking, the energy dependences of both of the quantities are 
significant compared to those in the SB1 and MB1 settings. 
Notable exceptions are the solution V 
(both $\sin^2 2\theta_{13}^{ \text{V} }$ and $D_{ \text{V} }^{(1)}$), 
and possibly 
$\sin^2 2\theta_{13}^{ \text{III} }$ and 
$\sin^2 2\theta_{13}^{ \text{VII} }$, 
all except for the low energy region $E \lesssim 10$ GeV. 
Unless there is a sensitivity to the low-energy region it would be difficult 
to resolve the degeneracy, in particular V, by the spectrum informations. 
Therefore, it is extremely important to lower the threshold into $E \lesssim 10$ GeV 
to resolve the degeneracy by spectrum analysis. 
An extensive effort toward this direction is made and the task is in progress 
\cite{ISS-detector}.

%

It is possible to understand, at least qualitatively, 
smallness of the difference from the true solution 
and lack of strong energy dependence of the solutions V and VII. 
Because the $\theta_{23}$ perturbation theory also applies to NF setting, 
it can be expected that the difference between the true solution and the clone 
one V is small. The similar statement holds for the solution VII given the smallness 
of the energy dependent term in $\sin^2 2\theta_{13}^{ \text{III} }$. 
Then, the question is why $\sin^2 2\theta_{13}^{ \text{III} }$ is small and 
lacks the significant energy dependence despite that the matter perturbation 
theory is not valid for NF setting. 
Qualitatively, the answer is that pinning to a small value due to the fact that 
the assumed true value itself is small, and lack of energy dependence 
because of fast reach to the asymptotic behavior discussed in Sec.~\ref{asymptotic}.

One notices that the intrinsic solution II in NF setting has the similar features as 
the one in MB2 setting, as can be seen by comparing 
Figs.~\ref{RN-nufact} and \ref{DN-nufact} to Figs.~\ref{R2} and \ref{D2}. 
It is because the value of $\theta_{13}$ taken is small and the atmospheric term 
is comparable to the solar term. 
Parallelism is not so complete in the other types of degeneracies, but 
some features can be understood in analogy to the case of MB2 setting.\footnote{
An example is that there is the region that $R_{ \text{V} }$ is large 
in small $\theta_{13}$. 
This is for the same reason of the case in MB2 setting, 
the difference between the true and V of order $\sim \epsilon_{ \text{oct} } Y_{\pm}$, 
is negligible compare with $s_{1}$. 
Another example is much stronger energy dependence of $s_{ \text{II} }$ in NF setting 
than SB1's, which is reminiscent of the feature of MB2 setting. 
It comes from larger effect of the solar-scale oscillation term. 
}
%
In doing so the difference due to the much wider no-solution region of the 
sign-$\Delta m_{31}^2$ degeneracy due to long baseline must, of course, 
be taken into account.

\section{Parameter Degeneracy in T-Conjugate Measurement }
\label{T-conjugate}

We analyze in this section the problem of parameter degeneracy 
in T-conjugate measurement in neutrino oscillation. 
Though measurement of T violation does not appear to be feasible immediately 
understanding its structure may be interesting theoretically. 
The topics was first treated in \cite{MNP2}, but we make the structure 
of the degeneracy more transparent in harmony with the symmetry argument in 
Sec.~\ref{invariance}. In fact, the structure of the degeneracy with T-conjugate 
measurement is one of the simplest one among the cases discussed 
due to the symmetry. It is also an ideal tool to obtain the vacuum limit.

\subsection{The intrinsic degeneracy in T-conjugate measurement}
\label{T-intrinsic}

The intrinsic degeneracy solutions ($s_{i}$, $\delta_{i}$) (i=1, 2) 
are defined in $\nu_{\mu} \rightarrow \nu_e$ channel by 
\begin{eqnarray} 
P - Z &=& 
X_{\pm}s_{1}^2 + 
Y_{\pm} s_{1} \left( 
\cos \delta_{1} \cos \Delta_{31} \mp \sin \delta_{1} \sin \Delta_{31} 
\right), 
\nonumber \\
P - Z &=& 
X_{\pm}s_{2}^2 + 
Y_{\pm} s_{2} \left( 
\cos \delta_{2} \cos \Delta_{31} \mp \sin \delta_{2} \sin \Delta_{31} 
\right), 
\label{T-intrinsic-def1}
\end{eqnarray}
and in T-conjugate channel by 
\begin{eqnarray} 
P^{T} - Z &=& 
X_{\pm}s_{1}^2 + 
Y_{\pm} s_{1} \left( 
\cos \delta_{1} \cos \Delta_{31} \pm \sin \delta_{1} \sin \Delta_{31} 
\right),  
\nonumber \\
P^{T} - Z &=& 
X_{\pm}s_{2}^2 + 
Y_{\pm} s_{2} \left( 
\cos \delta_{2} \cos \Delta_{31} \pm \sin \delta_{2} \sin \Delta_{31} 
\right). 
\label{T-intrinsic-def2}
\end{eqnarray}
By subtracting two equations in (\ref{T-intrinsic-def1}) and (\ref{T-intrinsic-def2}), 
respectively, and then subtracting and adding the resultant two equations, 
we obtain (assuming $Y_{\pm} \sin \Delta_{31} \neq 0$)   
\begin{eqnarray} 
s_{2} \sin \delta_{2} &=& s_{1} \sin \delta_{1}, 
\nonumber \\
s_{2} \cos \delta_{2} &=& s_{1} \cos \delta_{1} + 
\frac{ X_{\pm} }{ Y_{\pm} \cos \Delta_{31} } 
(s_{1}^2 -  s_{2}^2 ). 
\label{T-intrinsic-delta} 
\end{eqnarray}
Inserting (\ref{T-intrinsic-delta}) into 
$\cos^2 \delta_{2} + \sin^2 \delta_{2} = 1$ yields the equation for $s_{2}$ in a form 
$(s_{2}^2 - s_{1}^2) (s_{2}^2 - s_{ \text{II} }^2) = 0$, which admits the intrinsic 
degeneracy solution 
\begin{eqnarray} 
s_{ \text{II} } &=& \sqrt { s_{1}^2 + 2 \left( \frac{Y_{\pm} \cos \Delta_{31}}{X_{\pm}} \right) s_{1} \cos \delta_{1} + 
 \left( \frac{Y_{\pm} \cos \Delta_{31}}{X_{\pm}} \right)^2 }.  
\label{T-intrinsic-s-solution}
\end{eqnarray}
Given the solution $s_{2}=s_{ \text{II} }$ the solution for $\delta_{2}$ can be obtained 
by using  (\ref{T-intrinsic-delta}) as 
\begin{eqnarray} 
s_{ \text{II} } \sin \delta_{ \text{II} } &=& s_{1} \sin \delta_{1}, 
\nonumber \\
s_{ \text{II} } \cos \delta_{ \text{II} } &=& - 
\left(
s_{1} \cos \delta_{1} + \frac{Y_{\pm} \cos \Delta_{31} }{X_{\pm}}  
\right). 
\label{T-intrinsic-delta-solution}
\end{eqnarray}
By further expanding by $\frac{Y_{\pm}}{X} $, assuming it small, 
the solution obtained in \cite{MNP2} is reproduced; 
\begin{eqnarray} 
s_{ \text{II} } \simeq s_{1} + \frac{Y_{\pm} \cos \Delta_{31} }{X_{\pm}}  
\cos \delta_{1}. 
\label{app-solution}
\end{eqnarray}

\subsection{The sign-$\Delta m^2$ degeneracy in T-conjugate measurement}
\label{T-signdm2-old}

As we learned in Sec.~\ref{invariance} the symmetry argument tells us 
that there exists the sign-$\Delta m^2$ degeneracy in T-conjugate measurement. 
In this subsection we verify it by deriving explicit solutions 
without recourse to the symmetry argument. 
We denote $s_{13}$ variable for the opposite-sign $\Delta m^2_{31}$ solution 
as $(s_{3}, \delta_{3})$, whose two (as we prove) solutions will be denoted as 
$(s_{ \text{III} }, \delta_{ \text{III} } )$ and $(s_{ \text{IV} }, \delta_{ \text{IV} } )$.

The sign-$\Delta m^2_{31}$ degeneracy is defined by the following two sets of equations: 
\begin{eqnarray} 
P - Z &=& 
X_{\pm}s_{1}^2 + 
Y_{\pm} s_{1} \left( 
\cos \delta_{1} \cos \Delta_{31} \mp \sin \delta_{1} \sin \Delta_{31} 
\right), 
\nonumber \\
P - Z &=& 
X_{\mp} s_{3}^2 + Y_{\mp} s_{3} \left( 
\cos \delta_{3} \cos \Delta_{31} \pm \sin \delta_{3} \sin \Delta_{31} 
\right), 
\label{T-sign-def-a} 
\end{eqnarray}
and
\begin{eqnarray} 
P^{T} - Z &=& 
X_{\pm}s_{1}^2 + 
Y_{\pm} s_{1} \left( 
\cos \delta_{1} \cos \Delta_{31} \pm \sin \delta_{1} \sin \Delta_{31} 
\right), 
\nonumber \\
P^{T} - Z &=& 
X_{\mp} s_{3}^2 + Y_{\mp} s_{3} \left( 
\cos \delta_{3} \cos \Delta_{31} \mp \sin \delta_{3} \sin \Delta_{31} 
\right). 
\label{T-sign-def-b}
\end{eqnarray}
By the similar procedure as in the previous subsection we obtain 
(assuming $Y_{\pm} \sin \Delta_{31} \neq 0$) 
\begin{eqnarray}
s_{3} \sin \delta_{3} &=& - 
\left( \frac{Y_{\pm} }{ Y_{\mp} } \right) s_{1} \sin \delta_{1}, 
\nonumber \\
s_{3} \cos \delta_{3} &=& 
\left( \frac{Y_{\pm} }{ Y_{\mp} } \right) s_{1} \cos \delta_{1} + 
\frac{ 1 }{ Y_{\mp} \cos \Delta_{31} }
\left( X_{\pm} s_{1}^2 - X_{\mp} s_{3}^2 \right). 
\label{T-sign-delta}
\end{eqnarray}
Inserting (\ref{T-sign-delta}) into 
$\cos^2 \delta_{3} + \sin^2 \delta_{3} = 1$ leads to the equation for $s_{3}^2$ as 
\begin{eqnarray} 
\left( s_{3}^2 - \frac{ X_{\pm} }{ X_{\mp} } s_{1}^2 \right) 
\left( s_{3}^2 - \frac{ X_{\pm} }{ X_{\mp} } s_{ \text{II} }^2 \right) 
 = 0, 
\label{T-sign-s-equation} 
\end{eqnarray}
where $s_{ \text{II} }$ is defined in (\ref{T-intrinsic-s-solution}). 
We note that the relation (\ref{identity}) is essential to reduce the equation in 
(\ref{T-sign-s-equation}) to the current form. 
%
%
The solutions for CP phase $\delta$ can be obtained by inserting the solutions  
of (\ref{T-sign-s-equation}), its positive root, into (\ref{T-sign-delta}) with use 
of (\ref{identity}), and for $\delta_{ \text{IV} }$ together with (\ref{T-intrinsic-delta}). 
The solutions of $\delta$ together with the $\theta_{13}$ solutions 
of the sign-$\Delta m^2_{31}$ degeneracy are summarized as below:  
\begin{eqnarray} 
s_{ \text{III} } &=& \sqrt{ \frac{ X_{\pm} }{ X_{\mp} } } s_{1},  
\hspace{6mm}
\delta_{ \text{III} } = \pi - \delta_{1}, 
\nonumber \\ 
s_{ \text{IV} } &=& \sqrt{ \frac{ X_{\pm} }{ X_{\mp} } } s_{ \text{II} },  
\hspace{6mm}
\delta_{ \text{IV} } = \pi - \delta_{ \text{II} }. 
\label{T-s-delta-solution} 
\end{eqnarray}
The solutions (\ref{T-s-delta-solution}) are in perfect agreement with the 
expectation based on invariance of the oscillation probability given in 
Sec.~\ref{invariance}. 
Figure~\ref{biP-plot-T} clearly exhibits the structure obtained in (\ref{T-s-delta-solution}).

\subsection{The octant degeneracy in T-conjugate measurement}
\label{T-octant}

The $\theta_{23}$ octant degeneracy in T-conjugate measurement is defined by
\begin{eqnarray} 
P  &=& 
X^{\text{true}}_{\pm}s_{1}^2 + 
Y_{\pm} s_{1} \left( 
\cos \delta_{1} \cos \Delta_{31} \mp \sin \delta_{1} \sin \Delta_{31} 
\right) +
Z^{\text{true}},
\nonumber \\
P &=& 
X^{\text{false}}_{\pm}s_{5}^2 + 
Y_{\pm} s_{5} \left( 
\cos \delta_{5} \cos \Delta_{31} \mp \sin \delta_{5} \sin \Delta_{31} 
\right) +
Z^{\text{false}},
\\
P^{T}&=& 
X^{\text{true}}_{\pm}s_{1}^2 + 
Y_{\pm} s_{1} \left( 
\cos \delta_{1} \cos \Delta_{31} \pm \sin \delta_{1} \sin \Delta_{31} 
\right) +
Z^{\text{true}},
\nonumber \\
P^{T} &=& 
X^{\text{false}}_{\pm}s_{5}^2 + 
Y_{\pm} s_{5} \left( 
\cos \delta_{5} \cos \Delta_{31} \pm \sin \delta_{5} \sin \Delta_{31} 
\right) +
Z^{\text{false}}.
\label{T-octant-def}
\end{eqnarray}
From these equations, we obtain
\begin{eqnarray}
s_{5} \cos \delta_{5} 
&=&
s_{1} \cos \delta_{1} + 
\frac{1}{Y_{\pm} \cos \Delta_{31}} 
\left[ 
X^{\text{true}}_{\pm} s_{1}^2 - X^{\text{false}}_{\pm} s_{5}^2 + Z^{\text{true}} - Z^{\text{false}} \right] 
\nonumber \\
s_{5} \sin \delta_{5}
&=&
s_{1} \sin \delta_{1}. 
\label{T-oct-intr-delta-solution}
\end{eqnarray}
The relation $\cos^2 \delta_{5} + \sin^2 \delta_{5} = 1$ gives the quadratic equation 
for $s_{5}^2$ which results in the similar solution 
\begin{eqnarray}
s_{\text{V}, \text{VI} }^2
&=&
\frac{1}{ ( X^{\text{false}}_{\pm})^2 } 
\left[ 
X^{\text{false}}_{\pm} 
\left\{  X^{\text{true}}_{\pm} s_{1}^2 + 
Y_{\pm}  \cos \Delta_{31} s_{1} \cos \delta_{1} + 
\left(Z^{\text{true}} - Z^{\text{false}} \right) 
\right\} 
\right.
\nonumber \\
& & \left. \hspace{10mm} 
+ \frac{1}{2} (Y_{\pm} \cos \Delta_{31})^2 
\hspace{1mm} \mp \hspace{1mm} 
d_{\pm}^{\text{T-oct-intr}} \sqrt{ \frac{ D_{\pm}^{\text{T-oct-intr}} }{ (d_{\pm}^{\text{T-oct-intr}})^2 }}
\right], 
\label{T-oct-intr-s-solution}
\end{eqnarray}
where the upper $-$ (lower $+$) sign is for $s_{\text{V}}$ ($s_{\text{VI}}$). 
The functions $D_{\pm}^{\text{T-oct-intr}}$ and $d_{\pm}^{\text{T-oct-intr}}$ 
are defined by 
\begin{eqnarray}
&& D_{\pm}^{\text{T-oct-intr}} 
=
\left( Y_{\pm} \cos \Delta_{31} \right)^2 
\nonumber \\
&\times&
\left[
\left( X^{\text{false}}_{\pm} s_{1} \cos \delta_{1} + \frac{1}{2} Y_{\pm} \cos \Delta_{31} 
\right)^2 + 
X^{\text{false}}_{\pm} 
\biggl\{
\left( Z^{\text{true}} - Z^{\text{false}} \right) - 
(X^{\text{true}}_{\pm} - X^{\text{false}}_{\pm}) s_{1}^2 
\biggr\}
\right], 
\nonumber \\
&& d_{\pm}^{\text{T-oct-intr}} = 
\lim_{\theta_{23} \rightarrow \pi/4} D_{\pm}^{\text{T-oct-intr}}  
=
Y_{\pm} \cos \Delta_{31} 
\left( X_{\pm} s_{1} \cos \delta_{1} + \frac{1}{2} Y_{\pm} \cos \Delta_{31} 
\right), 
\label{T-oct-D-d-def}
\end{eqnarray}
where $X_{\pm}$ and $Y_{\pm}$ in the last line is meant to be those 
at $\theta_{23} = \pi/4$.
Once the solutions $s_{\text{V}}$ and $s_{\text{VI}}$ are known 
one can readily obtain $\delta_{\text{V}}$ and $\delta_{\text{VI}}$ 
by inserting the $s_{13}$ solutions into (\ref{T-oct-intr-delta-solution}).

Though, we do not discuss in any detail, it must be obvious that the solutions 
($s_{\text{VII}}, \delta_{\text{VII}}$) and ($s_{\text{VIII}}, \delta_{\text{VIII}}$), 
the ones with both the $\Delta m^2_{31}$-sign and the octant flips, 
are given by using the solutions obtained in this subsection by 
the same type of equation as (\ref{T-s-delta-solution}): 
$s_{ \text{VII} } = \sqrt{ X_{\pm}^{ \text{false} } / X_{\mp}^{ \text{false} }  } s_{ \text{V} }$,
$s_{ \text{VIII} } = \sqrt{ X_{\pm}^{ \text{false} } / X_{\mp}^{ \text{false} }  } s_{ \text{VI} }$,
$\delta_{ \text{VII} }  = \pi - \delta_{ \text{V} }$, and 
$\delta_{ \text{VIII} } = \pi - \delta_{ \text{VI} }$. 
Alternatively, they can be also obtained by the octant flip mapping, implicitly given in 
(\ref{T-oct-intr-s-solution}) and (\ref{T-oct-intr-delta-solution}), 
from ($s_{\text{III}}, \delta_{\text{III}}$) and ($s_{\text{IV}}, \delta_{\text{IV}}$) 
obtained in the previous section.

\subsection{Parameter Degeneracy in Vacuum}
\label{vacuum}

Though it is pedagogically useful to work out the parameter degeneracy in vacuum 
we just give the results by taking the vacuum limit in the degeneracy solutions 
obtained for T conjugate measurement. 
We need the results in vacuum to define our convention we used to specify unambiguously the solutions for the sign-$\Delta m^2_{31}$ degeneracy 
in Sec.~\ref{convention}.\footnote{
Notice that the mass hierarchy does matter in the discussion of parameter 
degeneracy in vacuum though it might be thought contrary. 
In fact, it is known that the hierarchy can be determined by measuring 
the sign of the solar-atmospheric interference term in vacuum \cite{MNPZ07}. 
}

In fact, it is straightforward to observe that the expressions of the degeneracy 
solutions in vacuum: 
They are almost identical to those obtained in this section; 
The only necessary step is to take the vacuum limit 
\begin{eqnarray}
\lim_{a \rightarrow 0} X_{\pm} &=& X_{vac} \equiv 4 s^2_{23} \sin^2 \Delta_{31}, 
\nonumber  \\
\lim_{a \rightarrow 0} Y_{\pm} &=& \pm Y_{vac} \equiv \pm 2 \sin{2\theta_{12}} \sin{2\theta_{23}}  \Delta_{21} \sin \Delta_{31}, 
\nonumber  \\
\lim_{a \rightarrow 0} Z &=& Z_{ \text{vac} } \equiv c^2_{23} \sin^2{2\theta_{12}}, 
\label{XYZ-vac}
\end{eqnarray}
in the solutions 
(\ref{T-intrinsic-s-solution}), 
(\ref{T-intrinsic-delta-solution}), and 
(\ref{T-s-delta-solution}). 
Their explicit forms are given in Appendix D of  \cite{NSI-perturbation}.
The degeneracy solutions which involve $\theta_{23}$ octant flip can also be 
obtained by taking the same limit in the solutions obtained in Sec.~\ref{T-octant}.

\section{Parameter Degeneracy with Golden and Silver Channels}
\label{Golden-Silver}

We discuss the parameter degeneracy for a given measurement in 
the $\nu_{e} \rightarrow \nu_{\mu}$ (golden) and the 
$\nu_{e} \rightarrow \nu_{\tau}$ (silver) channels. 
The oscillation probability in the former and the latter channels are 
given by $P^{T}$ in (\ref{PmueT}) and $P^{S}$ in (\ref{Petau}), respectively. 
We note that detection of $\nu_{\tau}$ requires the energy at least above $\tau$ 
production threshold, and therefore most probably, neutrino factory would be 
the appropriate place for the silver channel \cite{silver}.

\subsection{The intrinsic degeneracy in Golden-Silver measurement}
\label{GS-intrinsic}

\begin{figure}[bhtp]
\begin{center}
\vglue 0.3cm
\includegraphics[bb=0 0 240 233 , clip, width=0.36\textwidth]{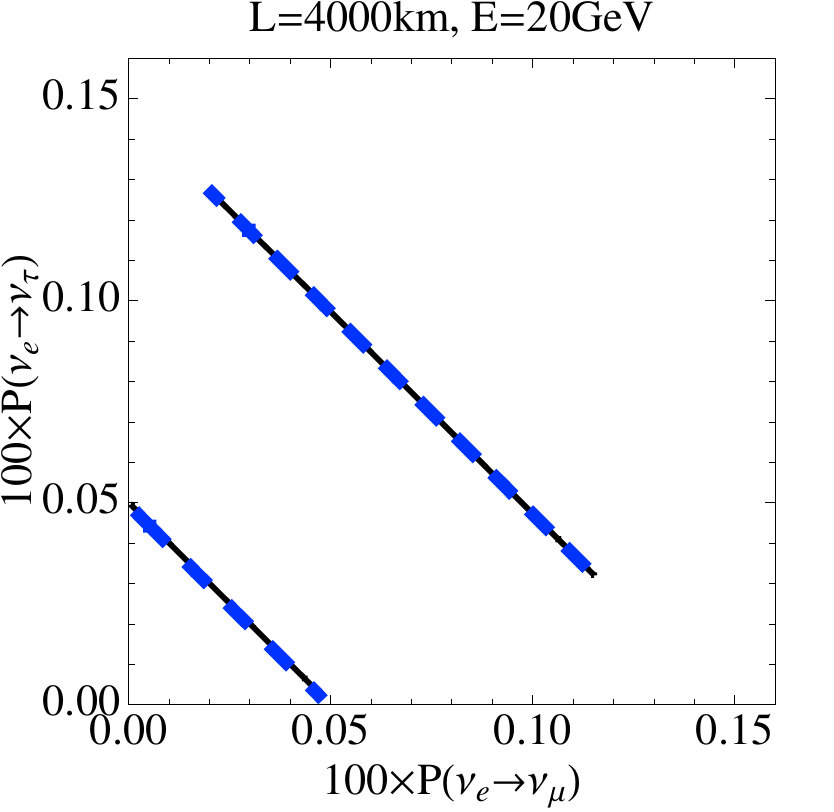}
\end{center}
\vglue -0.3cm
\caption{
$P^{T} - P^{S}$ (golden-silver) bi-probability plot for a neutrino factory setting. 
Black solid line : $\sin^2 2\theta_{13}=0.004 (0.001)$ with normal hierarchy
for larger (smaller) oscillation probability. 
Blue dashed line : $\sin^2 2\theta_{13} = 0.0158 (0.0039)$ with inverted hierarchy 
for lager (smaller) oscillation probability. 
Notice that black and blue lines are overlapping. 
$\theta_{23}$ is taken to be 42 degree. 
}
\label{biP-plot-GS}
\end{figure}

The intrinsic degeneracy is defined by 
\begin{eqnarray} 
P^{T} - Z &=& 
X_{\pm}s_{1}^2 + 
Y_{\pm} s_{1} \left( 
\cos \delta_{1} \cos \Delta_{31} \pm \sin \delta_{1} \sin \Delta_{31} 
\right) 
\nonumber \\
P^{T} - Z &=& 
X_{\pm}s_{2}^2 + 
Y_{\pm} s_{2} \left( 
\cos \delta_{2} \cos \Delta_{31} \pm \sin \delta_{2} \sin \Delta_{31} 
\right) 
\label{GS-intrinsic-def1}
\end{eqnarray}
and
\begin{eqnarray} 
P^{S} - \tan^2{\theta_{23}} Z &=& 
\cot^2{\theta_{23}} X_{\pm}s_{1}^2 - 
Y_{\pm} s_{1} \left( 
\cos \delta_{1} \cos \Delta_{31} \pm \sin \delta_{1} \sin \Delta_{31} 
\right) 
\nonumber \\
P^{S} - \tan^2{\theta_{23}} Z &=& 
\cot^2{\theta_{23}} X_{\pm}s_{2}^2 - 
Y_{\pm} s_{2} \left( 
\cos \delta_{2} \cos \Delta_{31} \pm \sin \delta_{2} \sin \Delta_{31} 
\right) 
\label{GS-intrinsic-def2}
\end{eqnarray}
By subtracting two equations in (\ref{GS-intrinsic-def1}) and (\ref{GS-intrinsic-def2}) 
we obtain 
\begin{eqnarray} 
X_{\pm} ( s_{1}^2 -  s_{2}^2 ) + Y_{\pm} S_{I} = 0, 
\hspace{10mm}
X_{\pm} ( s_{1}^2 -  s_{2}^2 ) - \tan^2{\theta_{23}} Y_{\pm} S_{I} = 0, 
\label{GS-intrinsic1}
\end{eqnarray}
where
\begin{eqnarray} 
S_{I \pm} \equiv 
\left[ 
\cos \Delta_{31} 
( s_{1} \cos \delta_{1} - s_{2} \cos \delta_{2} ) \pm 
\sin \Delta_{31} 
( s_{1} \sin \delta_{1} - s_{2} \sin \delta_{2} ) 
\right] 
\label{SIpm-def}
\end{eqnarray}
Then, it follows that $s_{ \text{II} } = s_{1}$ 
(using the label $s_{ \text{II} }$ for the intrinsic solution) 
and $S_{I} =0$ assuming that $Y_{\pm} \neq 0$. 
The former result is, of course, expected by the feature of ``shrunk ellipse'' 
in the bi-probability plot given in Fig.~\ref{biP-plot-GS}. 
Using $s_{ \text{II} } = s_{1}$ the equation $S_{I} =0$ has a solution, 
apart from the trivial solution $\delta_{2} = \delta_{1}$, as 
\begin{eqnarray} 
\cos \delta_{ \text{II} } = 
\cos ( \delta_{1} \mp 2 \Delta_{31} ), 
\hspace{10mm}
\sin \delta_{ \text{II} } = 
- \sin ( \delta_{1} \mp 2 \Delta_{31} )
\label{GS-cos-sin-delta-intrinsic}
\end{eqnarray}
which implies that 
\begin{eqnarray} 
\delta_{ \text{II} } = 2\pi- ( \delta_{1} \mp 2 \Delta_{31}) 
\hspace{6mm}
(\text{mod.}~2\pi).
\label{GS-intrinsic-delta-solution}
\end{eqnarray}
The structure of the solutions of $\delta$ should be obvious from the form of 
the oscillation probabilities in (\ref{GS-intrinsic-def1}) and (\ref{GS-intrinsic-def2}) 
which is reflected to the feature of shrunk ellipse in Fig.~\ref{biP-plot-GS}; 
The two degenerate solutions must have the same values of 
$\cos (\delta \mp  \Delta_{31})$, and hence 
$\delta_{ \text{II} } \mp  \Delta_{31} = 2\pi - (\delta_{1} \mp  \Delta_{31})$.

\subsection{The sign-$\Delta m^2$ degeneracy in Golden-Silver measurement}
\label{GS-signdm2}

The sign-$\Delta m^2$ degeneracy is defined by 
\begin{eqnarray} 
P^{T} - Z &=& 
X_{\pm}s_{1}^2 + 
Y_{\pm} s_{1} \left( 
\cos \delta_{1} \cos \Delta_{31} \pm \sin \delta_{1} \sin \Delta_{31} 
\right) 
\nonumber \\
P^{T} - Z &=& 
X_{\mp}s_{3}^2 +  
Y_{\mp} s_{3} \left( 
\cos \delta_{3} \cos \Delta_{31} \mp \sin \delta_{3} \sin \Delta_{31} 
\right) 
\label{GS-sign-def1}
\end{eqnarray}
\begin{eqnarray} 
P^{S} - \tan^2{\theta_{23}} Z &=& 
\cot^2{\theta_{23}} X_{\pm}s_{1}^2 - 
Y_{\pm} s_{1} \left( 
\cos \delta_{1} \cos \Delta_{31} \pm \sin \delta_{1} \sin \Delta_{31} 
\right) 
\nonumber \\
P^{S} - \tan^2{\theta_{23}} Z &=& 
\cot^2{\theta_{23}} X_{\mp}s_{3}^2 - 
Y_{\mp} s_{3} \left( 
\cos \delta_{3} \cos \Delta_{31} \mp \sin \delta_{3} \sin \Delta_{31} 
\right) 
\label{GS-sign-def2}
\end{eqnarray}
By subtracting two equations in (\ref{GS-sign-def1}) and (\ref{GS-sign-def2}) 
we obtain
\begin{eqnarray} 
X_{\pm} s_{1}^2 -  X_{\mp} s_{3}^2 + S_{S} = 0, 
\hspace{10mm}
X_{\pm} s_{1}^2 -  X_{\mp} s_{3}^2 -  \tan^2{\theta_{23}} S_{S} = 0, 
\label{GS-sign1}
\end{eqnarray}
where
\begin{eqnarray} 
S_{S} \equiv \cos \Delta_{31} 
\left( Y_{\pm} s_{1} \cos \delta_{1} -  Y_{\mp} s_{3} \cos \delta_{3} \right) 
&\pm&
\sin \Delta_{31} 
\left( Y_{\pm} s_{1} \sin \delta_{1} + Y_{\mp} s_{3} \sin \delta_{3} \right).
\label{GS-sign2}
\end{eqnarray}
Then, (\ref{GS-sign1}) implies that the two separate factors both have to vanish: 
\begin{eqnarray} 
s_{ \text{III} } &=& \sqrt{ \frac{ X_{\pm} }{ X_{\mp} } } s_{1}, 
\hspace{10mm}
S_{S} = 0, 
\label{GS-sign3}
\end{eqnarray}
where we use the notations $(s_{ \text{III} }, \delta_{ \text{III} })$ and 
$(s_{ \text{IV} }, \delta_{ \text{IV} })$ for the sign-$\Delta m^2_{31}$ 
degeneracy solutions. 
The former result is in agreement with the symmetry argument given in 
Sec.~\ref{invariance}. 
Because $s_{ \text{II} } = s_{1}$, $s_{ \text{IV} }=s_{ \text{III} }$ holds. 

By using the above $s_{3}$ solution and the relation (\ref{identity}) in 
Sec.~\ref{probability}, it is easy to show that the equation $S_{S} = 0$ 
can be converted to the same equation as (\ref{GS-cos-sin-delta-intrinsic}) 
(after obvious replacement of $\delta_{2}$ to $\delta_{3}$)
apart from the sign change of $\cos \delta_{3}$ term. 
It means that $\pi - \delta_{3}$ obeys exactly the same equation as 
(\ref{GS-intrinsic-delta-solution}). 
Then, we obtain the solutions 
%
\begin{eqnarray} 
\delta_{ \text{III} } &=& \pi - \delta_{1}, 
\nonumber \\
\delta_{ \text{IV} } &=& \pi - \delta_{ \text{II} } =  -\pi + ( \delta_{1} \mp 2 \Delta_{31})
\hspace{6mm}
(\text{mod.}~2\pi).
\label{GS-delta-intrinsic}
\end{eqnarray}
The structure of the solution in one-to-one correspondence to the 
intrinsic degeneracy solutions is perfectly consistent with the symmetry 
argument in Sec.~\ref{invariance}.

\subsection{The octant degeneracy in Golden-Silver measurement}
\label{GS-octant}

Now, we discuss the $\theta_{23}$ octant degeneracy.
We will see that simplicity of the golden-silver setting prevails in it. 
The $\theta_{23}$ octant degeneracy solutions ($s_{5}, \delta_{5}$) satisfy 
\begin{eqnarray} 
P^{T} &=& 
X^{\text{true}}_{\pm}s_{1}^2 + 
Y_{\pm} s_{1} \left( 
\cos \delta_{1} \cos \Delta_{31} \pm \sin \delta_{1} \sin \Delta_{31} 
\right) +
Z^{\text{true}}, 
\nonumber \\
P^{T} &=& 
X^{\text{false}}_{\pm}s_{5}^2 + 
Y_{\pm} s_{5} \left( 
\cos \delta_{5} \cos \Delta_{31} \pm \sin \delta_{5} \sin \Delta_{31} 
\right) +
Z^{\text{false}}, 
\label{GS-octant-def1}
\end{eqnarray}
and 
\begin{eqnarray} 
P^{S} &=& 
\cot^2 \theta_{23} X^{\text{true}}_{\pm}s_{1}^2 - 
Y_{\pm} s_{1} \left( 
\cos \delta_{1} \cos \Delta_{31} \pm \sin \delta_{1} \sin \Delta_{31} 
\right) +
\tan^2 \theta_{23} Z^{\text{true}}, 
\nonumber \\
P^{S} &=& 
\tan^2 \theta_{23} X^{\text{false}}_{\pm}s_{5}^2 - 
Y_{\pm} s_{5} \left( 
\cos \delta_{5} \cos \Delta_{31} \pm \sin \delta_{5} \sin \Delta_{31} 
\right) +
\cot^2 \theta_{23} Z^{\text{false}}.  
\label{GS-octant-def2}
\end{eqnarray}
Using (\ref{Xtrue-Xfalse}), 
(\ref{GS-octant-def2}) can be 
written as 
\begin{eqnarray} 
P^{S} &=& 
X^{\text{false}}_{\pm}s_{1}^2 - 
Y_{\pm} s_{1} \left( 
\cos \delta_{1} \cos \Delta_{31} \pm \sin \delta_{1} \sin \Delta_{31} 
\right) +
Z^{\text{false}}, 
\nonumber \\
P^{S} &=& 
X^{\text{true}}_{\pm}s_{5}^2 - 
Y_{\pm} s_{5} \left( 
\cos \delta_{5} \cos \Delta_{31} \pm \sin \delta_{2} \sin \Delta_{31} 
\right) +
Z^{\text{true}}.  
\label{GS-octant-def3}
\end{eqnarray}
By subtracting two equations in (\ref{GS-octant-def1}) and (\ref{GS-octant-def3}) 
with the same octant labels one can easily obtain 
the solution of $s_{5}$ as $s_{5}^2 = s_{1}^2$ which leads to
\begin{eqnarray}
s_{\text{V}} =  s_{\text{VI}} =  s_{1}.
\end{eqnarray}
To obtain $\delta_{5}$, we proceed as usual which leads to the result 
\begin{eqnarray}
\cos (\delta_{\text{V}} \mp \Delta_{31})
=
\cos (\delta_{1} \mp \Delta_{31}) +
\frac{1}{s_{1} Y_{\pm}} 
\left[  \left(X^{\text{true}}_{\pm} - X^{\text{false}}_{\pm} \right) s_{1}^2 + \left(Z^{\text{true}} - Z^{\text{false}} \right) \right]. 
\label{GS-octant-delta-solution5}
\end{eqnarray}
As dictated by the general argument, $s_{\text{VI}}$ must be given by 
$s_{\text{V}}$ as above, and $\delta_{\text{VI}}$ as 
\begin{eqnarray} 
\delta_{ \text{VI} } = 2\pi- ( \delta_{V} \mp 2 \Delta_{31}) 
\hspace{6mm}
(\text{mod.}~2\pi),
\label{GS-octant-delta-solution6}
\end{eqnarray}
using (\ref{GS-intrinsic-delta-solution}) because they are the intrinsic degeneracy pair. 
In fact, it is easy to see that $\delta_{ \text{V} }$ and $\delta_{ \text{VI} }$ are 
the two solutions which satisfy 
$\cos (\delta_{\text{V}} \mp \Delta_{31}) = \cos (\delta_{\text{VI}} \mp \Delta_{31})$ 
in (\ref{GS-octant-delta-solution5}).

As in the case of T-conjugate measurement described in Sec.~\ref{T-octant}, 
the solutions ($s_{ \text{VII} }, \delta_{ \text{VII} }$) and 
($s_{ \text{VII} }, \delta_{ \text{VII} }$), the ones with octant as well as 
the $\Delta m^2_{31}$-sign flips, can be obtained from the above ones as 
\begin{eqnarray} 
s_{ \text{VII} } &=& s_{ \text{VIII} } = \sqrt{ \frac{ X_{\pm}^{ \text{false} } }{ X_{\mp}^{ \text{false} } } } s_{ \text{V} }, 
\nonumber \\
\delta_{ \text{VII} } &=& \pi - \delta_{ \text{V} }, 
\hspace{8mm}
\delta_{ \text{VIII} } = \pi - \delta_{ \text{VI} }.
\label{GS-oct-sign-s-delta-solution}
\end{eqnarray}
%


\section{Parameter Degeneracy in CPT-conjugate Measurement}
\label{CPT-conjugate}

We discuss in this section the problem of parameter degeneracy in 
CPT violation measurement. 
CPT-violation observable was considered to be useful to resolve the mass 
hierarchy because it gives the probability difference $P - P^{CPT}$ which is 
largest among the similar quantities \cite{MNP3,parke-beta}.

\subsection{The intrinsic degeneracy in CPT-conjugate measurement}
\label{CPT-intrinsic}

\begin{figure}[bhtp]
\begin{center}
\vglue 0.5cm
\includegraphics[bb=0 0 240 226 , clip, width=0.36\textwidth]{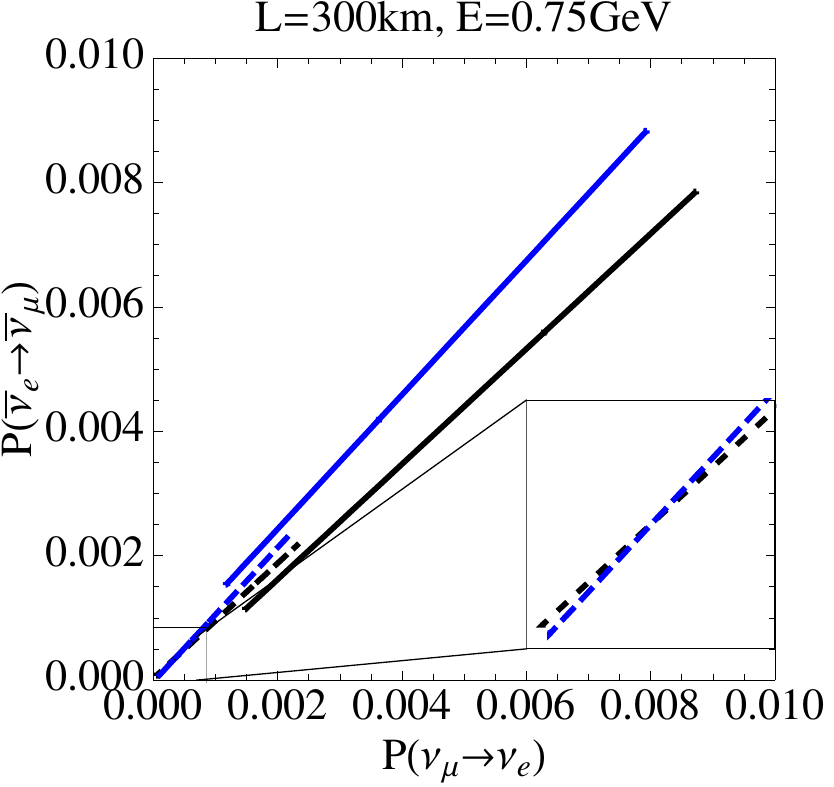}
\end{center}
\vglue -0.5cm
\caption{
$P - P^{CPT}$ bi-probability plot. 
The black and the blue solid (dashed) lines, which correspond respectively to 
the normal and the inverted hierarchies, are for $\sin^2 2\theta_{13}=0.01 (0.001)$. 
The region with small probabilities $P \leq 0.0007$ is magnified into the sub-panel 
to show more clearly the crossing of the two shrunk ellipses in the region. 
}
\label{biP-plot-CPT}
\end{figure}

With expression of the oscillation probabilities in (\ref{Pmue}) 
and (\ref{PmueCPT}), 
the intrinsic degeneracy solutions ($s_{i}$, $\delta_{i}$) (i=1, 2) 
in CPT-conjugate measurement 
are defined with $\nu_{\mu} \rightarrow \nu_e$ channel by 
\begin{eqnarray} 
P - Z &=& 
X_{\pm}s_{1}^2 + 
Y_{\pm} s_{1} \left( 
\cos \delta_{1} \cos \Delta_{31} \mp \sin \delta_{1} \sin \Delta_{31} 
\right), 
\nonumber \\
P - Z &=& 
X_{\pm}s_{2}^2 + 
Y_{\pm} s_{2} \left( 
\cos \delta_{2} \cos \Delta_{31} \mp \sin \delta_{2} \sin \Delta_{31} 
\right), 
\label{CPT-intrinsic-def1}
\end{eqnarray}
and in CPT-conjugate $\bar{\nu}_{e} \rightarrow  \bar{\nu}_{\mu}$ channel by 
\begin{eqnarray} 
P^{CPT} - Z &=& 
X_{\mp}s_{1}^2 - 
Y_{\mp} s_{1} \left( 
\cos \delta_{1} \cos \Delta_{31} \mp \sin \delta_{1} \sin \Delta_{31} 
\right), 
\nonumber \\
P^{CPT} - Z &=& 
X_{\mp}s_{2}^2 - 
Y_{\mp} s_{2} \left( 
\cos \delta_{2} \cos \Delta_{31} \mp \sin \delta_{2} \sin \Delta_{31} 
\right).  
\label{CPT-intrinsic-def2}
\end{eqnarray}
By subtracting two equations in (\ref{CPT-intrinsic-def1}) and 
(\ref{CPT-intrinsic-def2}) respectively, and subtracting and adding the resultant 
two equations we obtain, 
assuming that $C^{(+)} \neq 0$, 
\begin{eqnarray} 
S_{I \mp} = 0, 
\hspace{6mm}
2 (s_{1}^2 - s_{2}^2)  \pm C^{(-)} S_{I \mp} = 0. 
\label{CPT-intrinsic2}
\end{eqnarray}
where $C^{(\pm)}$ and $S_{I \pm}$ are defined in (\ref{Cpm-def}) and 
(\ref{SIpm-def}), respectively. 
We then obtain $s_{ \text{II} }= s_{1}$, that is, the intrinsic degeneracy solution of 
$\theta_{13}$ for CPT conjugate measurement is identical to the true one, 
in agreement with the expectation of the bi-probability plot. 
See Fig.~\ref{biP-plot-CPT}. 
It is obvious that the solution of the first equation (\ref{CPT-intrinsic2}) is given by 
\begin{eqnarray} 
\delta_{ \text{II} } = 2\pi- ( \delta_{1} \pm 2 \Delta_{31}) 
\hspace{6mm}
(\text{mod.}~2\pi).
\label{CPT-delta-intrinsic}
\end{eqnarray}
%

\subsection{The sign-$\Delta m^2$ degeneracy in CPT-conjugate measurement}
\label{CPT-signdm2}

We now discuss the sign-$\Delta m^2_{31}$ degeneracy in CPT-conjugate measurement. 
The true input solution ($s_{1}, \delta_{1}$) and the opposite $\Delta m^2_{31}$-sign 
clone solution ($s_{3}, \delta_{3}$) satisfy the following equations.  
In the $\nu_{\mu} \rightarrow \nu_{e}$ channel, 
\begin{eqnarray} 
P - Z &=& 
X_{\pm}s_{1}^2 + 
Y_{\pm} s_{1} \left( 
\cos \delta_{1} \cos \Delta_{31} \mp \sin \delta_{1} \sin \Delta_{31} 
\right), 
\nonumber \\
P - Z &=& 
X_{\mp}s_{3}^2 + 
Y_{\mp} s_{3} \left( 
\cos \delta_{3} \cos \Delta_{31} \pm \sin \delta_{3} \sin \Delta_{31} 
\right), 
\label{CPT-flipped-sign-def1}
\end{eqnarray}
and in CPT-conjugate channel 
\begin{eqnarray} 
P^{CPT} - Z &=& 
X_{\mp} s_{1}^2 - 
Y_{\mp} s_{1} \left( 
\cos \delta_{1} \cos \Delta_{31} \mp \sin \delta_{1} \sin \Delta_{31} 
\right), 
\nonumber \\
P^{CPT} - Z &=& 
X_{\pm} s_{3}^2 - 
Y_{\pm} s_{3} \left( 
\cos \delta_{3} \cos \Delta_{31} \pm \sin \delta_{3} \sin \Delta_{31} 
\right). 
\label{CPT-flipped-sign-def2}
\end{eqnarray}
By subtracting two equations in (\ref{CPT-flipped-sign-def1}) and 
(\ref{CPT-flipped-sign-def2}), respectively, and then subtracting and 
adding the resultant two equations, we obtain
\begin{eqnarray} 
\frac{T_{1 \pm}^{CPT} }{ C^{(+)} }
&-&  s_{3} \cos \delta_{3}  \cos \Delta_{31} 
\mp s_{3} \sin \delta_{3}  \sin \Delta_{31} = 0, 
\label{CPT-flipped-sign7} 
\\
\frac{ T_{2 \pm}^{CPT} - 2 s_{3}^2 } { C^{(-)}  }
&\pm& s_{3} \cos \delta_{3}  \cos \Delta_{31} 
+ s_{3} \sin \delta_{3}  \sin \Delta_{31} = 0. 
\label{CPT-flipped-sign8}
\end{eqnarray}
where we have defined 
\begin{eqnarray} 
T_{1 \pm}^{CPT} &\equiv& 
\pm E^{(-)} s_{1}^2  + 
D^{(+)} [ s_{1} \cos \delta_{1} \cos \Delta_{31} \mp s_{1} \sin \delta_{1} \sin \Delta_{31} ] 
\nonumber \\
T_{2 \pm}^{CPT} &\equiv& 
E^{(+)} s_{1}^2  \pm D^{(-)} 
[ s_{1} \cos \delta_{1} \cos \Delta_{31} \mp s_{1} \sin \delta_{1} \sin \Delta_{31} ] 
\label{T12-def}
\end{eqnarray}
where $C^{(\pm)}$ is defined in (\ref{Cpm-def}), while $D^{(\pm)}$ and $E^{(\pm)}$ 
are given in (\ref{DEpm-def}). 
%
From (\ref{CPT-flipped-sign7}) and (\ref{CPT-flipped-sign8}) it is straightforward 
to obtain the $s_{3}$ solution:
\begin{eqnarray} 
s_{ \text{III} } = s_{ \text{IV} } = \frac{1}{ \sqrt{2} } 
\sqrt{ T_{2 \pm}^{CPT} \pm 
\left( \frac{ C^{(-)} }{ C^{(+)} } \right) T_{1 \pm}^{CPT} }
\label{CPT-sign-s-solution}
\end{eqnarray}
where the $\pm$ sign is the hierarchy sign.  
Upon obtaining the $s_{3}$ solution one can readily obtain $\delta_{3}$ 
by solving (\ref{CPT-flipped-sign7}) for $\cos (\delta_{3} \mp \Delta_{31})$. 
The solutions read 
\begin{eqnarray} 
\delta_{ \text{III} } &=& \pm \Delta_{31} + \arccos \left( \frac{ T_{1 \pm}^{CPT} }{ C^{(+)} s_{ \text{III} } } \right) 
\hspace{5mm}
(\text{mod.} ~2\pi), 
\nonumber \\
\delta_{ \text{IV} } &=& 2\pi - \delta_{ \text{III} } \pm 2 \Delta_{31}. 
\label{CPT-sign-delta-solution}
\end{eqnarray}

\begin{figure}[bhtp]
\begin{center}
\includegraphics[bb=0 0 184 99 , clip, width=0.55\textwidth]{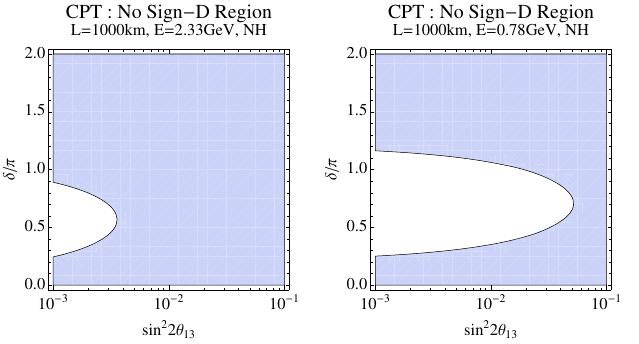}
\end{center}
\vglue -0.5cm
\caption{
Depicted as the shaded areas in the $\sin^2 2\theta_{13} - \delta/\pi$ space are 
the regions where no sign-$\Delta m^2_{31}$ degeneracy solution exists for 
CPT conjugate measurement for MB1 (left panel) and MB2 (right panel) settings. 
The true mass hierarchy is taken to be the normal one.
}
\label{no-solution-CPT}
\end{figure}

One might have suspected, from the feature of the bi-probability plot in 
Fig.~\ref{biP-plot-CPT}, that the sign-$\Delta m^2_{31}$ degeneracy 
solutions exist in a very limited region of small $\theta_{13}$. 
Therefore, we present in Fig.~\ref{no-solution-CPT} the region of no solution of the 
sign-$\Delta m^2_{31}$ degeneracy region by the shaded region. 

\subsection{The $\theta_{23}$ octant degeneracy in CPT-conjugate measurement}
\label{CPT-octant}

The $\theta_{23}$ octant degeneracy is defined by the following two sets of 
equations: 
\begin{eqnarray} 
P  &=& 
X^{\text{true}}_{\pm}s_{1}^2 + 
Y_{\pm} s_{1} \left( 
\cos \delta_{1} \cos \Delta_{31} \mp \sin \delta_{1} \sin \Delta_{31} 
\right) +
Z^{\text{true}}, 
\nonumber \\
P  &=& 
X^{\text{false}}_{\pm}s_{5}^2 + 
Y_{\pm} s_{5} \left( 
\cos \delta_{5} \cos \Delta_{31} \mp \sin \delta_{5} \sin \Delta_{31} 
\right) +
Z^{\text{false}}.  
\label{CPT-octant-def1}
\end{eqnarray}
%
%
\begin{eqnarray} 
P^{CPT}  &=& 
X^{\text{true}}_{\mp}s_{1}^2 - 
Y_{\mp} s_{1} \left( 
\cos \delta_{1} \cos \Delta_{31} \mp \sin \delta_{1} \sin \Delta_{31} 
\right) +
Z^{\text{true}}, 
\nonumber \\
P^{CPT}  &=& 
X^{\text{false}}_{\mp}s_{5}^2 - 
Y_{\mp} s_{5} \left( 
\cos \delta_{5} \cos \Delta_{31} \mp \sin \delta_{5} \sin \Delta_{31} 
\right) +
Z^{\text{false}}.
\label{CPT-octant-def2}
\end{eqnarray}
Following the similar procedure as before it is not difficult to obtain the 
equation which involve neither $\delta_{5}$ nor $\delta_{1}$. 
Then, by using (\ref{identity}) and (\ref{Xtrue-Xfalse}) we obtain 
\begin{eqnarray}
s_{\text{V}} = s_{\text{VI}} = 
\tan \theta_{23} \sqrt{ 
s_{1}^2 -  
\frac{Z^{\text{true} }  - Z^{\text{false} }  }{ \sqrt{X^{\text{true}}_{\pm} X^{\text{true}}_{\mp}} } }.
\label{CPT-octant-s-solution}
\end{eqnarray}
%
Then, the phase $\delta_{5}$ is determined as 
\begin{eqnarray}
s_{\text{V}} \cos( \delta_{\text{V}} \pm \Delta_{31} ) 
=
s_{1} \cos( \delta_{1} \pm \Delta_{31} ) 
+ \left( \frac{1}{Y_{\pm}} - \frac{1}{Y_{\mp}} \right) \left(Z^{\text{true}} - Z^{\text{false}} \right).
\label{CPT-octant-delta-solution5}
\end{eqnarray}
As in the previous section the intrinsic degeneracy partner $\delta_{\text{VI}}$ 
is given by using (\ref{CPT-delta-intrinsic}) as 
\begin{eqnarray} 
\delta_{ \text{VI} } = 2\pi- ( \delta_{V} \pm 2 \Delta_{31}) 
\hspace{6mm}
(\text{mod.}~2\pi). 
\label{CPT-octant-delta-solution6}
\end{eqnarray}

As in the previous cases, the solutions with octant as well as the 
$\Delta m^2_{31}$-sign flips, 
($s_{ \text{VII} }, \delta_{ \text{VII} }$) and ($s_{ \text{VII} }, \delta_{ \text{VII} }$), 
are given by the general argument as 
\begin{eqnarray} 
s_{ \text{VII} } &=& s_{ \text{VIII} } = 
\xi_{\pm}^{ \text{CPT} } \left( s_{ \text{V} }, \delta_{ \text{V} } \right)  
\nonumber \\
\delta_{ \text{VII} } &=& \eta_{\pm}^{ \text{CPT} } ( s_{ \text{V} }, \delta_{ \text{V} } ), 
\hspace{8mm}
\delta_{ \text{VIII} } = \eta_{\pm}^{ \text{CPT} } ( s_{ \text{VI} }, \delta_{ \text{VI} } ),
\label{CPT-oct-sign-s-delta-solution}
\end{eqnarray}
where $\xi_{\pm}^{ \text{CPT} }$ and $\eta_{\pm}^{ \text{CPT} }$ are defined in 
(\ref{CPT-sign-s-solution}) and the first line in (\ref{CPT-sign-delta-solution}), 
respectively, as a function of ($s_{1}, \delta_{1}$).

\section{Conclusion}
\label{conclusion}

In this paper, we have analyzed the problem of parameter degeneracy 
in various settings, CP-conjugate, T-conjugate, CPT-conjugate measurement, 
as well as combining the golden and the silver channels. 
Using the approximate form of the oscillation probabilities obtained by Cervera {\it et al.} 
we have derived, for the first time except for CP-conjugate setting, 
the exact analytic expressions of the eightfold degeneracy solutions in all these cases 
assuming $\theta_{23} \neq \pi/4$. 
We hope that the simple explicit expressions of the clone solutions nicely fill 
the ``hole'' of informations and help understand the nature of the degeneracy. 
Furthermore, they would prove to be useful if they can be implemented in an 
analysis codes such as \cite{Globes,M-cube} to facilitate the search for fake 
minima of the $\chi^2$. 
All in all, we expect that such solutions would help in correctly interpreting data 
to be taken in precision measurement in the future neutrino oscillation experiments.

We have presented a new view of the parameter degeneracy as invariance 
under the discrete mappings of the flavor mixing parameters including the 
mass hierarchies. 
The explicit forms of the mappings can be obtained by the symmetry respected 
by a pair of the oscillation probabilities alone, (\ref{transformation}) in 
Sec.~\ref{invariance}, for the sign-$\Delta m^2_{31}$ degeneracy in 
T-conjugate and the Golden-Silver measurement. 
In all the other cases, the explicit forms of the mappings are given by the 
analytic expressions of the degeneracy solutions. 
Nature of the degeneracy as the intrinsic degeneracy duplicated by 
the sign of $\Delta m^2_{31}$ and $\theta_{23}$ octant is now given its 
precise meaning as the mapping relations between each pair of degeneracy 
solutions given in (\ref{correspondence3}). 
The structure emerged, the one-to-one correspondence between the true and 
the degeneracy solutions which we call the solution network, is illustrated 
pictorially in Fig.~\ref{eightfold-relation}. 
We have also clarified the relationships between the degeneracy solutions 
for the given true mass hierarchies, normal or inverted, in Sec.~\ref{normal-inverted}.

The explicit analytic expressions of the eightfold degeneracy solutions are 
used to make plots of the difference between the true and the degeneracy 
solutions to give an overview of the degeneracy. 
The features of the degeneracy solutions are so profound, making the true 
overview of the degeneracy extremely difficult. 
However, we believe that we have illuminated some of the significant features 
by taking the three superbeam type settings as well as the one akin to neutrino factory.
In particular, the mild energy dependence of the difference between the true and 
the clone solutions indicate the robustness of the degeneracy against spectrum analysis. 
These plots illuminate which degeneracy is likely to be difficult to lift, hence it could be useful to design future experiments in preparation of the degeneracy to be 
met in the measurement. 
Such precise understanding of the parameter degeneracy would be a definitive 
help if future precision measurement could be contaminated by new effects 
outside of the standard three-flavor mixing of neutrinos.

\vspace{-0.3cm}
\begin{acknowledgments} 
  \vspace{-0.3cm} 
We thank Andrea Donini for the numerous useful informative correspondences. 
H.M. thanks Renata Zukanovich Funchal and Instituto de F\'{\i}sica, 
Universidade de S\~ao Paulo, for the hospitality extended to him 
during a visit Dec.~2009$-$Jan.~2010 where part of this work was carried out. 
This work has been supported in part by KAKENHI, Grant-in-Aid for
Scientific Research No.~19340062, and is supported by Grant-in-Aid for JSPS Fellows 
No.~209677, Japan Society for the Promotion of Science. 

\end{acknowledgments}


\appendix

\section{Matter Perturbation Theory of Sign-$\Delta m^2_{31}$ Degeneracy}
\label{matter-perturb}

Here, we present the approximate formulas of the sign-$\Delta m^2_{31}$ 
degeneracy solutions within the framework of matter perturbation theory 
\cite{AKS,MNprd98} 
which assumes $\frac{A}{\Delta_{31}} \ll 1$.\footnote{
It is known that the matter perturbation theory treatment of the parameter degeneracy 
gives rise to a transparent view of the degeneracy, which include e.g., 
decoupling between degeneracies \cite{resolve23,T2KK-2nd,NSI-perturbation}.
}
%
As can be seen in (\ref{A/Dm2}), the values of the ratio 
(assuming the matter density and $\Delta m^2_{31}$ referenced in the equation) 
are 
0.060, 0.066, and 0.20, respectively, for SB1, MB2, MB1 settings discussed in 
Sec.~\ref{overview}. 
Therefore, the condition for validity of matter perturbation theory holds in a good approximation for the former two settings. 
For MB1 setting the approximation may be modest but we may utilize it 
for a qualitative discussions. 

Having the analytic expressions of the sign-$\Delta m^2_{31}$ degeneracy solutions 
at hand (see Sec.~\ref{CP-signdm2}), it is straightforward to expand it in terms of 
the small parameter $\frac{A}{\Delta_{31}}$. 
We only present the results. 
To first order in $\frac{A}{\Delta_{31}}$ we obtain the following expressions. 
For $\theta_{13}$, 
\begin{eqnarray}
\sin^2 2\theta_{13}^{\text{III}}
&=&
\sin^2 2\theta_{13}^{\text{I}} 
\left[ 
1 +
\frac{A}{\Delta_{31}} \frac{ 4 \sin \delta_{1} (\Delta_{31} \cos \Delta_{31} - \sin \Delta_{31})
(2 s_{1} X_{\text{vac}} \cos \Delta_{31} \pm Y_{\text{vac}}  \cos \delta_{1})}
{ \sin^2 \Delta_{31} ( 2 s_{1} X_{\text{vac}}  \cos \delta_{1} \pm Y_{\text{vac}} \cos \Delta_{31} )}
\right] 
 \nonumber \\ 
\label{sign-s3-MP}
\end{eqnarray}
and for $\delta$, 
\begin{eqnarray} 
\cos \delta_{ \text{III} }
&=& - \cos \delta_{1} 
+ 4 \left( \frac{A}{\Delta_{31}} \right) 
\left[ \frac{\sin \delta_{1} (\Delta_{31} \cos \Delta_{31} - \sin \Delta_{31}) } {\sin^2 \Delta_{31} } 
\right] 
\nonumber \\
&& \hspace{12mm} \times 
\left[ 
\frac{ s_{1}^2 X_{\text{vac}} + Z_{\text{vac}} \cos (\delta_{1} - \Delta_{31}) \cos (\delta_{1} + \Delta_{31}) 
   \pm s_{1} Y_{\text{vac}} \cos \delta_{1} \cos \Delta_{31} }
   { \pm s_{1} Y_{\text{vac}}\cos \delta_{1}  + 2 Z_{\text{vac}}\cos \Delta_{31} } 
 \right], 
\nonumber \\
 \nonumber \\
\sin \delta_{ \text{III} }
&=& \sin \delta_{1} 
+ 4 \left( \frac{A}{\Delta_{31}} \right) 
\left[ 
 \frac{\cos \delta_{1} (\Delta_{31} \cos \Delta_{31} - \sin \Delta_{31}) } {\sin^2 \Delta_{31} } 
 \right] 
 \nonumber \\
&& \hspace{12mm} \times 
\left[ 
\frac{ s_{1}^2 X_{\text{vac}} + Z_{\text{vac}} \cos (\delta_{1} - \Delta_{31}) \cos (\delta_{1} + \Delta_{31}) 
   \pm s_{1} Y_{\text{vac}} \cos \delta_{1} \cos \Delta_{31} }
   { \pm s_{1} Y_{\text{vac}}\cos \delta_{1}  + 2 Z_{\text{vac}}\cos \Delta_{31} } 
\right], 
\nonumber \\
\label{sign-del3-MP}
\end{eqnarray}
where $X_{\text{vac}}$ etc. are defined in Sec.~\ref{vacuum}.

\section{Perturbation Theory of $\theta_{23}$ Octant Degeneracy}
\label{octant-perturb}

In order to understand features of $\theta_{23}$ octant degeneracy it is useful to 
have a perturbative framework assuming that deviation of $\theta_{23}$ 
from the maximal is small, 
\begin{eqnarray}
\theta_{23}=\frac{\pi}{4} + \epsilon_{\text{oct}} 
\hspace{8mm}
(\epsilon_{\text{oct}}  \ll 1). 
\label{th23-pert}
\end{eqnarray}
%
%
%
By expanding the octant degeneracy solution derived in Sec.~\ref{CP-octant} 
we obtain $\theta_{13}$ to first order in $\epsilon_{\text{oct}}$, as 
\begin{eqnarray}
&& \sin^2 2\theta_{13}^{\text{V} }
=
\sin^2 2\theta_{13}^{\text{I}}
( 1+ 4 \epsilon_{\text{oct}} ) \nonumber \\
&& + 16 \epsilon_{\text{oct}} Z \bigg[ 
\frac{ \pm 2 s_{1}^2 \sqrt{X_{\pm} X_{\mp}} \sin 2 \Delta_{31}  + 
s_{1} \{ Y_{\pm} \sin(\delta_{1} \pm \Delta_{31}) + Y_{\mp}  \sin(\delta_{1} \mp \Delta_{31}) \} }
{\sqrt{X_{\pm} X_{\mp}} \left[ 
s_{1} \{Y_{\pm} \sin(\delta_{1} \mp \Delta_{31})+Y_{\mp} \sin(\delta_{1} \pm \Delta_{31})  \} \mp 2 Z \sin 2\Delta_{31} \right] }
\bigg] 
\label{octant-s5-th23P} 
\end{eqnarray}
where $X_{\pm}$ etc. implies those evaluated at $\theta_{23} = \pi/4$. 
Similarly, we obtain for $\delta$
\begin{eqnarray}
&& \cos \delta_{\text{V}} = 
\cos \delta_{1} + 
2 \epsilon_{\text{oct}} \sin \delta_{1} \nonumber \\
&& \hspace{5mm} \times \left[
\frac{(s_{1}^2 X_{\mp} - Z) Y_{\pm}  \cos(\delta_{1} \pm \Delta_{31})
+ (s_{1}^2 X_{\pm} - Z) Y_{\mp} \cos(\delta_{1} \mp \Delta_{31})
- 2 s_{1} Z (X_{\pm} - X_{\mp})}
{ s_{1} \sqrt{ X_{\pm} X_{\mp} } 
\{ s_{1} Y_{\mp} \sin(\delta_{1} \pm \Delta_{31}) + s_{1} Y_{\pm} \sin(\delta_{1} \mp \Delta_{31})
 - 2 Z \sin 2\Delta_{31} \} } \right], \nonumber \\
 \nonumber \\
&& \sin \delta_{ \text{V} } = 
\sin \delta_{1} -
2 \epsilon_{\text{oct}} \cos \delta_{1} \nonumber \\
&& \hspace{5mm} \times \left[
\frac{(s_{1}^2 X_{\mp} - Z) Y_{\pm}  \cos(\delta_{1} \pm \Delta_{31})
+ (s_{1}^2 X_{\pm} - Z) Y_{\mp} \cos(\delta_{1} \mp \Delta_{31})
- 2 s_{1} Z (X_{\pm} - X_{\mp})}
{ s_{1} \sqrt{ X_{\pm} X_{\mp} } 
\{ s_{1} Y_{\mp} \sin(\delta_{1} \pm \Delta_{31}) + s_{1} Y_{\pm} \sin(\delta_{1} \mp \Delta_{31})
 - 2 Z \sin 2\Delta_{31} \} } \right].
 \nonumber \\
 \label{octant-delta5d-th23P}
\end{eqnarray}


\begin{thebibliography}{99}


\bibitem {MNS}
Z.~Maki, M.~Nakagawa and S.~Sakata,
Prog.\ Theor.\ Phys.\  {\bf 28}, 870 (1962).


\bibitem{atm}
T.~Kajita,
  New J.\ Phys.\  {\bf 6}, 194 (2004). 


\bibitem{solar}
A.~B.~McDonald,
New J.\ Phys.\  
  {\bf 6}, 121 (2004)
  [arXiv:astro-ph/0406253]. 
  

\bibitem{reactor} 
K.~Inoue,
New J.\ Phys.\  
{\bf 6}, 147 (2004).


\bibitem{K2K}
E.~Aliu {\it et al.}  [K2K Collaboration],
  Phys.\ Rev.\ Lett.\  {\bf 94}, 081802 (2005)
  [arXiv:hep-ex/0411038].
%
M.~H.~Ahn {\it et al.}  [K2K Collaboration],
  Phys.\ Rev.\  D {\bf 74}, 072003 (2006)
  [arXiv:hep-ex/0606032].


\bibitem{MINOS}
  P.~Adamson {\it et al.}  [MINOS Collaboration],
  Phys.\ Rev.\ Lett.\  {\bf 101}, 131802 (2008)
  [arXiv:0806.2237 [hep-ex]].


\bibitem {Kr2Det}
V.~Martemyanov, L.~Mikaelyan, V.~Sinev, V.~Kopeikin and Yu.~Kozlov,
  Phys.\ Atom.\ Nucl.\  {\bf 66}, 1934 (2003)
  [Yad.\ Fiz.\  {\bf 66}, 1982 (2003)]
  [arXiv:hep-ex/0211070].


\bibitem{MSYIS03}
H.~Minakata, H.~Sugiyama, O.~Yasuda, K.~Inoue and F.~Suekane,
  Phys.\ Rev.\  D {\bf 68}, 033017 (2003)
  [Erratum-ibid.\  D {\bf 70}, 059901 (2004)]
  [arXiv:hep-ph/0211111].


\bibitem {DCHOOZ} 
F.~Ardellier {\it et al.}  [Double Chooz Collaboration],
  arXiv:hep-ex/0606025; 

\bibitem {Daya-Bay} 
X.~Guo {\it et al.}  [Daya Bay Collaboration],
  arXiv:hep-ex/0701029; 

\bibitem {RENO} 
K.~K.~Joo  [RENO Collaboration],
  Nucl.\ Phys.\ Proc.\ Suppl.\  {\bf 168}, 125 (2007).

\bibitem {reactor-white} 
See also 
K.~Anderson {\it et al.},
  arXiv:hep-ex/0402041.


\bibitem {T2K}
Y.~Itow {\it et al.}, arXiv:hep-ex/0106019.\\
For an updated version, see:
http://neutrino.kek.jp/jhfnu/loi/loi.v2.030528.pdf


\bibitem {NOVA}
D.~Ayres {\it et al.}  [Nova Collaboration],
  arXiv:hep-ex/0503053. 


\bibitem {KM}
M. Kobayashi and T. Maskawa, Prog. Theor. Phys. 
{\bf 49}, 652 (1973).


\bibitem{KamLAND}
S.~Abe {\it et al.}  [KamLAND Collaboration],
  Phys.\ Rev.\ Lett.\  {\bf 100}, 221803 (2008)
  [arXiv:0801.4589 [hep-ex]].


\bibitem{SNO}
B.~Aharmim {\it et al.}  [SNO Collaboration],
  arXiv:0910.2984 [nucl-ex]. 


\bibitem{SKatm}
Y.~Ashie {\it et al.}  [Super-Kamiokande Collaboration],
  Phys.\ Rev.\ Lett.\  {\bf 93}, 101801 (2004)
  [arXiv:hep-ex/0404034].
%
Y.~Ashie {\it et al.}  [Super-Kamiokande Collaboration],
  Phys.\ Rev.\  D {\bf 71}, 112005 (2005)
  [arXiv:hep-ex/0501064].


\bibitem {CHOOZ}
M.~Apollonio {\it et al.} [CHOOZ Collaboration],
Eur.\ Phys.\ J.\ C {\bf 27}, 331 (2003)
[arXiv:hep-ex/0301017]; 
%
Phys.\ Lett.\ B {\bf 466}, 415 (1999)
[arXiv:hep-ex/9907037].
%

\bibitem {Palo-Verde}
The Palo Verde Collaboration,
F.~Boehm {\it et al.},
Phys.\ Rev.\ D {\bf 64}, 112001 (2001)
[arXiv:hep-ex/0107009]. 


\bibitem {K2K-bound}
M.~H.~Ahn {\it et al.} [K2K Collaboration],
Phys.\ Rev.\ Lett.\ {\bf 93}, 051801 (2004)
[arXiv:hep-ex/0402017].
%

\bibitem {MINOS-bound}
P.~Adamson {\it et al.}  [MINOS Collaboration],
  arXiv:0909.4996 [hep-ex].


\bibitem{intrinsic}
 J.~Burguet-Castell, M.~B.~Gavela, J.~J.~Gomez-Cadenas, P.~Hernandez and O.~Mena,
  Nucl.\ Phys.\  B {\bf 608}, 301 (2001)
  [arXiv:hep-ph/0103258].


\bibitem{MNjhep01}
 H.~Minakata and H.~Nunokawa,
  JHEP {\bf 0110}, 001 (2001)
  [arXiv:hep-ph/0108085].


\bibitem{octant}
G.~L.~Fogli and E.~Lisi,
  Phys.\ Rev.\  D {\bf 54}, 3667 (1996)
  [arXiv:hep-ph/9604415].


\bibitem{MNtaup01}
H.~Minakata and H.~Nunokawa,
  Nucl.\ Phys.\ Proc.\ Suppl.\  {\bf 110}, 404 (2002)
  [arXiv:hep-ph/0111131].


\bibitem{BMW02} 
V.~Barger, D.~Marfatia and K.~Whisnant,
  Phys.\ Rev.\  D {\bf 65}, 073023 (2002)
  [arXiv:hep-ph/0112119].


\bibitem{KMN02}
  T.~Kajita, H.~Minakata and H.~Nunokawa,
  Phys.\ Lett.\  B {\bf 528}, 245 (2002)
  [arXiv:hep-ph/0112345].


\bibitem{MNP2}
H.~Minakata, H.~Nunokawa and S.~J.~Parke,
  Phys.\ Rev.\  D {\bf 66}, 093012 (2002)
  [arXiv:hep-ph/0208163].


\bibitem{golden}
A.~Cervera, A.~Donini, M.~B.~Gavela, J.~J.~Gomez Cadenas, P.~Hernandez, O.~Mena and S.~Rigolin,
  Nucl.\ Phys.\  B {\bf 579}, 17 (2000)
  [Erratum-ibid.\  B {\bf 593}, 731 (2001)]
  [arXiv:hep-ph/0002108].
  
  
\bibitem{donini03}
A.~Donini, D.~Meloni and S.~Rigolin,
  JHEP {\bf 0406}, 011 (2004)
  [arXiv:hep-ph/0312072].


\bibitem{wolfenstein}
L.~Wolfenstein,
Phys.\ Rev.\ D {\bf 17}, 2369 (1978).


\bibitem{valle}
  J.~W.~F.~Valle,
  Phys.\ Lett.\  B {\bf 199} (1987) 432.


\bibitem{petcov}
  M.~M.~Guzzo, A.~Masiero and S.~T.~Petcov,
  Phys.\ Lett.\  B {\bf 260}, 154 (1991).
%

\bibitem{grossman}
Y.~Grossman,
  Phys.\ Lett.\  B {\bf 359}, 141 (1995)
  [arXiv:hep-ph/9507344].


\bibitem{berezhiani}
Z.~Berezhiani and A.~Rossi,
  Phys.\ Lett.\  B {\bf 535}, 207 (2002)
  [arXiv:hep-ph/0111137].


\bibitem{NOVE09-mina}
H.~Minakata,
  arXiv:0905.1387 [hep-ph].


\bibitem{NSI-perturbation}
  T.~Kikuchi, H.~Minakata and S.~Uchinami,
  JHEP {\bf 0903}, 114 (2009)
  [arXiv:0809.3312 [hep-ph]].


\bibitem{NSI-2phase}
  A.~M.~Gago, H.~Minakata, H.~Nunokawa, S.~Uchinami and R.~Zukanovich Funchal,
  JHEP {\bf 1001}, 049 (2010)
  [arXiv:0904.3360 [hep-ph]].


\bibitem{burguetC02}
J.~Burguet-Castell, M.~B.~Gavela, J.~J.~Gomez-Cadenas, P.~Hernandez and O.~Mena,
  Nucl.\ Phys.\  B {\bf 646}, 301 (2002)
  [arXiv:hep-ph/0207080].


\bibitem{huber02}
P.~Huber, M.~Lindner and W.~Winter,
  Nucl.\ Phys.\  B {\bf 645}, 3 (2002)
  [arXiv:hep-ph/0204352].
 
 
\bibitem{huber03}
P.~Huber, M.~Lindner and W.~Winter,
  Nucl.\ Phys.\  B {\bf 654}, 3 (2003)
  [arXiv:hep-ph/0211300].


\bibitem{huber05}
  P.~Huber, M.~Lindner and W.~Winter,
  JHEP {\bf 0505}, 020 (2005)
  [arXiv:hep-ph/0412199].


\bibitem{donini-nufact03}
A.~Donini,
  AIP Conf.\ Proc.\  {\bf 721}, 219 (2004)
  [arXiv:hep-ph/0310014].


\bibitem{autiero04}
  D.~Autiero {\it et al.},
  Eur.\ Phys.\ J.\  C {\bf 33}, 243 (2004)
  [arXiv:hep-ph/0305185].


\bibitem{donini04}
  A.~Donini, E.~Fernandez-Martinez, P.~Migliozzi, S.~Rigolin and L.~Scotto Lavina,
  Nucl.\ Phys.\  B {\bf 710}, 402 (2005)
  [arXiv:hep-ph/0406132].


\bibitem{burguetC04}
 J.~Burguet-Castell, D.~Casper, J.~J.~Gomez-Cadenas, P.~Hernandez and F.~Sanchez, 
 Nucl.\ Phys.\ B {\bf 695}, 217 (2004) [arXiv:hep-ph/0312068].


\bibitem{mena04} 
O.~Mena and S.~J.~Parke,
  Phys.\ Rev.\  D {\bf 70}, 093011 (2004)
  [arXiv:hep-ph/0408070].


\bibitem{mena05} 
O.~Mena Requejo, S.~Palomares-Ruiz and S.~Pascoli,
  Phys.\ Rev.\  D {\bf 72}, 053002 (2005)
  [arXiv:hep-ph/0504015].


\bibitem{mena06} 
O.~Mena, S.~Palomares-Ruiz and S.~Pascoli,
  Phys.\ Rev.\  D {\bf 73}, 073007 (2006)
  [arXiv:hep-ph/0510182].


\bibitem{MEMPHYS}
J.~E.~Campagne, M.~Maltoni, M.~Mezzetto and T.~Schwetz,
  JHEP {\bf 0704}, 003 (2007)
  [arXiv:hep-ph/0603172].


\bibitem{BNL}
D.~Beavis {\it et al.},
  arXiv:hep-ex/0205040;
%
M.~V.~Diwan {\it et al.},
  Phys.\ Rev.\ D {\bf 68}, 012002 (2003)
  [arXiv:hep-ph/0303081].


\bibitem{T2KK-1st}
M.~Ishitsuka, T.~Kajita, H.~Minakata and H.~Nunokawa,
  Phys.\ Rev.\  D {\bf 72}, 033003 (2005)
  [arXiv:hep-ph/0504026].


\bibitem{T2KK-2nd}
  T.~Kajita, H.~Minakata, S.~Nakayama and H.~Nunokawa,
  Phys.\ Rev.\  D {\bf 75}, 013006 (2007)
  [arXiv:hep-ph/0609286].


\bibitem{nufact-lowE1}
 S.~Geer, O.~Mena and S.~Pascoli,
  Phys.\ Rev.\  D {\bf 75}, 093001 (2007)
  [arXiv:hep-ph/0701258].


\bibitem{nufact-lowE2}
  A.~D.~Bross, M.~Ellis, S.~Geer, O.~Mena and S.~Pascoli,
  Phys.\ Rev.\  D {\bf 77}, 093012 (2008)
  [arXiv:0709.3889 [hep-ph]].


\bibitem{BNL-fermi}
V.~Barger, M.~Dierckxsens, M.~Diwan, P.~Huber, C.~Lewis, D.~Marfatia and B.~Viren,
  Phys.\ Rev.\  D {\bf 74}, 073004 (2006)
  [arXiv:hep-ph/0607177].


\bibitem{resolve23}
K.~Hiraide, H.~Minakata, T.~Nakaya, H.~Nunokawa, H.~Sugiyama, W.~J.~C.~Teves and R.~Zukanovich Funchal,
  Phys.\ Rev.\  D {\bf 73}, 093008 (2006)
  [arXiv:hep-ph/0601258].


\bibitem{peres-smi23}
O.~L.~G.~Peres and A.~Y.~Smirnov,
  Phys.\ Lett.\ B {\bf 456}, 204 (1999)
  [arXiv:hep-ph/9902312];
%
  Nucl.\ Phys.\ B {\bf 680}, 479 (2004)
  [arXiv:hep-ph/0309312];


\bibitem{concha23}
 M.~C.~Gonzalez-Garcia, M.~Maltoni and A.~Y.~Smirnov,
  Phys.\ Rev.\ D {\bf 70}, 093005 (2004)
  [arXiv:hep-ph/0408170].

  
\bibitem{choubey23}
S.~Choubey and P.~Roy,
  Phys.\ Rev.\ D {\bf 73}, 013006 (2006)
  [arXiv:hep-ph/0509197].


\bibitem{kajita-atm23}
M.~Shiozawa, T.~Kajita, S.~Nakayama, Y.~Obayashi, and 
K.~Okumura, 
in Proceedings of the RCCN International Workshop
on Sub-dominant Oscillation Effects in Atmospheric Neutrino
Experiments, Kashiwa, Japan, Dec. 2004, p.57;
%
T.~Kajita,
  Nucl.\ Phys.\ Proc.\ Suppl.\  {\bf 155}, 87 (2006).


\bibitem{huber23}
  P.~Huber, M.~Maltoni and T.~Schwetz,
  Phys.\ Rev.\  D {\bf 71}, 053006 (2005)
  [arXiv:hep-ph/0501037].


\bibitem{meloni08}
  D.~Meloni,
  Phys.\ Lett.\  B {\bf 664}, 279 (2008)
  [arXiv:0802.0086 [hep-ph]].


\bibitem{munich04}
E.~K.~Akhmedov, R.~Johansson, M.~Lindner, T.~Ohlsson and T.~Schwetz,
JHEP {\bf 0404}, 078 (2004)
[arXiv:hep-ph/0402175].



\bibitem {MNplb00}
H.~Minakata and H.~Nunokawa,
  Phys.\ Lett.\  B {\bf 495}, 369 (2000)
  [arXiv:hep-ph/0004114].


\bibitem {sato}
J.~Sato,
  Nucl.\ Instrum.\ Meth.\  A {\bf 472}, 434 (2001)
  [arXiv:hep-ph/0008056].


\bibitem {richter}
B.~Richter, arXiv:hep-ph/0008222.


\bibitem {Geer}
S.~Geer,
  Phys.\ Rev.\  D {\bf 57}, 6989 (1998)
  [Erratum-ibid.\  D {\bf 59}, 039903 (1999)]
  [arXiv:hep-ph/9712290]; 


\bibitem {De-Rujula}
A.~De Rujula, M.~B.~Gavela and P.~Hernandez,
  Nucl.\ Phys.\  B {\bf 547}, 21 (1999)
  [arXiv:hep-ph/9811390].


\bibitem{silver}
  A.~Donini, D.~Meloni and P.~Migliozzi,
  Nucl.\ Phys.\  B {\bf 646}, 321 (2002)
  [arXiv:hep-ph/0206034].


\bibitem{MNP1}
  H.~Minakata, H.~Nunokawa and S.~J.~Parke,
  Phys.\ Lett.\  B {\bf 537}, 249 (2002)
  [arXiv:hep-ph/0204171].


\bibitem{large-theta-P}
H.~Minakata,
Acta Phys.\ Polon.\  B {\bf 40}, 3023 (2009)
  [arXiv:0910.5545 [hep-ph]].


\bibitem{beta1}
P.~Zucchelli,
Phys.\ Lett.\ B {\bf 532}, 166 (2002).
%

\bibitem{beta2}
J.~Bouchez, M.~Lindroos and M.~Mezzetto,
  AIP Conf.\ Proc.\  {\bf 721}, 37 (2004)
  [arXiv:hep-ex/0310059].


\bibitem{MNnufact01}
H.~Minakata and H.~Nunokawa,
  Nucl.\ Instrum.\ Meth.\  A {\bf 503}, 218 (2001)
  [arXiv:hep-ph/0111130].


\bibitem{AKS}
J.~Arafune, M.~Koike and J.~Sato,
  Phys.\ Rev.\  D {\bf 56}, 3093 (1997)
  [Erratum-ibid.\  D {\bf 60}, 119905 (1999)]
  [arXiv:hep-ph/9703351].


\bibitem{MNprd98} 
H.~Minakata and H.~Nunokawa,
  Phys.\ Rev.\  D {\bf 57}, 4403 (1998)
  [arXiv:hep-ph/9705208].


\bibitem{uchinami-thesis}
S.~Uchinami,
 Dr. of Science Thesis, Tokyo Metropolitan University, online at\\
 http://musashi.phys.metro-u.ac.jp/PhD-underscore-Uchinami.pdf



\bibitem{kobayashi}
T.~Kobayashi, 
talk given at 8th TOKUTEI-RCCN Workshop on Neutrinos, November 9, Institute for Cosmic Ray Research, Chiba, Japan (2001), 
http://www-rccn.icrr.u-tokyo.ac.jp/nu-meeting/08/04-Kobayashi.pdf, 
and private communications.



\bibitem {MNplb97}
  H.~Minakata and H.~Nunokawa,
  Phys.\ Lett.\  B {\bf 413}, 369 (1997)
  [arXiv:hep-ph/9706281].


\bibitem{MU-version1}
  H.~Minakata and S.~Uchinami,
  arXiv:1001.4219v1 [hep-ph].


\bibitem{ISS-nufact}
A.~Bandyopadhyay {\it et al.}  [ISS Physics Working Group],
  Rept.\ Prog.\ Phys.\  {\bf 72}, 106201 (2009)
  [arXiv:0710.4947 [hep-ph]].


\bibitem{huber-winter}
 P.~Huber and W.~Winter,
  Phys.\ Rev.\  D {\bf 68}, 037301 (2003)
  [arXiv:hep-ph/0301257].


\bibitem{NSI-nufact} 
N.~Cipriano Ribeiro, H.~Minakata, H.~Nunokawa,
  S.~Uchinami and R.~Zukanovich Funchal,
  JHEP {\bf 0712}, 002 (2007)
  [arXiv:0709.1980 [hep-ph]].


\bibitem{kopp3}
J.~Kopp, T.~Ota and W.~Winter,
  Phys.\ Rev.\  D {\bf 78}, 053007 (2008)
  [arXiv:0804.2261 [hep-ph]].


\bibitem{ISS-detector}
  T.~Abe {\it et al.}  [ISS Detector Working Group],
  JINST {\bf 4}, T05001 (2009)
  [arXiv:0712.4129 [physics.ins-det]].


\bibitem{MNPZ07}
  H.~Minakata, H.~Nunokawa, S.~J.~Parke and R.~Zukanovich Funchal,
  Phys.\ Rev.\  D {\bf 76}, 053004 (2007)
  [Erratum-ibid.\  D {\bf 76}, 079901 (2007)]
  [arXiv:hep-ph/0701151].


\bibitem{MNP3}
H.~Minakata, H.~Nunokawa and S.~J.~Parke,
  Phys.\ Rev.\  D {\bf 68}, 013010 (2003)
  [arXiv:hep-ph/0301210].


\bibitem{parke-beta}
A.~Jansson, O.~Mena, S.~J.~Parke and N.~Saoulidou,
  Phys.\ Rev.\  D {\bf 78}, 053002 (2008)
  [arXiv:0711.1075 [hep-ph]].



\bibitem{Globes}
  P.~Huber, M.~Lindner and W.~Winter,
  Comput.\ Phys.\ Commun.\  {\bf 167}, 195 (2005)
  [arXiv:hep-ph/0407333].


\bibitem{M-cube}
  M.~Blennow and E.~Fernandez-Martinez,
  Comput.\ Phys.\ Commun.\  {\bf 181}, 227 (2010)
  [arXiv:0903.3985 [hep-ph]].


\end{thebibliography}
\end{document}